\documentclass[reprint,amsmath,amssymb, aps, pra]{revtex4-1}
\usepackage{graphicx}
\usepackage{epsfig}
\usepackage{color}
\usepackage{amsmath}
\usepackage{amsfonts}
\usepackage{mathrsfs}
\usepackage{txfonts}
\usepackage{color}
\usepackage{multirow}

\usepackage{hyperref}

\newcommand{\bbr}{{\mathbf{r}}}

\newcommand{\bse}{{\boldsymbol{e}}}
\newcommand{\bsg}{{\boldsymbol{g}}}
\newcommand{\bbk}{{\boldsymbol{k}}}

\newcommand{\bea}{\begin{eqnarray}}
\newcommand{\eea}{\end{eqnarray}}
\def\beq{\begin{equation}}
\def\eeq{\end{equation}}

\newcommand{\duuu}{f_{i\downarrow}^\dagger f_{j\uparrow}^\dagger f_{j\uparrow} f_{i\uparrow} }
\newcommand{\uddd}{f_{i\uparrow}^\dagger f_{j\downarrow}^\dagger f_{j\downarrow} f_{i\downarrow} }
\newcommand{\dddu}{f_{i\downarrow}^\dagger f_{j\downarrow}^\dagger f_{j\downarrow} f_{i\uparrow} }
\newcommand{\uuud}{f_{i\uparrow}^\dagger f_{j\uparrow}^\dagger f_{j\uparrow} f_{i\downarrow} }

\newcommand{\uduu}{f_{i\uparrow}^\dagger f_{j\downarrow}^\dagger f_{j\uparrow} f_{i\uparrow} }
\newcommand{\dudd}{f_{i\downarrow}^\dagger f_{j\uparrow}^\dagger f_{j\downarrow} f_{i\downarrow} }
\newcommand{\uudu}{f_{i\uparrow}^\dagger f_{j\uparrow}^\dagger f_{j\downarrow} f_{i\uparrow} }
\newcommand{\ddud}{f_{i\downarrow}^\dagger f_{j\downarrow}^\dagger f_{j\uparrow} f_{i\downarrow} }

\newcommand{\uudd}{f_{i\uparrow}^\dagger f_{j\uparrow}^\dagger f_{j\downarrow} f_{i\downarrow} }
\newcommand{\dduu}{f_{i\downarrow}^\dagger f_{j\downarrow}^\dagger f_{j\uparrow} f_{i\uparrow} }

\begin{document}
\title{Interplay of non-symmorphic symmetry and spin-orbit coupling in hyperkagome spin liquids:\\ Applications to Na$_4$Ir$_3$O$_8$}
\author{ Biao Huang$^{1,5}$, Yong Baek Kim$^{2,3,4}$, Yuan-Ming Lu$^1$}
\affiliation{$^1$Department of Physics, The Ohio State University, Columbus, OH 43210, USA\\
$^2$Department of Physics and Centre for Quantum Materials,
University of Toronto, Toronto, Ontario M5S 1A7, Canada\\
$^3$Canadian Institute for Advanced Research/Quantum Materials Program, Toronto, Ontario MSG 1Z8, Canada\\
$^4$School of Physics, Korea Institute for Advanced Study, Seoul 130-722, Korea\\
$^5$Department of Physics and Astronomy, University of Pittsburgh, Pittsburgh PA 15260, USA
}
\date{\today}
\begin{abstract}
Na$_4$Ir$_3$O$_8$ provides a material platform to study three-dimensional quantum spin liquids in the geometrically frustrated hyperkagome lattice of Ir$^{4+}$ ions. In this work, we consider quantum spin liquids on hyperkagome lattice for generic spin models, focusing on the effects of anisotropic spin interactions. In particular, we classify possible $\mathbb{Z}_2$ and $U(1)$ spin liquid states, following the projective symmetry group analysis in the slave-fermion representation. There are only three distinct $\mathbb{Z}_2$ spin liquids, together with 2 different $U(1)$ spin liquids. The non-symmorphic space group symmetry of hyperkagome lattice plays a vital role in simplifying the classification, forbidding ``$\pi$-flux'' or ``staggered-flux'' phases in contrast to symmorphic space groups. We further prove that both $U(1)$ states and one $Z_2$ state among all 3 are symmetry-protected gapless spin liquids, robust against any symmetry-preserving perturbations. Motivated by the ``spin-freezing'' behavior recently observed in Na$_4$Ir$_3$O$_8$ at low temperatures, we further investigate the nearest-neighbor spin model with dominant Heisenberg interaction subject to all possible anisotropic perturbations from spin-orbit couplings. We found a $U(1)$ spin liquid ground state with spinon fermi surfaces is energetically favored over $Z_2$ states. Among all spin-orbit coupling terms, we show that only Dzyaloshinskii-Moriya (DM) interaction can induce spin anisotropy in the ground state when perturbing from the isotropic Heisenberg limit. Our work paves the way for a systematic study of quantum spin liquids in various materials with a hyperkagome crystal structure.
\end{abstract}
\maketitle


\section{Introduction}
Quantum spin liquids are exotic states of matter that defy the traditional Landau's paradigm of symmetry breaking. They are featured by a disordered ground state that evades ordering down to zero temperature, and fractionalized excitations on top of it\cite{Balents2010,Lee2014a,Savary2016,Zhou2016}. In parallel to rapid theoretical progress in understanding these quantum phases, more and more candidate spin liquid materials were discovered and extensively studied thanks to advancements in material synthesis and experimental characterization techniques. Most of these spin liquid materials contain layered (quasi-)two-dimensional lattices of magnetic moments, while Na$_4$Ir$_3$O$_8$ was found to be one rare candidate of spin liquid materials featuring a three-dimensional hyperkagome lattice spanned by the Ir$^{4+}$ moments\cite{Okamoto2007,Singh2013}. Measurements of thermodynamic quantities point to a high frustration ratio, where the Curie-Weiss temperature $\Theta_{CW}\approx -650K$ despite no magnetic orders observed down to a few Kelvin. Its experimental discovery had prompted a series of theoretical studies of possible spin liquid states in this system\cite{Hopkinson2007,Lawler2008a,Zhou2008a,Lawler2008,Chen2013c,Singh2012,Buessen2016}.

Early works of quantum spin liquids in Na$_4$Ir$_3$O$_8$ mainly focus on the spin-isotropic Heisenberg model due to the following observation: when strong spin-orbit coupling of 5$d$ electrons in Ir outweights the crystal field, the resulting model consists of an effective $J_{\mbox{\scriptsize eff}}=1/2$ moment on each Ir ion, with dominant Heisenberg interactions over various small anisotropic interactions\cite{Chen2008}. Several studies on isotropic spin liquids in the slave-fermion representation finds agreement with thermodynamic measurements at higher temperatures on different aspects\cite{Zhou2008a,Lawler2008}.

More recently, two experimental studies of muon spin relaxation ($\mu$SR)\cite{Dally2014} and nuclear magnetic resonance (NMR)\cite{Shockley2015} discovered a new ``spin freezing'' behavior in Na$_4$Ir$_3$O$_8$ below $6\sim 7 K$, which may underscore the spin-anisotropic effects at low temperatures. This is characterized in the $\mu$SR experiment \cite{Dally2014} by a much larger magnetization in the field-cooled (FC) process compared with that in the zero-field-cooled (ZFC) process below 6$K$. Also, in the NMR experiment \cite{Shockley2015}, the spin relaxation rate $1/T_1$ drops by orders of magnitude below $6K$.  The origin of such spin freezing still begs further investigation: while the $\mu$SR results were interpreted as the evidence of a short-range ordered magnetic ground state at low temperature\cite{Dally2014}, the later NMR data contrasted such an ordering\cite{Shockley2015}, and indicated a disordered paramagnetic phase. But it is clear from both experiments that below $6\sim7$K, the spin dynamics of the system drastically slows down. Therefore, a microscopic mechanism that breaks the continuous spin-rotational symmetry, such as spin-orbit coupling, may hamper the spin relaxation and play vital roles in the spin freezing phenomena. As such, several recent works have studied the anisotropic effects by mapping out the classical magnetic phase diagrams\cite{Shindou2016,Mizoguchi2016} in the presence of anisotropic interactions. In this work, we will focus on the effects of quantum fluctuations and study quantum spin liquids with spin anisotropy\cite{You2012a,Shindou2011,Dodds2013}.

In addition to the progress of experiments, the hyperkagome lattice also draws theoretical attentions due to its nonsymmorphic space group symmetry --- a space group operation combining translation by a fraction of the unit-cell and a point group operation, such as screw or glide operations. Recent works\cite{Parameswaran2013,Watanabe2015} found that a non-symmorphic symmetry forbids systems with certain integer filling numbers to be an insulator, and therefore provide a road-map for the search of symmetry protected semimetals. For example the space group $P4_132$ of hyperkagome lattice features a 4-fold screw operation, and can only be a band insulator if the number of electrons per unit cell is a multiple of 8 due to time reversal symmetry. It is therefore intriguing to ask: does non-symmorphic space group\cite{Lee2016} provide any constraints on a gapped quantum spin liquid ground state? While $U(1)$ spin liquids are described by Gutzwiller-projected metals/insulators of spinons with a conserved ``spinon filling number'', $Z_2$ spin liquids correspond to projected superconductors of spinons with no number conservation. In this work we extend previous results on symmetry criteria for gapless $Z_2$ spin liquids\cite{Lu2016b} to the case with no spin conservation (due to spin-orbit couplings), addressing the stability of gapless spectrum in symmetric $Z_2$ spin liquids on hyperkagome lattice.

%
%

In view of these developments, we perform a systematic classification of symmetric $\mathbb{Z}_2$ and $U(1)$ spin liquids on hyperkagome lattice, using the projective symmetry group (PSG) analysis\cite{Wen2002} in the slave-fermion representation. The results are summarized in Table \ref{tab:psg} - \ref{tab:mftriplet}. This classification categorizes possible candidate wavefunctions of spin liquid states for a generic spin model on hyperkagome lattice, with or without spin-orbit couplings. We show that there are only 3 distinct $\mathbb{Z}_2$ spin liquids and 2 different $U(1)$ spin liquids compatible with spin-$1/2$ moments per site on the hyperkagome lattice. We show that among all three $Z_2$ spin liquids 2(a,b,c) in TABLE \ref{tab:psg}, only one state 2(b) has a gapless spectrum stable against any symmetry-preserving local perturbations. Meanwhile both $U(1)$ spin liquids have a robust gapless spectrum protected by space group and time reversal symmetries: $U1^0$ state features spinon fermi surfaces while $U1^1$ state hosts spinon Dirac cones at low energy. These results hold with or without global spin rotational symmetries, i.e. they apply to all cases irrespective of spin-orbit couplings.
We emphasize that the classification is not limited to a specific spin model, and only requires space group and time reversal symmetries of the hyperkagome lattice. Therefore, it can be used to analyze other candidate spin liquid materials, such as PbCuTe$_2$O$_6$\cite{Koteswararao2014,Khuntia2016}, where local spin moments also form the hyperkagome lattice. Moreover these results can help us narrow down the promising spin liquid candidates for real materials, given the experimental evidences for gapless excitations in both Na$_4$Ir$_3$O$_8$\cite{Okamoto2007,Singh2008} and PbCuTe$_2$O$_6$\cite{Khuntia2016}.

We further apply the above classification to study the possible spin liquid ground state in Na$_4$Ir$_3$O$_8$ with considerable spin-orbit couplings. Motivated by the $\mu$SR and NMR experiments mentioned earlier, we pay special attention to the effects of spin-anisotropic interactions, and investigate whether they induce spin-triplet couplings in the spin liquid state. For this purpose, we focus on a  nearest-neighbor spin model that reflects the experimental signatures of Na$_4$Ir$_3$O$_8$. As pointed out previously in Ref.\cite{Mizoguchi2016}, spin-anisotropic effects only become notable below $T\sim 6$K in contrast to a large Heisenberg interaction $J\sim 300$K. Thus, we mainly deal with small anisotropic interaction as perturbations to the dominant Heisenberg term; the anisotropic perturbations are allowed to take all possible forms constrained only by the crystal symmetry. How does these terms from spin-orbit couplings\cite{You2012a,Reuther2014} affect the properties of a spin liquid ground state? A mean-field calculation shows that even in the presence of anisotropic interactions, a $U(1)$ spin liquid without spinon pairing is favored over their neighboring $Z_2$ states, reminiscent of the isotropic limit where only Heisenberg interactions are present\cite{Lawler2008}. In particular, a gapless $U(1)$ spin liquid ($U1^0$ state) with fermi surfaces has lower energy than the Dirac spin liquid $U1^1$ state. Furthermore, most anisotropic interactions do not induce spin anisotropy in the ground state unless their strength is comparable to the Heisenberg interaction. In contrary, even an infinitesimal Dzyaloshinskii-Moriya (DM) interaction can induce spin anisotropy in the ground state, for a general reason lying in the mean-field energy functional. Therefore, we conclude that DM interaction is the most important factor to induce a spin-anisotropic spin liquid ground state in Na$_4$Ir$_3$O$_8$. We also show the distortion and spin textures of spinon Fermi surfaces induced by the DM interactions as compared to the isotropic spin liquid.

The rest of this paper is organized as follows. In section \ref{sec:classification} we classify distinct $Z_2$ and $U(1)$ spin liquids compatible with space group and time reversal symmetries of hyperkagome lattices, following PSG formalism\cite{Wen2002} in the slave-fermion representation. In section \ref{sec:z2sl} we analyze physical properties of these spin liquid states, in particular addressing the stability of gapless excitations in these states. In section \ref{sec:modelandMF} we discuss mean-field energetics of these spin liquid states for a generic nearest-neighbour spin model of Na$_4$Ir$_3$O$_8$, focusing on spin anisotropy in the spin liquid ground state induced by spin-orbit couplings. Finally we conclude in section \ref{sec:summary}.

We also briefly outline the contents of the Appendices. In Appendix \ref{app:group extension} we describe the space group symmetry and compute its $Z_2$ extension in regard to the classification of $Z_2$ spin liquids. Next we classify symmetric $Z_2$ (see Appendix \ref{app:z2 class}) and $U(1)$ (see Appendix \ref{app:u1 class}) spin liquids on hyperkagome lattice, and study their mean-field ansatz in Appendix \ref{app:mf ansatz}.

\begin{widetext}

\begin{figure}[h]
\begin{center}
\includegraphics[width=5.5cm]{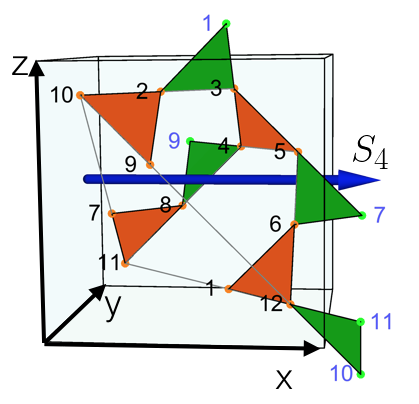}
\includegraphics[width=5cm]{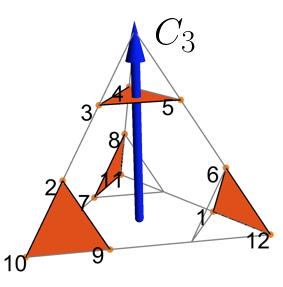}
\includegraphics[width=6cm]{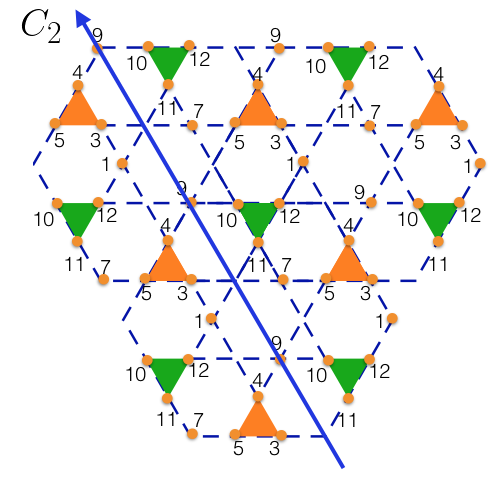}
\end{center}
\caption{\label{fig:sublattice}
Structure and symmetry of the hyperkagome lattice.
{\bf Left:} A cubic unit cell of hyperkagome lattice consists of 12 sublattices, where the sublattice displacements are shown in Table \ref{tab:sublattice}. There are 24 non-equivalent nearest neighbour bonds forming  8 corner-shared triangles per unit cell. Those staying within a unit cell are colored in orange while those connecting nearby unit cells are colored in green. The relative positions of nearest neighbour bonds are specified in Table \ref{tab:nn}. The four-fold  screw symmetry $S_4$ axis is parallel to $x$-axis and rotates each triangles to the neighbouring corner-shared ones along $x$ (up to translation by unit-cell). For instance, $(10,9, 2)\rightarrow (2, 3,1)\rightarrow (1, 6, 12) \rightarrow (12, 11, 10) \rightarrow (10, 9, 2)$, and similarly $(7,11,8)\rightarrow (8,9,4)\rightarrow(4,3,5) \rightarrow (5,6,7) \rightarrow (7,11,8)$. {\bf Middle:}  12 sublattices $s=1,\dots,12$ occupy corners of the three side-faces and the top of a truncated tetrahedron.  The $C_3$ rotation axis passes through the center of the tetrahedron, and is also the $(1,1,1)$ diagonal line of the cubic unit cell.    {\bf Right:} The triangles form four sets of non-parallel planes normal to the four diagonal lines of the cubic unit cell. Each plane has the shape of partially filled kagom\'{e} lattice. Here we show one of the four planes with (in our convention) the $C_2$ axis lying on it. Blue dashed lines are guides to the eyes showing the symmetry. Occupied sites are denoted by orange dots with site indices labeled. The $C_2$ axis passes through sublattice site 9.
}
\end{figure}
\end{widetext}

\section{Classifying symmetric spin liquids on hyperkagome lattice}\label{sec:classification}

\subsection{Crystal structure of hyperkagome lattice\label{sec:symm}}

Since we need to use the crystal structure of hyperkagome lattice extensively throughout this work, we summarize its property and related space symmetry operations in this section. The hyperkagome lattice has a cubic structure with 12 sublattices per unit cell\cite{Okamoto2007}, as shown in Fig. \ref{fig:sublattice}. The positions for the 12 sublattices are listed in Table \ref{tab:sublattice}. The 12 sublattice sites occupy the corners and the top of a truncated tetrahedron, and the tetrahedron fits into the cubic unit cell such that the $(1,1,1)$ diagonal line of the cubic unit cell serves as the central $C_3$ axis of the tetrahedron. Since the nearest neighbor (NN) bonds form a network of corner sharing regular triangles, hyperkagome lattice is not a bipartite lattice, giving rise to geometric frustrations for NN antiferromagnets on it. A pair of NN sites always come from different sublattices, and the relative locations of NN pairs are summarized in Table \ref{tab:nn}. The 24 independent NN bonds constitute 4 equilateral triangles within the unit cell (orange in Fig.\ref{fig:sublattice}), and another 4 equilateral triangles connecting nearby unit cells (green). Setting the lattice constant to be 1, all edges of the triangles have a length $1/4$. These triangles form 4 sets of non-parallel planes, each plane with the geometry of partially filled kagome lattice as shown in Fig. \ref{fig:sublattice}.

Now we look at point-group symmetry of the lattice. The hyperkagome lattice belongs to the space group $P4_132$\cite{Okamoto2007} (No. 213). In addition to the three translations for cubic unit cell
\bea
&(x,y,z)\overset{T_1}\longrightarrow(x+1,y,z),\\
&(x,y,z)\overset{T_2}\longrightarrow(x,y+1,z),\\
&(x,y,z)\overset{T_3}\longrightarrow(x,y,z+1)
\eea
the space group is generated by a 2-fold rotation along $(\frac{3}{8},\frac{3}{4}-x_2,x_2)$-axis:
\bea
(x,y,z)\overset{C_2}\longrightarrow(\frac{3}{4}-x, \frac{3}{4}-z, \frac{3}{4}-y)
\eea
a 3-fold rotation along $(1,1,1)$-axis:
\bea
(x,y,z)\overset{C_3}\longrightarrow(z,x,y)
\eea
and a 4-fold non-symmorphic screw: i.e. a $\pi/2$ rotation along $(x_1,\bar{\frac{1}{4}},\frac{1}{2})$-axis followed by a fractional translation of $(\frac{1}{4},0,0)$:
\bea
(x,y,z)\overset{S_4}\longrightarrow(\frac{1}{4}+x, \frac{1}{4}-z, \frac{3}{4}+y),~~~(S_4)^4=T_1.
\eea
Their actions on an arbitrary lattice site are summarized in Table \ref{tab:sublattice} and illustrated in Fig.\ref{fig:sublattice}. The hyperkagome lattice is non-centrosymmetric with neither inversion centers nor mirror planes, so it has a chiral octahedral point group $O$, consisting of 24 elements $\{C_2^{\nu_2}C_3^{\nu_3}S_4^{\nu_4}|\nu_2=0,1, \nu_3=0,1,2,\nu_4=0,1,2,3\}$ that can take one NN bond to the rest of the 24 inequivalent NN bonds. The commutation relations of these symmetry operations are given in Appendix \ref{appRotation}. In recent literatures, different conventions have been adopted regarding the labels of sublattice sites and coordinate-axis, as well as symmetry operations. We summarize the different conventions in Appendix \ref{appendix:C2} for comparison.

\begin{table}[h]
\caption{\label{tab:sublattice}
Sublattice coordinates and the action of rotations.
The coordinates of lattice sites are labeled by $\mathbf{x}=\mathbf{R}+{\bf r}_s\equiv (s; x,y,z)$, where $\{\mathbf{R}= (x,y,z)| x,y,z\in \mathbb{Z}\}$ denotes the cubic unit cell, and $\{ {\bf r}_s| s=1,\dots,12\}$ denotes the displacement of 12 sublattices. The length of the cubic unit cell is set to 1. This table summarizes how the symmetry operations $C_2, C_3, S_4$ take a site at $(s;x,y,z)$ to a new location $(s'; x', y', z')$ in the hyperkagome lattice.
}
\begin{tabular}{|c|c|c|c|c|}
\hline
$s$ & ${\bf r}_s$ & $C_2$ & $C_3$ & $S_4$ \\
\hline
1 & $(\frac{5}{8},\frac{3}{8},\frac{1}{8})$ &
$(7;-x,-z,-y)$ & $(7;z,x,y)$ &
$(12; x,-z,y+1)$ \\
\hline
2 & $(\frac{3}{8},\frac{3}{8},\frac{7}{8})$ &
$(8;-x,-z-1,-y)$ & $(6,z,x,y)$ &
$(1;x,-z-1,y+1)$ \\
\hline
3 & $(\frac{5}{8},\frac{5}{8},\frac{7}{8})$ &
$(11;-x,-z-1,-y)$ & $(5,z,x,y)$ &
$(6; x,-z-1,y+1)$ \\
\hline
4 & $(\frac{5}{8},\frac{7}{8},\frac{5}{8})$ &
$(10;-x,-z,-y-1)$ & $(3,z,x,y)$ &
$(5; x,-z-1,y+1)$ \\
\hline
5 & $(\frac{7}{8},\frac{5}{8},\frac{5}{8})$ &
$(12;-x-1,-z,-y)$ & $(4,z,x,y)$ &
\parbox{2cm}{$(7;x+1,-z-1,$\\$y+1)$} \\
\hline
6 & $(\frac{7}{8},\frac{3}{8},\frac{3}{8})$ &
$(6;-x-1,-z,-y)$ & $(8,z,x,y)$ &
\parbox{2cm}{$(11; x+1,-z-1,$\\$y+1)$} \\
\hline
7 & $(\frac{1}{8},\frac{5}{8},\frac{3}{8})$ &
$(1;-x,-z,-y)$ & $(9,z,x,y)$ &
$(8; x,-z-1,y+1)$ \\
\hline
8 & $(\frac{3}{8},\frac{7}{8},\frac{3}{8})$ &
$(2;-x,-z,-y-1)$ & $(2,z,x,y)$ &
$(4;x,-z-1,y+1)$ \\
\hline
9 & $(\frac{3}{8},\frac{1}{8},\frac{5}{8})$ &
$(9;-x,-z,-y)$ & $(1,z,x,y)$ &
$(3;x,-z-1,y)$ \\
\hline
10 & $(\frac{1}{8},\frac{1}{8},\frac{7}{8})$ &
$(4;-x,-z-1,-y)$ &$(12,z,x,y)$  &
$(2;x,-z-1,y)$ \\
\hline
11 & $(\frac{1}{8},\frac{7}{8},\frac{1}{8})$ &
$(3;-x,-z,-y-1)$ & $(10,z,x,y)$ &
$(9;x,-z,y+1)$ \\
\hline
12 & $(\frac{7}{8},\frac{1}{8},\frac{1}{8})$ &
$(5;-x-1,-z,-y)$ & $(11,z,x,y)$ &
$(10; x+1,-z,y)$ \\
\hline
\end{tabular}
\end{table}

\begin{table}
[h]
\caption{The relative positions for NN pairs $(s_1;{\bf x}_1)$ and $(s_2;{\bf x}_2)$.\label{tab:nn}}
\begin{tabular}{|c|r|r||r|r|}
\hline
$s_1 $ & \multicolumn{4}{c|}{$(s_2;\mathbf{x}_2-\mathbf{x}_1)$ (unit=$1/4$)} \\
\hline
 & \multicolumn{2}{c||}{Triangle 1} & \multicolumn{2}{c|}{Triangle 2}\\
\hline
1 & $(2;-,0,-)$ & $(3;0,+,-)$ & $(6;+,0,+)$ & $(12;+,-,0)$ \\
\hline
2 & $(1;+,0,+)$ & $(3;+,+,0)$ & $(9;0,-,-)$ & $(10;-,-,0) $\\
\hline
3 & $(1;0,-,+)$ & $(2;-,-,0)$ & $(4;0,+,-)$ & $(5;+,0,-)$ \\
\hline
4 & $(3;0,-,+)$ & $(5;+,-,0)$ & $(8;-,0,-)$ & $(9;-,+,0)$ \\
\hline
5 & $(3;-,0,+)$ & $(4;-,+,0)$ & $(6;0,-,-)$ & $(7;+,0,-)$ \\
\hline
6 & $(1;-,0,-)$ & $(5;0,+,+)$ & $(12;0,-,-)$ & $(7;+,+,0)$ \\
\hline
7 & $(8;+,+,0)$ & $(11;0,+,-)$ & $(5;-,0,+)$ & $(6;-,-,0)$ \\
\hline
8 & $(4;+,0,+)$ & $(7;-,-,0)$ & $(11;-,0,-)$ & $(9;0,+,+)$ \\
\hline
9 & $(2;0,+,+)$ & $(10;-,0,+)$ & $(4;+,-,0)$ & $(8;0,-,-)$ \\
\hline
10 & $(2;+,+,0)$ & $(9;+,0,-)$ & $(11;0,-,+)$ & $(12;-,0,+)$ \\
\hline
11 & $(7;0,-,+)$ & $(8;+,0,+)$ & $(10;0,+,-)$ & $(12;-,+,0)$ \\
\hline
12 & $(1;-,+,0)$ & $(6;0,+,+)$ & $(10;+,0,-)$ & $(11;+,-,0)$ \\
\hline
\end{tabular}
\end{table}

%
%
%
%
%

\subsection{Slave-fermion representation of symmetric spin liquids}

Next we construct mean-field Hamiltonians of spin liquid states preserving the symmetries discussed in the previous section, using the slave fermion representation and associated projective symmetry group classification\cite{Wen2002}. Motivated by significant spin-orbit couplings in Na$_4$Ir$_3$O$_8$, below we discuss a general formalism of PSG construction in the presence of local spin anisotropy.

In the slave-fermion representation, we write the local spin operator at site $i$ as
\begin{equation}\label{spin1}
\mathbf{S}_i = (\psi_i^\dagger)_{\alpha} \frac{\boldsymbol{\sigma}_{\alpha\beta}}{2} (\psi_i)_\beta,
\end{equation}
where $\sigma$'s are Pauli matrices, and the spinor $\psi_i = (f_{i\uparrow}, f_{i\downarrow})^T$ involves the slave fermions
\begin{equation}
\{f_{i\alpha}, f_{j\beta}^\dagger\} = \delta_{ij}\delta_{\alpha\beta}, \qquad
\{f_{i\alpha}, f_{j\beta}\} = 0 = \{f_{i\alpha}^\dagger, f_{j\beta }^\dagger\}.
\end{equation}
Here $i,j$ are site indices and $\alpha,\beta$ are spin indices. In the following we will always omit the spin indices to lighten notations. For the convenience of later analysis, we introduce
\begin{equation}\label{fermionic spinon}
\Psi_i = \left(
\begin{array}{cc}
f_{i\uparrow} & f_{i\downarrow}^\dagger\\
f_{i\downarrow} & -f_{i\uparrow}^\dagger
\end{array}
\right) = (\psi_i, {\cal T}\psi_i^\ast),
\end{equation}
where ${\cal T} = i\sigma_yK$ is the time-reversal operator and $K$ is complex conjugation. Then the spin operator can be rewritten as
\begin{equation}\label{spin2}
\mathbf{S}_i = \frac{1}{4}\mbox{Tr} ( \Psi_i^\dagger \boldsymbol{\sigma}\Psi_i)
\end{equation}
From the representation (\ref{spin2}), we immediately see that an SU(2) rotation $W_i=e^{-i\boldsymbol{\phi}\cdot \tau/2}$ acting on the right
\begin{equation}\label{gauget}
\Psi_i\rightarrow \Psi_i W_i\qquad \mbox{(gauge rotation with Pauli matrices $\vec\tau$)}
\end{equation}
leaves the physical spin operator $\mathbf{S}_i$ invariant, while an SU(2) rotation $R=e^{-i\boldsymbol{\theta}\cdot \sigma/2}$ acting on the left
\begin{equation}\label{rotation}
\Psi_i\rightarrow R^\dagger\Psi_i\qquad \mbox{(spin rotation with Pauli matrices $\vec\sigma$)}
\end{equation}
rotates the spin operator $\mathbf{S}_i$ by an angle $\boldsymbol{\theta}$. The invariance of physical spin operator $\mathbf{S}_i$ under (\ref{gauget}) indicates a local SU(2) gauge redundancy in the parton construction, crucial for the projective symmetry group defined later. Physically, such a gauge redundancy is due to the lack of ``charge'' degree of freedom for spin operators, so destroying a $\downarrow$-spinon ($f_{\downarrow}$) is equivalent to creating a $\uparrow$-spinon ($f^\dagger_{\uparrow}$). Mathematically,  the Hilbert space is enlarged by $f$-operators to include empty and double-occupancy on each site, which is absent in the physical Hilbert space $(\cos\frac{\theta_i}{2}e^{-i\varphi_i/2}f_{i\uparrow}^\dagger + \sin\frac{\theta_i}{2}e^{i\varphi_i/2}f_{i\downarrow}^\dagger)|0\rangle$ spanned by $\mathbf{S}_i$ operators in (\ref{spin2}). That  means we have the onsite constraint
\begin{equation}\label{single occupancy constraint}
f_{i\uparrow}^\dagger f_{i\uparrow} + f_{i\downarrow}^\dagger f_{i\downarrow} = 1.
\end{equation}
or written in a neat form invariant under $SU(2)$ gauge rotations,
\begin{equation}
\mbox{Tr} (\Psi_i^\dagger \Psi_i \boldsymbol\tau) = 0.
\end{equation}
Therefore in the constrained Hilbert space we have the gauge redundancy (\ref{gauget}).

The slave fermion representation allows for a mean field decomposition of Hamiltonians describing spin-spin interaction without invoking magnetic ordering. The most  general form of a mean field Hamiltonian is (see Appendix \ref{appendix:MFterms} for expansions in terms of $f$-operators)
\begin{eqnarray}\nonumber
H_{MF} &=& H_0 + H_x + H_y + H_z,\\ \nonumber
H_0 &=& \mbox{Tr}(\Psi_i u_{ij}^{(0)} \Psi_j^\dagger),\qquad
H_{x} = \mbox{Tr} (\sigma_{x} \Psi_i u_{ij}^{(x)} \Psi_j^\dagger),\\ \label{hmfPsi1}
H_{y} &=& \mbox{Tr} (\sigma_{y} \Psi_i u_{ij}^{(y)} \Psi_j^\dagger),\qquad
H_{z} =  \mbox{Tr} (\sigma_{z} \Psi_i u_{ij}^{(z)} \Psi_j^\dagger).
\end{eqnarray}
where the mean field amplitudes are
\begin{equation}\label{st}
u_{ij}^{(0)} = i b^{(0)}_{ij} + \boldsymbol{a}^{(0)}_{ij}\cdot \boldsymbol{\sigma},\qquad
u_{ij}^{(x,y,z)} = b^{(x,y,z)}_{ij} + i\boldsymbol{a}^{(x,y,z)}_{ij}\cdot \boldsymbol{\sigma}
\end{equation}
and $b^{(0,x,y,z)}_{ij}, \boldsymbol{a}^{(0,x,y,z)}_{ij}$ are real numbers serving as mean field parameters.
 The summation over sites $\sum_{i,j}$ is omitted to lighten the notations. The Hermiticity of the Hamiltonian guarantees Eq. (\ref{st}) and
\begin{eqnarray}\label{absymmsinglet}
&& b_{ji}^{(0)} = - b_{ij}^{(0)}, \qquad
\boldsymbol{a}_{ji}^{(0)} = \boldsymbol{a}_{ij}^{(0)}\\ \label{absymmtriplet}
&& b_{ji}^{(x,y,z)} = b_{ij}^{(x,y,z)},\qquad
\boldsymbol{a}_{ji}^{(x,y,z)} = - \boldsymbol{a}_{ij}^{(x,y,z)}.
\end{eqnarray}
So for onsite terms we immediately have
\begin{equation}\label{noonsite}
b_{ii}^{(0)} = \boldsymbol{a}_{ii}^{(x,y,z)}=0.
\end{equation}
 Note here the mean field Hamiltonian is split into singlet $H_0$ and triplet $H_{x,y,z}$ parts.
The singlet part $H_0$ is invariant under $SU(2)$ spin rotations as we can see from (\ref{rotation}) and (\ref{hmfPsi1}). It involves only the spin singlet hoppings $(f_{i\uparrow}^\dagger f_{j\uparrow} + f_{i\downarrow}^\dagger f_{j\downarrow})$ and singlet pairings $(f_{i\uparrow}^\dagger f_{j\downarrow}^\dagger - f_{i\downarrow}^\dagger f_{j\uparrow}^\dagger)$. On the other hand, $H_{x,y,z}$ generically breaks continuous spin rotational symmetry. $H_z$ contains spin-dependent hopping $(f_{i\uparrow}^\dagger f_{j\uparrow} - f_{i\downarrow}^\dagger f_{j\downarrow}  )$ and triplet pairing $(f_{i\uparrow}^\dagger f_{j\downarrow}^\dagger + f_{i\downarrow}^\dagger f_{j\uparrow}^\dagger )$, and $H_x, H_y$ involve spin-flip hopping $(f_{i\uparrow}^\dagger f_{j\downarrow} \pm f_{i\downarrow}^\dagger f_{j\uparrow})$ and  triplet pairing $(f_{i\uparrow}^\dagger f_{j\uparrow}^\dagger \pm f_{i\downarrow}^\dagger f_{j\downarrow}^\dagger)$. Each $H_A$ ($A=x,y,z$) contains four real parameters for each bond $(i,j)$, and a general mean-field Hamiltonian $H_{MF}$ involves 16 real parameters for each bond.

\subsection{Classification of symmetric $U(1)$ and $Z_2$ spin liquids}

Since symmetric spin liquids are described by projected superconductors of fermionic spinons in the slave fermion representation, it's crucial to understand how symmetry acts on the fermionic spinons $\Psi_i$ in (\ref{fermionic spinon}). In particular due to the $SU(2)$ gauge redundancy in the slave fermion construction (\ref{spin2}) of spin-$1/2$, each symmetry operation $U$ in the symmetry group $SG$ can be followed by gauge rotation $\{G_U(i)\in SU(2)\}$. More precisely, the mean-field Hamiltonian for fermionic spinons is only invariant under a combination of physical symmetry operation $U$ and associated gauge rotation $\{G_U(i)\}$:
\bea
G_U\hat U H_{MF}\hat U^{-1}G^{-1}_U=H_{MF}.
\eea
In particular, fractionalized fermionic spinons may not form a linear representation of the symmetry group, instead they can transform \emph{projectively} under symmetry operations\cite{Essin2013,Barkeshli2014,Tarantino2016}. As shown in Ref.\cite{Wen2002}, the projective symmetry operations on fermionic spinons can be systematically classified by their \emph{projective symmetry group (PSG)} in the slave fermion representation.

We therefore classify the PSGs for symmetric $Z_2$ and $U(1)$ spin liquids on hyperkagome lattice. We also construct their mean-field Hamiltonians (\ref{hmfPsi1}), that is invariant under projective symmetry operations of time reversal $\cal T$, 3 translations for cubic unit cells $T_1, T_2, T_3$, and 24 point group operations generated by $C_2, C_3, S_4$ as discussed previously
\begin{equation}\label{totalsymmetry}
  U={\cal T}^{\nu_{\cal T}} T_1^{\nu_x} T_2^{\nu_y} T_3^{\nu_z}  C_2^{\nu_2}C_3^{\nu_3}
  S_4^{\nu_4}
 \end{equation}
 where $\nu_{\cal T}\in\mathbb{Z}_2$, $\nu_{x,y,z}\in\mathbb{Z}$, $\nu_2\in \mathbb{Z}_2$,  $\nu_3\in \mathbb{Z}_3$, $\nu_4\in \mathbb{Z}_4$.
The detailed calculations can be found in Appendix \ref{app:z2 class}-\ref{app:u1 class}, while we summarize the PSG solutions in Table \ref{tab:psg}. There exists a convenient gauge where {\em all gauge rotations are independent of unit-cell or sublattice indices}, which simplifies further analysis. The gauge rotations associated with translations and 3-fold rotation can be chosen to be all trivial
\begin{equation}
G_{T_i}({\bf r},s)=G_{C_3}({\bf r},s)=1,~~~i=1,2,3.
\end{equation}
while the gauge rotations for 2-fold rotation $G_{C_2}({\bf r},s)\equiv g_{C_2}$ and 4-fold screw $G_{S_4}({\bf r},s)\equiv g_{S_4}$ have the same form of
\begin{equation}
g_{C_2}=g_{S_4}\in \mbox{SU(2)}
\end{equation}
Meanwhile anti-unitary time reversal symmetry is implemented by
\bea
\Psi_{{\bf r},s}\overset{\cal T}\longrightarrow\sigma_2\Psi_{{\bf r},s}\tau_2\cdot G_{\cal T}({\bf r},s)
\eea
with gauge rotation
\bea
G_{\cal T}({\bf r},s)\equiv g_{\cal T}\in\mbox{SU(2)}.
\eea
As shown in Appendix \ref{app:u1 class} there are 2 different $U(1)$ spin liquids in the PSG classification, labeled as $U1^0$ and $U1^1$ states. They correspond to $U(1)$-uniform ($U1^0$) and $U(1)$-staggered ($U1^1$) phases studied in Ref.\cite{Lawler2008}.

Meanwhile the solutions to PSG equations for symmetric $Z_2$ spin liquids must satisfy the following conditions (for details see Appendix \ref{app:z2 class})
\begin{equation}
g_{\cal T}^2=\eta_{\cal T}, \quad g_{C_2}^2=\eta_2, \quad g_{\cal T}g_{C_2} = \eta_{2\cal T} g_{C_2}g_{\cal T}
\end{equation}
and we find 5 gauge inequivalent solutions as listed in Table \ref{tab:psg}. Among them, 2 unphysical solutions with $g_{\cal T}=1$ are incompatible with spin-$1/2$ fermionic spinons (Kramers doublets), and always lead to vanishing mean-field ansatz. Therefore these 5 algebraic solutions only lead to 3 distinct symmetric $Z_2$ spin liquids on hyperkagome lattice: they are states 2(a), 2(b) and 2(c) in TABLE \ref{tab:psg}.

Such a simple classification result with only 2 symmetric $U(1)$ states and 3 symmetric $Z_2$ states owes a lot to the presence of nonsymmorphic screw symmetry $S_4$. In particular, as proved in Appendix \ref{app:z2 class}-\ref{app:u1 class}, screw operation $S_4$ generally rules out a large class of $Z_2$ spin liquids known as ``$\pi$-flux phases''\cite{Wen2002}, where $\pi$ flux is threaded in each unit cell. Similarly a class of $U(1)$ spin liquids known as ``staggered flux phases''\cite{Wen2002} with alternating flux along one direction are also incompatible with screw $S_4$.

%

\begin{figure}
\includegraphics[width=0.9\columnwidth]{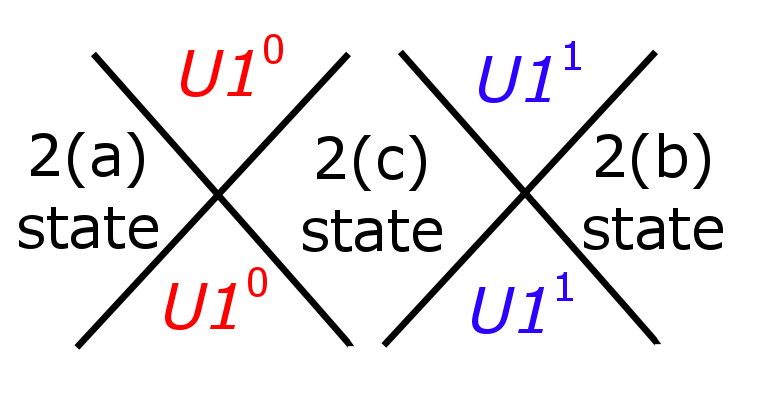}
\caption{(color online) Schematic ``global'' phase diagram of different symmetric spin liquids on hyperkagome lattice. Two $Z_2$ states 2(a) and 2(c) are in the neighborhood of $U(1)$ spin liquid $U1^0$ state (called $U(1)$-uniform state in Ref.\cite{Lawler2008}), while 2(b) and 2(c) states are in proximity to $U(1)$ spin liquid $U1^1$ state (called $U(1)$-staggered state in Ref.\cite{Lawler2008}).}
\label{fig:sketch phase diagram}
\end{figure}

Before discussing physical properties of these spin liquids, we clarify the relation between the symmetric $U(1)$ and $Z_2$ spin liquids on hyperkagome lattice. Quite generally, a $Z_2$ spin liquids can be viewed as a descendant of another $U(1)$ spin liquid, by breaking the gauge group from $U(1)$ down to $Z_2$ via the Anderson-Higgs mechanism. Such a continuous phase transition is driven by condensing the Cooper pairs of fermionic spinons in the slave fermion representation. In our case of hyperkagome spin liquids, each $U(1)$ spin liquid hosts 2 such neighboring $Z_2$ states: $U1^0$ state (``$U(1)$-uniform'' state in Ref.\cite{Lawler2008}) is proximate to two $Z_2$ states 2(a) and 2(c), while $U1^1$ state (``$U(1)$-staggered'' state in Ref.\cite{Lawler2008}) is proximate to two $Z_2$ states 2(b) and 2(c). Note that 2(c) state is in the neighborhood of both $U(1)$ states, as shown in TABLE \ref{tab:psg}. A schematic ``global'' phase diagram demonstrating the relation between different spin liquids is shown in FIG. \ref{fig:sketch phase diagram}.

In this work we only classify symmetric $Z_2$ and $U(1)$ spin liquids within the slave-fermion representation. There are other symmetric $Z_2$ and $U(1)$ spin liquids on hyperkagome lattice beyond the description of slave-fermion representation, for example the Schwinger-boson states studied in Ref.\cite{Lawler2008a}. We also want to mention that a large class of symmetric $Z_2$ spin liquids featuring strong spin-orbit couplings, the ``Majorana spin liquids''\cite{OBrien2016} realized in Kitaev-type models\cite{Kitaev2006}, can all be constructed and described in the slave-fermion representations\cite{Burnell2011}. However, these Majorana spin liquids preserves time reversal symmetry only on a bipartite lattice\cite{Chen2012a} and hence do not exist in the non-bipartite hyperkagome lattice.

\section{Properties of hyperkagome spin liquids\label{sec:z2sl}}

\subsection{Gapped vs. gapless states}\label{subsec:gapless proof}

Since these spin liquids preserve all symmetries of the system, one significant feature of them is the presence/absence of a gap for low-energy excitations on top of their ground states. This issue can be fully determined with the knowledge of symmetry operations on fermionic spinons in these states, as we will show below.

We start with symmetric $Z_2$ spin liquids, whose mean-field ansatz describes a superconductor of fermionic spinons. First of all, it's straightforward to show that an on-site singlet pairing term is allowed by both 2(a) and 2(c) states, which generally can lead to a gapped spinon spectrum. Therefore 2(a) and 2(c) states are generically gapped as shown in TABLE \ref{tab:psg}. On the other hand, since each $C_2$ axis crosses a single site in each unit cell, such an on-site pairing is forbidden in 2(b) state by 2-fold rotation associated with gauge rotation $g_{C_2}=i\tau_2$. Below we prove that mean-field Hamiltonian (\ref{hmfPsi1}) must be gapless for state 2(b).

In the Nambu basis of $\Phi_i = (f_{i\uparrow}, f_{i\downarrow}^\dagger, f_{i\downarrow}, -f_{i\uparrow}^\dagger)$, mean-field ansatz (\ref{hmfPsi1}) can be written and diagonalized in ${\bf k}$-space as
\bea
\hat H_{MF}=\sum_{i,j}\Phi_i^\dagger h^{MF}_{i,j}\Phi_j=\sum_{\bf k}\Gamma_{\bf k}^\dagger\Lambda_{\bf k}\Gamma_{\bf k}
\eea
where $\Lambda_{\bf k}$ is a $48\times48$ diagonal matrix describing band dispersions, and $\{\Gamma_{\bf k}\}$ are the eigenmodes. Due to particle-hole symmetry in such a Bogoliubov-de Gennes (BdG) Hamiltonian, we only fill half (24) of the total 48 BdG bands in $\{\Lambda_{\bf k}\}$. However, these bands are generally not separated from each other, instead they are entangled due to certain symmetries. In our case of space group $P4_132$, non-symmorphic screw symmetry $S_4$ dictates that bands always appear in quadruplets that cannot be disentangled\cite{Fang2015a,Shiozaki2015,Alexandradinata2016,Varjas2016}. Meanwhile, time reversal symmetry ${\cal T}$ leads to an extra 2-fold band degeneracy at high symmetry points in ${\bf k}$-space. Finally, the combination $C_2\cdot{\cal T}$ of 2-fold rotation and time reversal is an anti-unitary symmetry satisfying $(C_2\cdot{\cal T})^2=-1$ in state 2(b), giving rise to an extra 2-fold degeneracy in certain high-symmetry planes perpendicular to $C_2$ axis. Therefore, energy bands in the BdG Hamiltonian of state 2(b) always appear in a multiplet of $16=4\times2\times2$, that cannot be disentangled without breaking symmetry. Hence all gapped ground states must fill a multiple of 16 bands, and it's impossible for a gapped symmetric superconductor to fill only 24 bands. Therefore we proved the gaplessness of spinon mean-field Hamiltonian for state 2(b).

\begin{widetext}
	
	\begin{figure}
		\begin{center}
			\includegraphics[width=18cm]{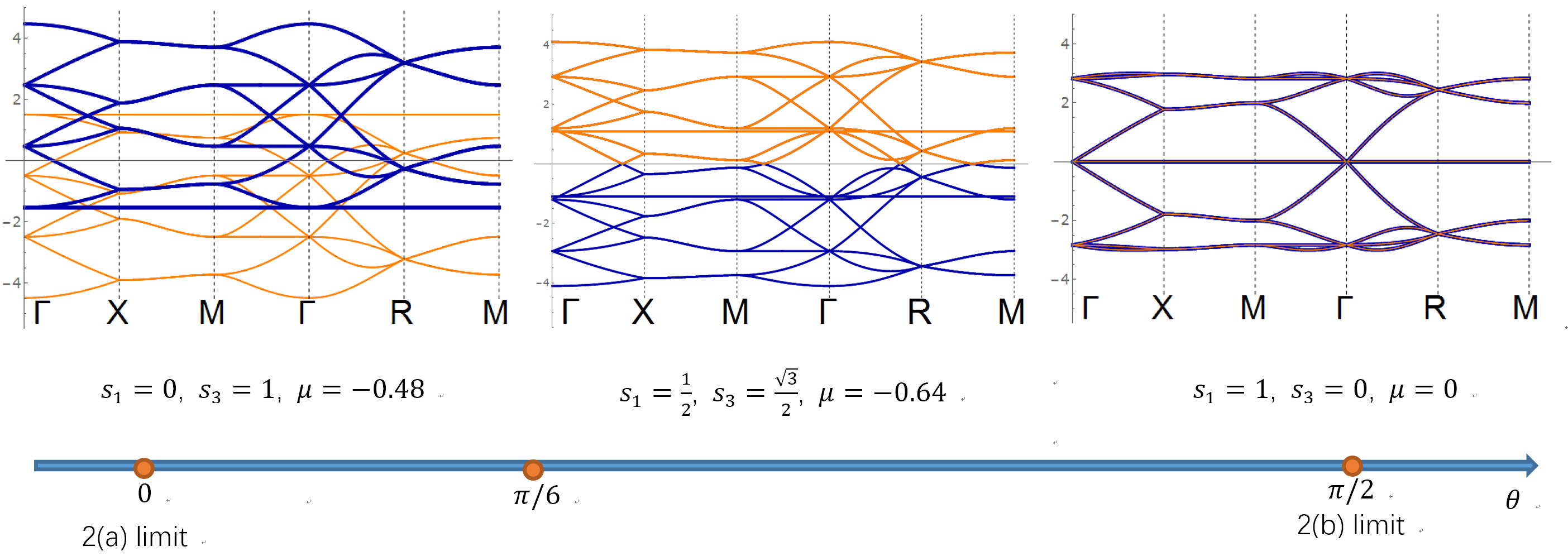}
		\end{center}
		\caption{\label{fig:singletdispersion}The dispersion of fermionic spinons in isotropic $Z_2$ spin liquid state 2(c) with $SU(2)$ spin rotational symmetry and up to NN mean-field amplitudes. Here $\Gamma=(0,0,0), X=(\pi,0,0), M=(\pi,\pi,0), R=(\pi,\pi,\pi)$ denotes the high-symmetry points in ${\bf k}$-space. The blue/orange colors denote the particle/hole redundancy of the BdG Hamiltonian. The only free parameter is the ratio between pairing and hopping $\theta \equiv \arcsin(s_1/s_3)$. The $\theta=0$ limit corresponds to case 2(a), a $U(1)$ spin liquid featuring spinon Fermi surfaces (labeled $U1^0$ state, or $U(1)$-uniform state in Ref.\cite{Lawler2008}) with uniform NN hoppings. The $\theta=\pi/2$ limit corresponds to case 2(b), another $U(1)$ spin liquid featuring spinon Dirac cones at fermi level (labeled $U1^1$ state, or $U(1)$-staggered state in Ref.\cite{Lawler2008}) with staggered pairing amplitudes on different triangles. In the intermediate regime $0<\theta<\pi/2$, it is a symmetric $\mathbb{Z}_2$ spin liquid with gapless spectrum. Note that 4-fold degenerate flat-bands exist in all cases.}
	\end{figure}

\end{widetext}

Next we analyze the two $U(1)$ spin liquids, whose mean-field ansatz only contains hopping terms in the slave fermion representation. In $U1^0$ state, as shown in Appendix \ref{app:u1 class}, the fermionic spinons transform in the same fashion as usual electrons under the whole space group and time reversal symmetries. Due to their single occupancy constraint (\ref{single occupancy constraint}), in each unit cell we have 12 spinons on average. However as shown in Ref.\cite{Watanabe2015}, any gapped (short-range-entangled) insulator in space group $P4_132$ with time reversal symmetry\cite{Watanabe2015} must have a filling number that's multiple of 8 per unit cell, incompatible with our filling number 12. Therefore a symmetric $U1^0$ state must support gapless spinon excitations in the bulk, such as the spinon fermi surface in the left panel of FIG. \ref{fig:singletdispersion}.

The other $U(1)$ spin liquid labeled $U1^1$ state, on the other hand, have different implementations for $C_2$ and $S_4$ operations compared to $U1^0$ state. As shown in Appendix \ref{app:u1 class}, both $C_2$ and $S_4$ operations are followed by gauge rotation $i\tau_{2}$ (up to a global $U(1)$ gauge rotation $e^{i\phi\tau_3}$), corresponding to the particle-hole transformation $f_\sigma\rightarrow f_{\sigma}^\dagger$ on fermionic spinons. Regarding this, we can simply write down the mean-field Hamiltonian of $U1^1$ state in the Nambu basis, and exactly the same argument for $Z_2$ state 2(b) discussed previously proves the gaplessness of $U1^1$ state.

Although the above analysis on 2(b) and $U1^1$ states is based on band structure of mean-field spinon Hamiltonian (\ref{hmfPsi1}), their gaplessness can be shown to remain valid even in the presence of arbitrary short-range interactions between spinons\cite{Lu2016b,Qi2016}, using an argument based on entanglement spectrum of a gapped state\cite{Watanabe2015}. Therefore among 3 symmetric $Z_2$ states classified on hyperkagome lattice, 2(b) is the only stable gapless $Z_2$ spin liquid. Meanwhile, both $U(1)$ spin liquids ($U1^0$ and $U1^1$ states) are stable gapless spin liquids.

\subsection{Mean-field spectrum of fermionic spinons}

 \begin{table}
\caption{\label{tab:psg}
Summary of 5 algebraic PSGs in slave fermion representation for symmetry group $SG=P4_132\times Z_2^{\cal T}$ on hyperkagome lattice. Only 3 of them with ${\cal T}^2=\eta_{\cal T}=-1$ are physical solutions with spin-$1/2$ fermionic spinons, leading to 3 distinct symmetric $Z_2$ spin liquids on hyperkaogme lattice: 2(a), 2(b), 2(c) states. These 3 symmetric $Z_2$ states are in the neighborhood of 2 symmetric $U(1)$ spin liquids: $U1^0$ and $U1^1$ states. For details see Appendix \ref{app:z2 class}-\ref{app:u1 class}.}
\begin{center}
\begin{tabular}{|c|c|c|c|c|c|c|c|c|}
\hline
Label& $\eta_{\cal T}$ & $\eta_{2}$ & $\eta_{2\cal T}$ & $g_{C_2}=g_{S_4}$   & $g_{\cal T}$ & Physical?& \multirow{2}{1.4cm}{$U(1)$ root states}&\multirow{2}{1cm}{Stably gapless?} \\
&&&&&&&&\\
 \hline
 1(a)
 &+1 & +1 & +1 & 1 &    1 & No & x & x
 \\ \hline
 1(b) 
 & +1 & -1 & +1 &  $i\tau_3$ & 1 & No & x & x
 \\ \hline
 2(a) 
&  -1 & +1 & +1 & 1 &  $i\tau_2$ & Yes & $U1^0$ & No
 \\ \hline
 2(b)  
 &-1 & -1 & +1  &  $i\tau_2$ & $i\tau_2$ & Yes & $U1^1$ & Yes
 \\ \hline
 2(c) 
 & -1 & -1 & -1 &  $i\tau_3$ & $i\tau_2$ &Yes & $U1^{0}$~\text{and}~$U1^1$ & No
 \\ \hline
\end{tabular}
\end{center}
\end{table}

\begin{widetext}

\begin{figure}
\begin{center}
\includegraphics[width=4.3cm]{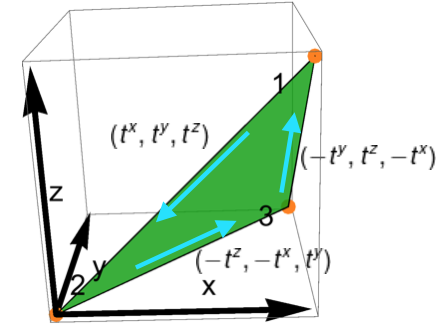}
\includegraphics[width=4cm]{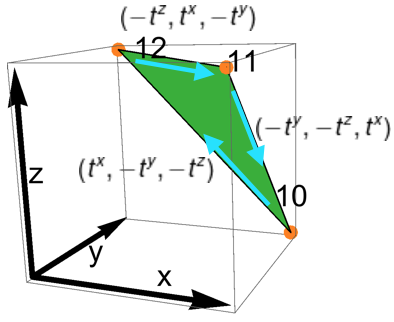}
\includegraphics[width=3.5cm]{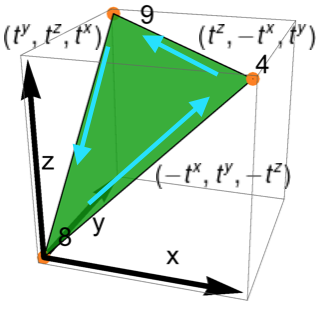}
\includegraphics[width=4cm]{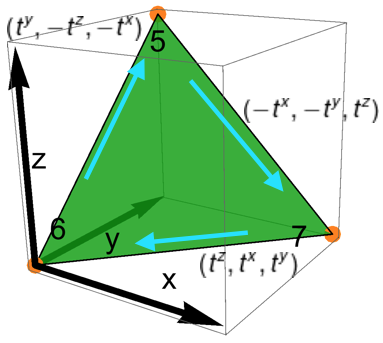}\\
\includegraphics[width=4.3cm]{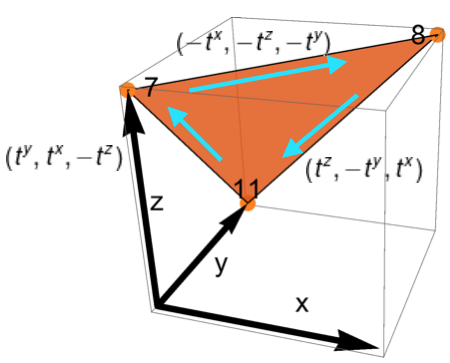}
\includegraphics[width=4cm]{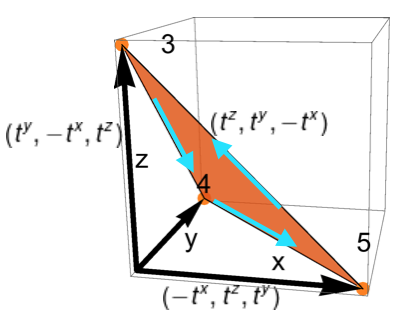}
\includegraphics[width=4cm]{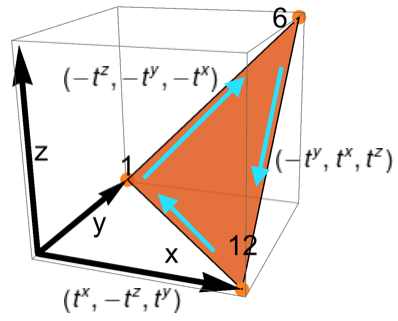}
\includegraphics[width=5cm]{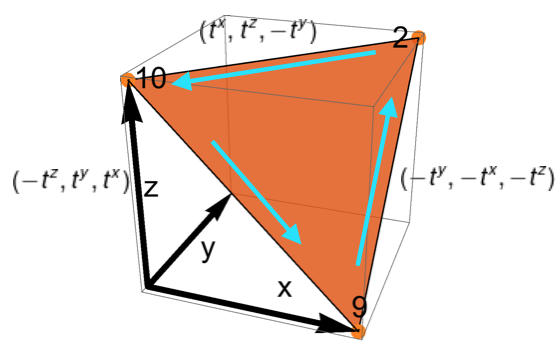}
\end{center}
\caption{The bond-dependent mean-field amplitudes $(t^{\mu}, t^{\nu}, t^\rho)$ for triplet terms  in Table \ref{tab:mftriplet}. Here we have chosen the bond $(1,2)$ to have amplitudes $(t^x, t^y, t^z)$, then all the other bonds are related by a combination of SO(3) and SU(2) rotations in  (\ref{psgconditiont}). Here we specify the results of SO(3) rotations, while the SU(2) rotation by $G_U$ in different cases are specified in Table \ref{tab:mftriplet}. The positive directions are denoted by blue arrows.  \label{fig:bondxyz}
}
\end{figure}

\end{widetext}

Due to space group symmetries, among all NN bonds, we only need to write down the mean-field amplitude $u_{ij}$ in one bond, and all the other bonds are generated by symmetry from condition (\ref{psgcondition}) for singlet terms, and (\ref{psgconditiont}) for triplet terms. The details of symmetry constraints on mean-field amplitudes are analyzed in Appendix \ref{app:mf ansatz} and we present the results below. For unphysical PSG solutions 1(a) and 1(b) where $g_{\cal T}=1$, we immediately see that time-reversal (\ref{timereversalConstr}) leads to the constraint $u_{ij}=-u_{ij}$ for both singlet and triplet terms, hence no mean field realization in these cases. For physical $Z_2$ states 2(a,b,c) with $g_{\cal T}=i\tau_2$,
we list the spin-singlet amplitudes in Table \ref{tab:mfsinglet}, and the triplet terms in Table \ref{tab:mftriplet} and Fig. \ref{fig:bondxyz}. Here we focus on the bonds connecting up to nearest neighboring sites. For further neighbouring bonds, similar analysis can be done using the PSG solutions in Table \ref{tab:psg}.

\begin{table}
\caption{\label{tab:mfsinglet}
Singlet mean-field amplitudes $u^{(0)}_{ij}$ up to nearest neighbour (NN) bonds. Here $\tau_i$'s are Pauli matrices. The onsite terms are identical for all sites. The NN bonds are divided into two groups, where $u_\beta^{(0)}$ denotes the bonds in the triangles (1,2,3), (9,8,4), (7,6,5), (12,11,10),
 and $u_\gamma^{(A)}$ denotes the bonds in the triangles (1,6,12), (3,4,5), (10,9,2), (7,8,11). That is, two corner-shared triangles belong to different groups.
 Note that we have $u_{ij}^{(0)} =u_{ji}^{(0)}$ from Eq. (\ref{absymmsinglet}).
}
\begin{center}
\begin{tabular}{|c|c|c|c|c|}
\hline
Cases & Onsite $u_\alpha^{(0)}$ & N.N.  $u_\beta^{(0)}$ & N.N. $u_{\gamma}^{(0)}$ & Type\\
\hline
2(a) & $\Delta \tau_1 + \mu \tau_3$ & $s_1 \tau_1 +s_3\tau_3$ & $s_1\tau_1 + s_3\tau_3$ & U(1) FS\\
\hline
2(b) & 0 & $s_1\tau_1 +s_3\tau_3$ & $-s_1\tau_1- s_3\tau_3$ & U(1) Dirac\\
\hline
2(c) & $\mu \tau_3$ & $s_1\tau_1 + s_3\tau_3$ & $-s_1\tau_1 + s_3\tau_3$ & $\mathbb{Z}_2$ gapless
 \\
\hline
\end{tabular}
\end{center}
\end{table}

\begin{table}
\caption{\label{tab:mftriplet}
Triplet mean field amplitudes $u^{(x,y,z)}_{ij}$ up to nearest neighbour bonds. Here $\tau_i$'s are Pauli matrices. The onsite terms vanishes identically.  Note from Equation (\ref{absymmtriplet}) that $u_{ij}^{(x,y,z)} =-u_{ji}^{(x,y,z)}$, unlike the singlet case. So the NN bonds are divided into two groups of {\em oriented} triangles (I) (1,2,3), (9,8,4), (7,6,5), (12,11,10), and (II) (1,6,12), (3,4,5), (10,9,2), (7,8,11), where $1\rightarrow2\rightarrow3$ and etc. are regarded as positive directions (see Fig. \ref{fig:bondxyz}). $u_\beta^{(x,y,z)}, u_\gamma^{(x,y,z)}$ denote the bonds in groups (I) and (II) respectively. Further, $t^{x,y,z}$ in different bonds are mixed. For instance, we choose $(1,2)$ bond to have $(t^x_1, t^y_1, t^z_1)$ for $t^{\mu,\nu,\rho}_1$ and $(t^x_3, t^y_3, t^z_3)$ for $ t^{\mu,\nu,\rho}_3$, then the $t^{\mu,\nu,\rho}_1, t^{\mu,\nu,\rho}_3$ in other bonds are specified in Fig.\ref{fig:bondxyz}.
}
\begin{center}
\begin{tabular}{|c|c|c|c|}
\hline
Cases  & Onsite $u_\alpha^{(x,y,z)}$ & N.N. $u_\beta^{(x,y,z)}$ & N.N. $u_{\gamma}^{(x,y,z)}$ \\
\hline
2(a) & 0 & $t^{\mu,\nu,\rho}_1 \tau_1 + t^{\mu,\nu,\rho}_3\tau_3$ & $t^{\mu,\nu,\rho}_1\tau_1 + t^{\mu,\nu,\rho}_3\tau_3$\\
\hline
2(b) &  0& $t^{\mu,\nu,\rho}_1\tau_1 +t^{\mu,\nu,\rho}_3\tau_3$ & $-t^{\mu,\nu,\rho}_1\tau_1- t^{\mu,\nu,\rho}_3\tau_3$\\
\hline
2(c) &  0& $t^{\mu,\nu,\rho}_1\tau_1 + t^{\mu,\nu,\rho}_3\tau_3$ & $-t^{\mu,\nu,\rho}_1\tau_1 + t^{\mu,\nu,\rho}_3\tau_3$
 \\
\hline
\end{tabular}
\end{center}
\end{table}

Below we analyze the mean-field dispersions of fermionic spinons, and especially the low-energy excitations near the nodal surfaces. To do so, we rewrite the mean-field Hamiltonian (\ref{hmfPsi1}) in the basis $\Phi_i = (f_{i\uparrow}, f_{i\downarrow}^\dagger, f_{i\downarrow}, -f_{i\uparrow}^\dagger)^T$ as	a 4$\times$4 matrix
\begin{equation}\label{mf ham}
H_{MF} = \sum_{ij}\Phi_i^\dagger \left(
\begin{array}{cc}
-u_{ij}^{(0)} + u_{ij}^{(z)} & u_{ij}^{(x)}-iu_{ij}^{(y)}\\
u_{ij}^{(x)}+iu_{ij}^{(y)} & -u_{ij}^{(0)} - u_{ij}^{(z)}
\end{array}
\right) \Phi_j.
\end{equation}
Making a Fourier transform into $\mathbf{k}$-space, it will be a 48$\times$48 matrix due to the 12 sublattice, 2 spin and 2 particle-hole indices.

\subsubsection{Isotropic case with $SU(2)$ spin rotational symmetry}

We first discuss the $SU(2)_{\text{spin}}$-invariant case with only singlet terms (Table \ref{tab:mfsinglet}), corresponding to the isotropic limit.
A further self-consistency calculation shows that at half-filling, $\Delta\approx 0.48 s_1, \mu\approx 0.48s_3$ for the onsite term in 2(a). Thus, case 2(a) with up to NN mean-field amplitudes leads to a one-parameter family of Hamiltonians $\{ H^{(2a)}[\varphi]\}$ with NN terms
\begin{equation}
\mbox{2(a):} \qquad s_1=A\sin\varphi, \quad s_3=A\cos\varphi
\end{equation}
and onsite terms
 \begin{equation}
\mbox{2(a):}\qquad \Delta\approx 0.48A\sin\varphi,\quad \mu\approx 0.48A\cos\varphi
  \end{equation}

Similarly, case 2(b) also leads to a one-parameter family of Hamiltonians $\{ H^{(2b)}[\varphi]\}$ with NN terms
\begin{equation}
\mbox{2(b):}\qquad s_1=B\sin\varphi, \quad s_3=B\cos\varphi.
 \end{equation}
and vanishing on-site terms due to symmetry constaints.

For both 2(a) and 2(b) states, since the ratios of hopping and pairing amplitudes are the same in all NN bonds, one can always perform a global gauge rotation and eliminate e.g. the pairing ($\tau_1$) term so that $\varphi=s_1=0$. These correspond to their root $U(1)$ spin liquids with NN amplitudes only: $U1^0$ state (i.e. $U(1)$-uniform state in Ref.\cite{Lawler2008}) for case 2(a), and $U1^1$ state (i.e. $U(1)$-staggered state in Ref.\cite{Lawler2008}) for case 2(b).  More generally such a global gauge rotation can fix $\varphi$ to any desired value with the same spectrum. Therefore we cannot realize a $Z_2$ spin liquid with up to NN mean-field amplitudes for states 2(a) and 2(b).

In comparison, the mean-field Hamiltonians of 2(a) case has the same (unit) hopping amplitudes on all NN bonds, while bonds in neighbouring triangles have opposite (unit) hopping amplitudes in case 2(b). Their spinon dispersions $\varepsilon_\bbk^{(2a)}, \varepsilon_\bbk^{(2b)}$ are plotted by the blue lines in Fig.\ref{fig:singletdispersion} for $\theta=0$ and $\theta=\pi/2$ respectively. We see that case 2(a) corresponds to a $U(1)$ spin liquid ($U1^0$ state) with spinon Fermi surfaces, while 2(b) case leads to $U1^1$ state featuring spinon Dirac cones at the fermi energy.

Finally in case 2(c), the NN hopping/pairing ratios are not the same on different bonds, and they are hence independent variables that cannot be fixed by a global gauge rotation. The spectrum of mean-field Hamiltonian $\{ H^{(2c)}[\theta]\}$ depends on the parameter $\theta\equiv \arcsin(s_1/s_3)$ in this case. In particular, in two limits where only hopping ($\theta=0$) or pairing $(\theta=\pi/2$) terms are present, the Hamiltonian of state 2(c) reduces to those of 2(a) and 2(b) cases respectively,
\begin{equation}
\begin{array}{l}
H^{(2c)}[\theta=0]=H^{(2a)}\rightarrow U1^0~\text{state}\\
H^{(2c)}[\theta=\pi/2]=H^{(2b)}\rightarrow U1^1~\text{state}
\end{array}
\end{equation}
In the intermediate regime $0<\theta<\pi/2$, case 2(c) describes a $\mathbb{Z}_2$ spin liquid with both hopping and pairing terms, whose typical dispersion is illustrated in Fig.\ref{fig:singletdispersion} for $\theta=\pi/6$. In summary, in the isotropic limit with $SU(2)$ spin rotational symmetry, only the 2(c) case can realize a $Z_2$ spin liquid with up to NN mean-field amplitudes, and it bridges the two root $U(1)$ spin liquids ($U1^0$ and $U1^1$ states) for 2(a) and 2(b) cases. In fact, this state was investigated in Ref.\cite{Lawler2008}, where a variational Monte-Carlo study of Gutzwiller projected wavefunctions found that the $U1^0$ spin liquid state with $\theta=0$ has the lowest variational energy.

\subsubsection{Anisotropic case with spin-orbit couplings}

Now let's consider the effect of spin-orbit couplings which breaks spin rotational symmetry, and include the spin-triplet terms in the mean-field ansatz. The spin-triplet terms (Table \ref{tab:mftriplet}) involve much more parameters than singlet ones and contain far more features. Before focusing on a special class of symmetric spin liquids motivated by an anisotropic spin model describing Na$_4$Ir$_3$O$_8$ in section \ref{sec:modelandMF}, here we discuss some general features of the anisotropic spin liquids in comparison to the singlet ones.

For singlets terms, we choose the gauge so that $s_1=0$ for 2(a) and 2(b) cases as discussed earlier. For the NN triplet amplitudes $(t^{x,y,z}_1, t^{x,y,z}_3)$ on certain bond (Table\ref{tab:mftriplet}), one can choose a suitable spin quantization axis such that the only non-vanishing parameters are $t^z_1,  t^x_3, t^y_3, t^z_3$ for all the three cases 2(a,b,c). Then we can summarize the free parameters and the properties for the three states in the presence of spin anisotropy:
\begin{itemize}
\item 2(a): On-site amplitudes $\Delta, \mu$, NN singlet amplitude $s_3$ and NN triplet amplitudes $t^z_1,   t^x_3, t^y_3, t^z_3$. A $\mathbb{Z}_2$ spin liquid can be realized as long as $t^z_1\ne 0$ i.e. with NN triplet pairing terms, otherwise it corresponds to a $U(1)$ spin liquid $U1^0$ state. The NN triplet pairings can open up a gap on the spinon fermi surface.
\item 2(b): No on-site amplitudes allowed, NN singlet amplitude $s_3$ and NN triplet amplitudes $t^z_1,  t^x_3, t^y_3, t^z_3$. Similarly, when $t^z_1\ne 0$ we realize a $\mathbb{Z}_2$ spin liquid with NN triplet pairings; otherwise it is a $U(1)$ spin liquid $U1^1$ state. The inclusion of triplet terms will change the linear Dirac-type dispersion in the singlet case, but generically can never open up a gap in the spinon spectrum.

\item 2(c): On-site amplitude $\mu$, NN singlet amplitudes $s_1, s_3$ and  NN triplet amplitudes $t^z_1,  t^x_3, t^y_3, t^z_3$. Generally 2(c) state is separated from 2(a) state in the phase diagram by an intermediate $U(1)$ spin liquid $U1^0$ state, and from 2(b) state by an intermediate $U1^1$ state. In the absence of pairing terms i.e. $s_1=t^z_1=0$, the 2(c) state reduces to a symmetric $U(1)$ spin liquid. This is the situation we will focus on in the next section.
\end{itemize}
Generally for all three states, the inclusion of triplet terms will lift the spin degeneracy in spinon spectrum, splitting each spinon band into two bands. In particular, the 4-fold degenerate flat bands will become dispersive and non-degenerate.
We found that state 2(b) always has a gapless spinon spectrum while states 2(a) and 2(c) can be gapped in certain parameter ranges, consistent with the general proof provided in section \ref{subsec:gapless proof}. Hence state 2(b) provides an interesting example of stable gapless $Z_2$ spin liquids\cite{Lu2016b}, whose gapless excitations are protected by only space group and time reversal symmetries with no spin conservations.

\section{The Spin Model for $\mbox{Na}_4\mbox{Ir}_3\mbox{O}_8$ and Its Energetics\label{sec:modelandMF}}

In previous sections, we have classified symmetric spin liquid states by their projective symmetry groups in the slave-fermion representation. Now we consider energetics of physical spin models, and focus on whether certain spin Hamiltonian will give rise to the spin-triplet mean-field amplitudes in the spin liquid states. For this purpose, we will only compare energy among different spin liquid states, and leave the energetic competition between spin liquids and various magnetically ordered states\cite{Shindou2016,Mizoguchi2016} to future works. Also, we will restrain ourselves to nearest neighbour (NN) interactions only, as the NN Heisenberg model on hyperkagome lattice may already lead to a disordered spin liquid ground state\cite{Lawler2008,Kimchi2014a}. At first sight, one may think that as long as the physical Hamiltonian breaks the spin-rotational symmetry, the spin-anisotropic states will automatically be favored energetically. However, we shall see that this needs not to be the case, due to the strong suppression of spin anisotropy by Heisenberg-type interactions.

As discussed in the introduction, we consider a Hamiltonian with a dominant NN Heisenberg term with various spin-anisotropic perturbations\cite{Chen2008,Micklitz2010a,Mizoguchi2016},
\begin{equation}\label{heisenberg}
 H = H_J+H',\quad H_J = J\sum_{\langle i,j\rangle}\mathbf{S}_i\cdot \mathbf{S}_j.
 \end{equation}
The most general perturbation Hamiltonian can be decomposed into the Dzyaloshinskii-Moriya (DM) term $H_D$, Kitaev term $H_K$, and symmetric exchange anisotropic (SEA) term $H_\Gamma$,
 \begin{eqnarray}\nonumber
 H' &=& H_D + H_K + H_\Gamma\\ \nonumber
&=&\sum_{\langle i,j\rangle} \mathbf{D}_{ij}\cdot(\mathbf{S}_i\times \mathbf{S}_j) + \sum_{\langle i,j\rangle,A} K_{ij}^A S_i^A S_j^A +  \sum_{\langle i,j\rangle, B\ne C} \Gamma_{ij}^{BC} S_i^B S_j^C. \\\label{perturbation}
 \end{eqnarray}
 where $|D_{ij}^A|, |K_{ij}^A|, |\Gamma_{ij}^{AB}|$ may be much smaller than $J$, and $A, B, C$ are Cartesian coordinates as before.
 Each term involves three independent interaction parameters on each bond $\langle i,j\rangle$.   Now we make use of the symmetry of hyperkagome lattice to relate interaction parameters in different bonds, especially the space group generated by  24 ``point group'' operations $C_2^{\nu_2}C_3^{\nu_3}S_4^{\nu_4}$ and 3 unit-cell translations. Since physical spins $\{\mathbf{S}_i\}$ also rotate under point group operations in a spin-orbit coupled system, we have the symmetry constraints on the physical Hamiltonian
 \begin{equation}\label{hphysicalsymm}
 H[S^A_i] = UHU^\dagger = H[\sum_{B}O_{AB}^{(U)} S^B_{U(i)}],
 \end{equation}
 where $O^{(U)}$ are SO(3) matrices (\ref{so3matrices}) associated with point group operation $U$. Therefore we are left with totally 9 parameters $\{D^A,K^A,\Gamma^A\}$ for all NN couplings:
\begin{eqnarray}\nonumber
 H' &=&\sum_A\Big[D^A \sum_{\langle i,j\rangle} (S_i^{B} S_j^{C}-S_i^{C} S_j^{B} )+ K^A \sum_{\langle i,j\rangle } S_i^{A} S_j^{A} \\ \label{fullenergy}
 && \quad + \Gamma^A \sum_{\langle i,j\rangle} (S_i^{B} S_j^{C}+ S_i^{C}S_j^{B})\Big].
 \end{eqnarray}

In a mean-field analysis of the above spin model in slave fermion representation, each NN bond contribute the same energy due to symmetry and hence we will focus on one band, say the bond $(2,3)$ connecting sublattice No.2 and No.3 from now on.
We make a further simplification in the discussion below that we only consider $D^z, K^z, \Gamma^z\ne 0$ on bond $(2,3)$, resulting in the following mean-field energy  (see Appendix \ref{appendix:MFcal})
  \begin{eqnarray}\nonumber
{\cal E} = \frac{\langle H\rangle}{N_{\mbox{\scriptsize site}}} &=& 2\left(J\langle \mathbf{S}_i \cdot \mathbf{S}_j\rangle
+
D\langle S_i^xS_j^y - S_i^yS_j^x\rangle \right. \\ \label{energyfunc}
&&\left. +
K \langle S_i^z S_j^z \rangle
+
\Gamma \langle S_i^xS_j^y + S_i^yS_j^x \rangle\right)_{(i\in 2,j\in 3)}
\end{eqnarray}
Such a simplification has been adopted in recent literatures \cite{Shindou2016,Mizoguchi2016} when considering an idealized Na$_4$Ir$_3$O$_8$ structure.  However, the essential physics we will discuss below does not rely on this assumption, and one can directly generalize this discussion to include all 9 interaction parameters in (\ref{fullenergy}) for reasons elaborated in Appendix \ref{appendix:MFcal}.

     \begin{figure}
  [h]
  \includegraphics[width=6cm]{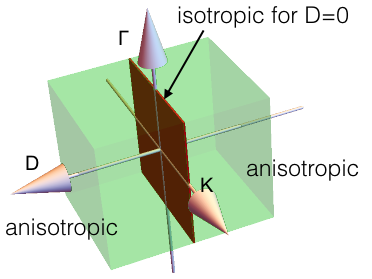}
  \caption{\label{fig:phasediag} Phase diagram for a dominant Heisenberg interaction ($J\equiv 1$) with perturbations of DM, Kitaev, and symmetric exchange anisotropy (SEA) types in mean-field energy (\ref{energyfunc}). We consider the regime where perturbations are small $|D|, |K|, |\Gamma|\le 0.2$. While an infinitesimal DM interaction can induce anisotropy in the spin liquid ground state, the Kitaev and SEA terms both need a finite strength $K, \Gamma$ comparable to Heisenberg $J$ to induce spin anisotropy.
  }
  \end{figure}
We have performed variational calculation of the mean-field energy (\ref{energyfunc}) for all three $Z_2$ spin liquid states 2(a,b,c) in Table \ref{tab:mfsinglet} and \ref{tab:mftriplet}. We set the Heisenberg interaction strength $J$ as unity, and consider perturbations $|D|, |K|, |\Gamma| \le 0.2$ in (\ref{energyfunc}). We first note that state 2(b) has much higher energy than 2(a) and 2(c) in the isotropic spin-singlet case\cite{Lawler2008}. As we only consider perturbative effects of anisotropic terms, we mainly focus on 2(a) and 2(c) states here.

The variational energy of singlet state 2(c) for isotropic Heisenberg interactions had been calculated in Ref.\cite{Lawler2008}, both at mean-field level and for projected wavefunctions using variational Monte-Carlo method. The ground state is a $U(1)$ spin liquid ($U1^0$ state) with spherical Fermi surfaces at the center and corners of the Brillouin zone\cite{Zhou2008a}, which is connected to two $\mathbb{Z}_2$ spin liquids 2(a) or 2(c) via continuous phase transitions (by turning up spinon pairing terms). Our variational calculation shows that even in the presence of anisotropic interactions, the $U1^0$ (i.e. $U(1)$-uniform) state is still favored energetically over both $\mathbb{Z}_2$ spin liquids 2(a) and 2(c). In other words, the spinon pairing amplitudes $s_1, t^x_1, t^y_1, t^z_1$ vanish identically in mean-field ground state for all parameter regimes we consider. In the absence of pairings, 2(a) and 2(c) states reduce to the same $U1^0$ spin liquid state as we discussed earlier. Then the question is whether a weak spin-orbit coupling ($|D|, |K|, |\Gamma|\ll J\equiv 1$) can induce an anisotropic $U(1)$ spin liquid ground state ($t^x_3, t^y_3, t^z_3\ne0$ in Table \ref{tab:mftriplet}). The variational results show a dramatic difference in the effects of DM interactions from those of Kitaev and SEA types, as shown in Fig.\ref{fig:phasediag}. An infinitesimal DM interaction is sufficient to induce anisotropy in the mean-field ground state, where the magnitude of anisotropy is proportional to the strength of DM interaction. For Kitaev or SEA types of perturbations, on the other hand, the spin liquid ground state remains spin rotational invariant unless $|K|, |\Gamma|$ become comparable to $J$.

To better understand the big difference between DM and other types of anisotropic interactions, below we analyze the mean-field energy (\ref{energyfunc}). The  mean-field decompositions consist of the (isotropic) singlet part $(\rho_s, \Delta_s)$ and (anisotropic) triplet parts $(\rho_{x,y,z}, \Delta_{x,y,z})$. The hopping amplitudes read
\begin{eqnarray}
&& \rho_s = \frac{\langle f_{i\uparrow}^\dagger f_{j\uparrow} \rangle + \langle f_{i\downarrow}^\dagger f_{j\downarrow}\rangle}{2},\quad
\rho_z = \frac{\langle f_{i\uparrow}^\dagger f_{j\uparrow} \rangle - \langle f_{i\downarrow}^\dagger f_{j\downarrow}\rangle}{2i},\\
&& \rho_x = \frac{\langle f_{i\uparrow}^\dagger f_{j\downarrow}\rangle - \langle f_{i\downarrow}^\dagger f_{j\uparrow}\rangle}{2i}, \quad
\rho_y = \frac{\langle f_{i\uparrow}^\dagger f_{j\downarrow}\rangle + \langle f_{i\downarrow}^\dagger f_{j\uparrow}\rangle}{2};
\end{eqnarray}
and pairing amplitudes are
\begin{eqnarray}
&& \Delta_s = \frac{\langle f_{i\uparrow}^\dagger f_{j\downarrow}^\dagger\rangle - \langle f_{i\downarrow}^\dagger f_{j\uparrow}^\dagger\rangle}{2},\quad
\Delta_z = \frac{\langle f_{i\uparrow}^\dagger f_{j\downarrow}^\dagger\rangle + \langle f_{i\downarrow}^\dagger f_{j\uparrow}^\dagger\rangle}{2i},\\
&& \Delta_x = \frac{\langle f_{i\uparrow}^\dagger f_{j\uparrow}^\dagger \rangle + \langle f_{i\downarrow}^\dagger f_{j\downarrow}^\dagger\rangle}{2},\quad
\Delta_y = \frac{\langle f_{i\uparrow}^\dagger f_{j\uparrow}^\dagger \rangle - \langle f_{i\downarrow}^\dagger f_{j\downarrow}^\dagger\rangle}{2i},
\end{eqnarray}
where the time-reversal symmetry guarantees $\rho_{s,x,y,z}, \Delta_{s,x,y,z}$ to be real numbers (see Appendix \ref{appendix:MFcal} for details). Here $(\rho_x, \rho_y, \rho_z)$ as well as $(\Delta_x, \Delta_y,\Delta_z)$ are chosen such that they rotate as SO(3) vectors under $SU(2)$ spin rotations. These amplitudes are related to the mean-field parameters in Table \ref{tab:mfsinglet}-\ref{tab:mftriplet} through consistent mean-field equations, whose details can be found in Appendix \ref{appendix:MFcal}. The mean-field amplitudes for different NN bonds $\langle i,j \rangle$ are illustrated in Fig. \ref{fig:bondxyz}. Here focusing  on one bond $(i=2, j=3)$, we can write the mean-field energy per site as
\begin{eqnarray}
{\cal E}& \equiv & \frac{\langle H\rangle}{N_{\mbox{\scriptsize site}}} = {\cal E}_J + {\cal E}_D + {\cal E}_K + {\cal E}_\Gamma\\ \label{mfEJmain}
{\cal E}_J &=& -J\left[
3(\rho_s^2+ \Delta_s^2) -\rho_z^2 - \rho_x^2-\rho_y^2 - \Delta_z^2-\Delta_x^2-\Delta_y^2
\right]
\\
{\cal E}_D &=& 4D\left[ \Delta_s \Delta_z + \rho_s \rho_z
\right]
\\
{\cal E}_K &=& K \left[
\Delta_x^2+\Delta_y^2 - \Delta_s^2 - \Delta_z^2 + \rho_x^2+\rho_y^2 - \rho_s^2-\rho_z^2
\right]
\\
{\cal E}_\Gamma &=& 4\Gamma \left[
\Delta_x\Delta_y + \rho_x\rho_y
\right]
\end{eqnarray}
From the expression (\ref{mfEJmain}) it is clear that the Heisenberg term will favor singlet amplitudes and suppress all triplet amplitudes. Due to the dominant role of Heisenberg interaction $J\gg |D|, |K|, |\Gamma| $, the ground state has a strong tendency towards having an isotropic spin liquid ground state. However, the DM interaction can survive the suppression of spin anisotropy by coupling the triplet amplitudes $\rho_z, \Delta_z$ with singlet amplitudes $\rho_s, \Delta_s$. Since Heisenberg energy only depends quadratically on the triplet amplitudes while DM energy depends linearly on triplet terms, an arbitrary small $D/J$ will lead to non-zero triplet amplitudes in the presence of a finite singlet amplitude. Though the above analysis considers only one component $D^z, K^z, \Gamma^z$ in the anisotropic interactions, the conclusion remains valid even when all 9 parameters $\{D^A, K^A, \Gamma^A\}$ are considered in the spin Hamiltonian. Thus, it is expected that for the real material Na$_4$Ir$_3$O$_8$ where all anisotropic perturbations are present\cite{Mizoguchi2016}, it is up to strength of DM interactions to decide how anisotropic the spin liquid ground state is.

\begin{figure}
[h]
\begin{center}
\includegraphics[width=4cm]{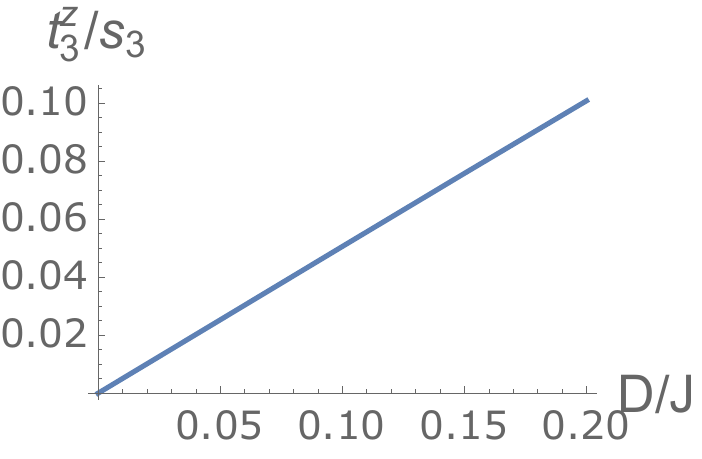}
\includegraphics[width=4.5cm]{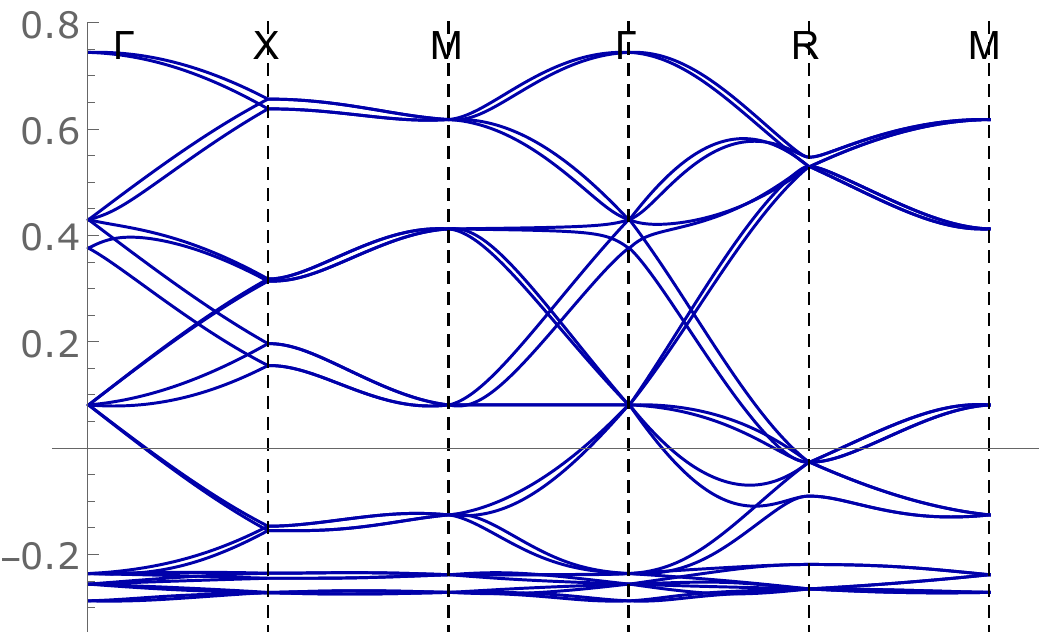}
\end{center}
\caption{{\bf Left:} Mean field variational results showing that the DM interaction $D $ induces triplet amplitude $t^z_3$ (in bond $(2,3)$, amplitudes in other bonds can be referred from Fig. \ref{fig:bondxyz}). The dependence is almost linear for small $D/J$. {\bf Right:} Spinon dispersion for $D/J=0.2$, where mean-field energy is minimized at $ s_3 =  0.16584, t^y_3= 0.01672, \mu=-0.079$.  \label{fig:dmvariation}
}
\end{figure}

Due to the important role of DM terms, we examine more carefully the spin liquid states in the presence of only Heisenberg and DM interactions. Again, for simplicity we assume an idealized crystal structure where only one DM term per bond is present, i.e. the model in (\ref{energyfunc}). The variational calculation gives an almost linear dependence of the triplet-singlet amplitude ratio $t^z_3/s_3$ on the strength of DM interaction for small $D/J$, as shown in Fig.\ref{fig:dmvariation}. The corresponding dispersion of the spin-anisotropic $U(1)$ spin liquid $U1^0$ state at $D/J=0.2$ is also shown in Fig.\ref{fig:dmvariation}. Compared with Fig.\ref{fig:singletdispersion}(vi) we clearly see the splitting of bands due to spin-orbit coupling. In particular, the 4-fold degeneracy of flat-bands in isotropic case is fully lifted and all bands disperse.

\begin{figure}
[h]
\parbox{4cm}{
\includegraphics[width=4cm]{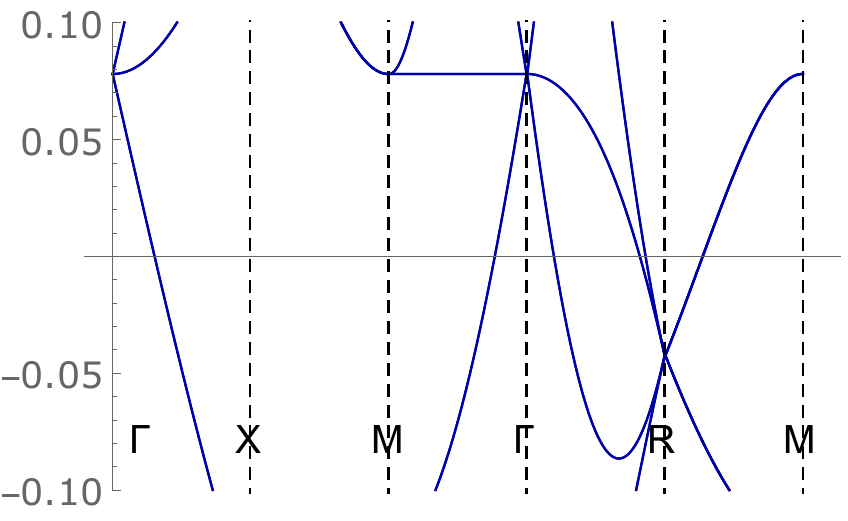}\\
\includegraphics[width=4cm]{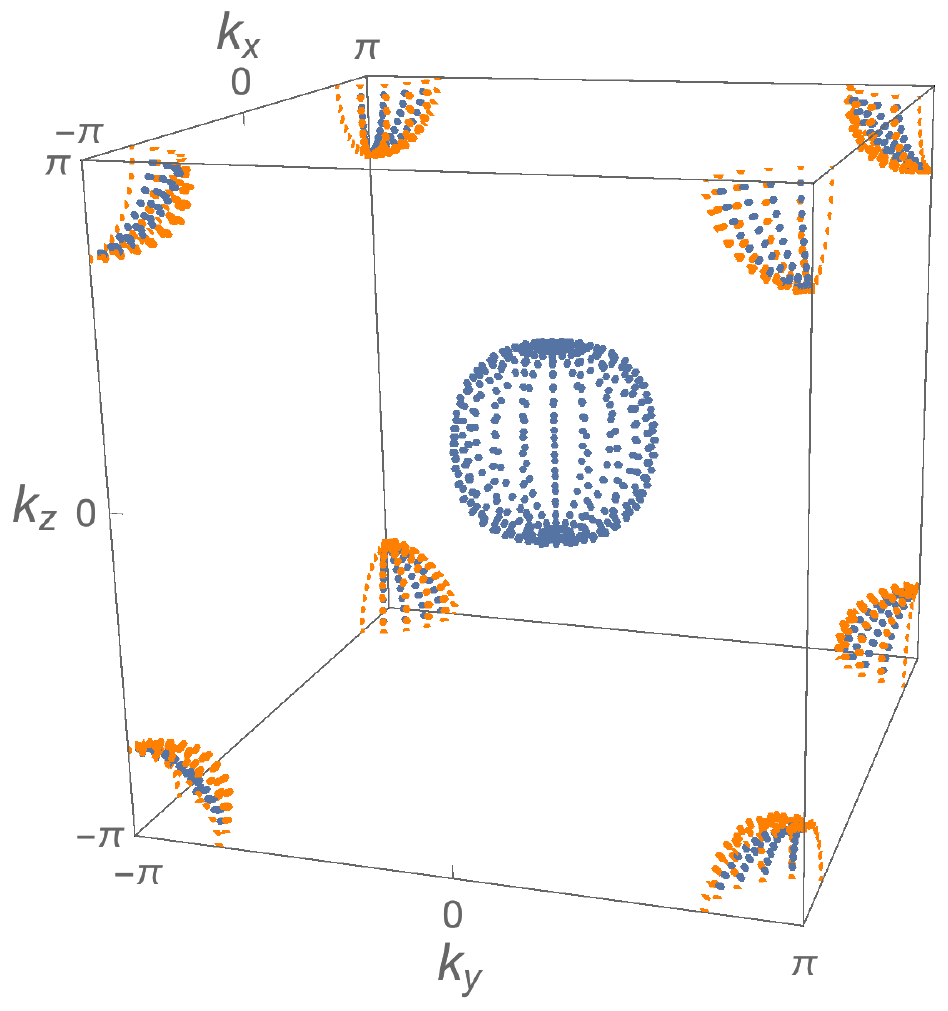}
\begin{center}
(i)
\end{center}
}\quad
\parbox{4cm}{
\includegraphics[width=4cm]{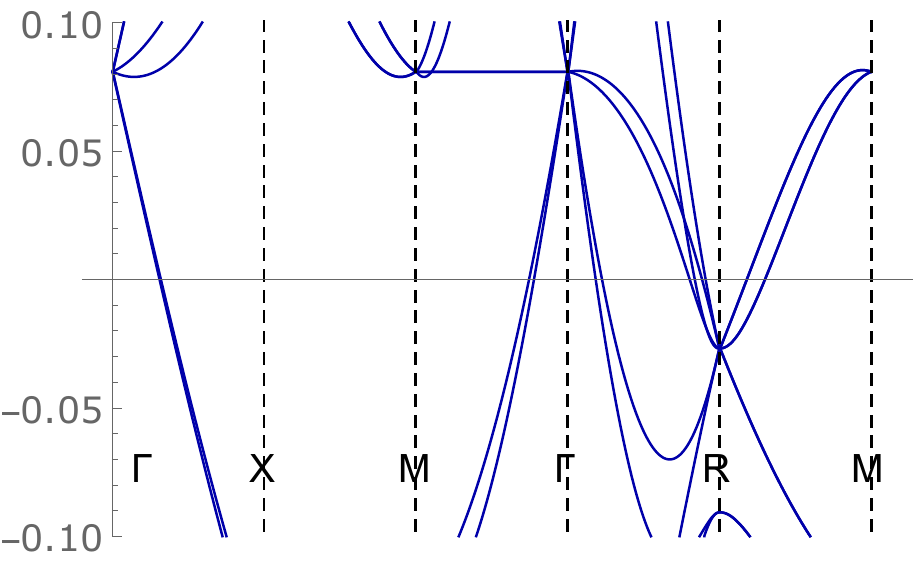}
\includegraphics[width=4cm]{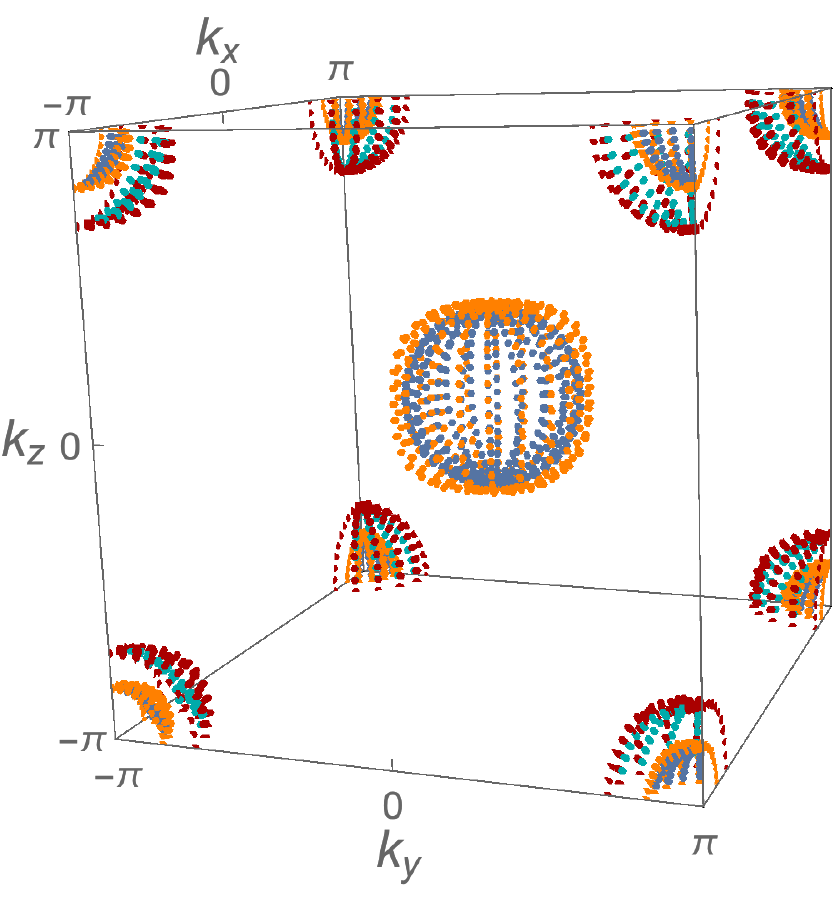}
\begin{center}
(ii)
\end{center}
}
\caption{The spinon Fermi surfaces of energetically favored $U(1)$ spin liquids. (i): Spin-isotropic (singlet) $U(1)$ spin liquid with $s_3= 0.16456, \mu=-0.078$.
(ii): Spin-anisotropic $U(1)$ spin liquid induced by DM interaction $D/J=0.2$, where variational calculation gives the mean-field amplitudes $ s_3 =  0.16584, t^y_3= 0.01672, \mu=-0.079$. The fermi surfaces at the corner belong to 9-12th bands, while the two Fermi surfaces at the BZ center belong to 13th (orange) and 14th (blue) bands.
 \label{fig:fsballs}
}
\end{figure}

\begin{figure}
[h]
\begin{center}
\parbox{2.5cm}{
\includegraphics[width=2.5cm]{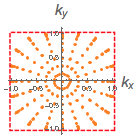}\\
\includegraphics[width=2.5cm]{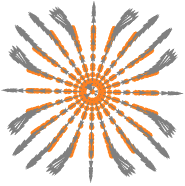} \\
\begin{center}
Top
\end{center}
}
\parbox{2.5cm}{
\includegraphics[width=2.5cm]{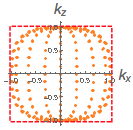}\\
\includegraphics[width=2.5cm]{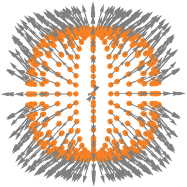}
\begin{center}
Front
\end{center}
}
\parbox{2.5cm}{
\includegraphics[width=2.5cm]{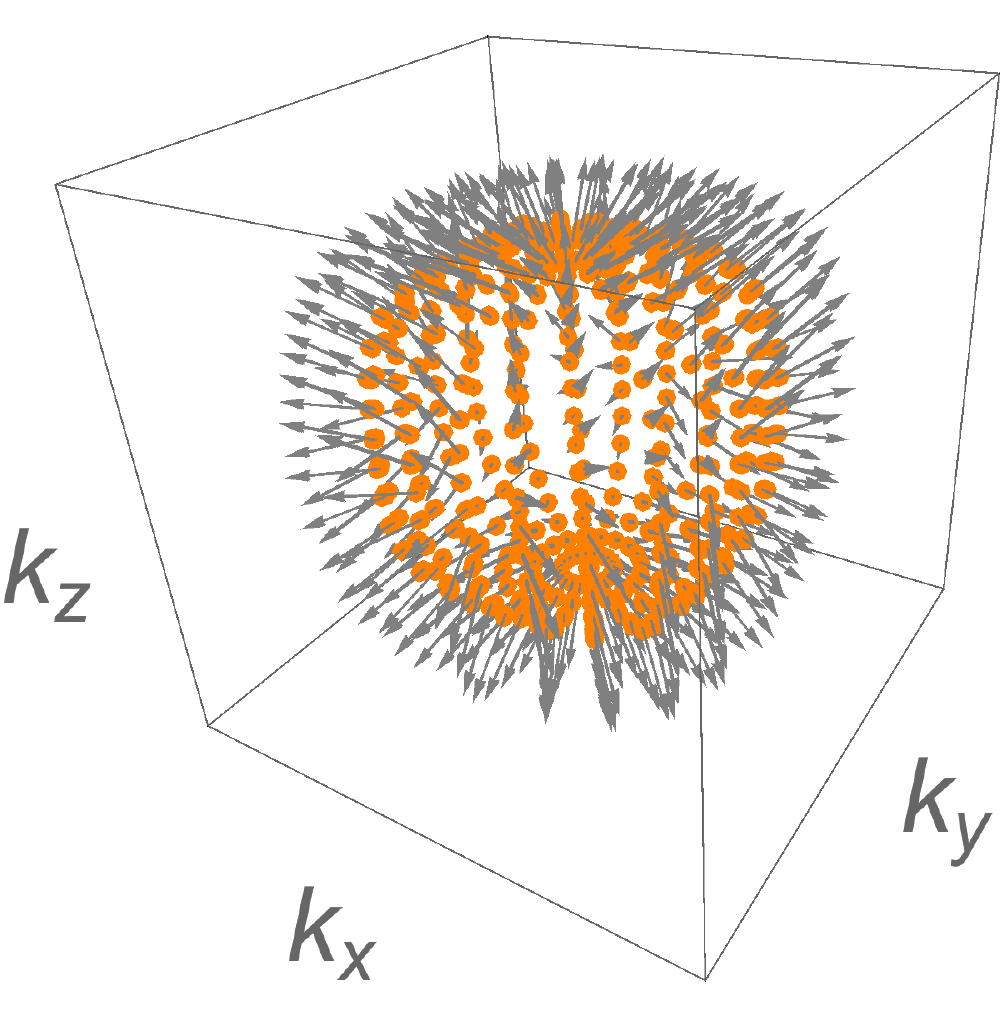}
}
\end{center}
\caption{Distorted shapes and spin textures of the spinon fermi surface in the presence of DM interactions $D/J=0.2$ from spin-orbit couplings. Here we plot the 13-th band in Fig.\ref{fig:fsballs} (ii), shown by orange dots surrounding the center of Brillouin zone. Arrows denote the spin textures on the fermi surface, with clear signatures of spin-momentum locking. The dashed red boxes here are guides to the eye, showing the spherical fermi surface in isotropic limit is distorted towards a cubic shape in the anisotropic case. \label{fig:fstextures}
}
\end{figure}

Finally we take a closer look at the Fermi surfaces of the anisotropic $U(1)$ spin liquid ground state in the presence of DM interactions. The Fermi surfaces in the spin-isotropic limit have been discussed in earlier literature\cite{Zhou2008a}, where each Fermi surface has double spin-degeneracy.  When triplet amplitudes are turned on, the spin degeneracy of the Fermi surfaces are completely lifted, resulting in 6 Fermi surfaces with spin textures shown in Fig.\ref{fig:fsballs} and \ref{fig:fstextures}. The 9-12th bands contribute four Fermi surfaces at the corners of Brillouin zone (BZ), and the 13-14th bands contribute the Fermi surfaces at the BZ center. Fig.\ref{fig:fstextures} clearly demonstrates the spin-momentum locking on the non-degenerate spinon fermi surface, due to spin-orbit coupling in this non-centrosymmetric crystal.

%


\section{Discussion and Outlook \label{sec:summary}}

In this work, we classify symmetric $\mathbb{Z}_2$ and $U(1)$ spin liquids on a hyperkagome lattice in the slave-fermion representation. Non-symmorphic space group provides strong symmetry constraints to spinon symmetry fractionalization on hyperkagome lattice, and we found only three $Z_2$ spin liquids and two $U(1)$ spin liquids as shown in TABLE \ref{tab:psg}. The small number of possible spin liquid states should greatly facilitate further analysis of their properties and more accurate calculations of energies in the future, such as a variational Monte-Carlo study of the Gutzwiller projected wavefunctions. We show that both $U(1)$ states have a stable gapless spectrum protected by space group and time reversal symmetries\cite{Parameswaran2013,Watanabe2015}. Meanwhile among all 3 $Z_2$ states, we show that states 2(a) and 2(c) are generically gapped, while state 2(b) is a symmetry-protected gapless $Z_2$ spin liquid\cite{Lu2016b,Qi2016}. As shown schematically in FIG. \ref{fig:sketch phase diagram}, two $Z_2$ spin liquids 2(a) and 2(c) states share the same ``root'' $U(1)$ spin liquid $U1^0$ state, while 2(b) and 2(c) share another root state $U1^1$ state.

{Experimentally, specific heat, magnetic susceptibility and thermal conductivity measurements\cite{Okamoto2007,Singh2013} all pointed to a gapless spin liquid ground state in Na$_4$Ir$_3$O$_8$. Therefore our classification and analysis on gapped vs. gapless spectrum can already narrow down the list of promising spin liquids to 3 candidates, including two $U(1)$ spin liquids ($U1^0$ and $U1^1$ states) and one $Z_2$ spin liquid 2(b) state. Which one of them is realized as the spin liquid ground state in Na$_4$Ir$_3$O$_8$ is further determined by their energetics. Our mean-field calculations suggest that $U1^0$ state is energetically favored over other candidates.}

%
%
%

Motivated by the recent experiments\cite{Dally2014,Shockley2015} on Na$_4$Ir$_3$O$_8$ uncovering the low-energy spin anisotropy properties, we examined the effects of anisotropic interactions on symmetric spin liquid states. The spin anisotropy that breaks spin rotational symmetry on the spin liquid ground states is studied at the mean field level, and we found a drastic difference between DM interactions and other types of anisotropic interactions (Kitaev and symmetric anisotropy). While an infinitesimal DM interaction will always induce spin-triplet couplings in the mean-field ansatz, the other anisotropic interactions of Kitaev and symmetric anisotropy types need a finite threshold comparable to the Heisenberg interaction to break spin rotational symmetry in mean-field ground state. In the mean-field energy analysis, we attribute such a difference to the strong suppression of spin anisotropy by Heisenberg interaction, and that only DM interaction can couple singlet and triplet amplitudes bilinearly in the energy functional that survives the suppression of spin anisotropy. It is interesting to see whether higher order quantum fluctuations would inherit or modify such a mean-field picture in future works.

Finally, we point out that hyperkagome lattice structure is also found in the material PbCuTe$_2$O$_6$ in recent experiments\cite{Koteswararao2014,Khuntia2016}. {The three-dimensional network of Cu$^{2+}$ constitutes a hyperkagome lattice, where both thermodynamic\cite{Koteswararao2014} and $\mu$SR, NMR\cite{Khuntia2016} experiments suggest a spin liquid ground state, a gapless one as indicated by NMR data\cite{Khuntia2016}. Therefore, our classification and gapless analysis of spin liquid states on hyperkagome lattices pave the road for further understanding the property of PbCuTe$_2$O$_6$ in the future.}

\section*{Acknowledgement}
This work was supported by NSF grant, DMR-1309615 at the Ohio State University, U.S. ARO (W911NF-11-1-0230) and AFOSR (FA9550-16-1-0006) at the University of Pittsburg (BH), the start-up funds at the Ohio State University (YML), the NSERC of Canada, Canadian Institute for Advanced Research, the Center for Quantum Materials at the University of Toronto (YBK).

\appendix

\section{Symmetry group of hyperkagome lattice and its $Z_2$ extension\label{app:group extension}}

\subsection{Conventions of site labeling and symmetry operations\label{appendix:C2}}
Here we make connections of our site labeling of sublattices and the symmetry operations to previous literatures. Our sublattice site labelings are consistent with those in \cite{Chen2008,Shindou2016}, while the $\hat x$-$\hat y$ axes in those literatures are rotated to $(-\hat y)$-$(\hat x)$ directions.
In many literatures \cite{Chen2008,Shindou2016,Mizoguchi2016}, instead of the 4-fold screw rotation $S_4$, a set of 12 two-fold rotations are used to describe the symmetry of hyperkagome lattice. Each sublattice sites are associated with one 2-fold rotation axis passing through it. In Table \ref{localc2} we express those 2-fold rotations in our convention, in terms of the point group operations followed by translations $U=T_1^{n_1}T_2^{n_2}T_3^{n_3} C_2^{\nu_2} C_3^{\nu_3} S_4^{\nu_4}$.
\begin{table}
[h]
\caption{All 2-fold rotations expressed in our convention. Using the point group symmetry operations, we can form 12 two-fold rotations $U^2=\bse$. Each $U$ has the rotation axis passing through one of the sublattice sites $s=1,\dots, 12$. \label{localc2}}
\begin{tabular}{|c|c|c|c|c|c|c|}
\hline
$s$ & 1 & 2 & 3 & 4 & 5 & 6 \\
\hline
$U$ & $C_2C_3$ & $T_3C_2C_3^2$ & $T_1T_3^{-1}C_2S_4^2$ & $T_1^{-1}T_3C_3S_4$ & $T_1T_2T_3^{-1}C_3^2S_4^3$ & $T_1C_2$ \\
\hline \hline
$s$ & 7 & 8 & 9 & 10 & 11 & 12 \\
\hline
$U$ & $C_2C_3^2$ & $T_2C_2C_3$ & $C_2$ & $C_3^2S_4^3$ & $T_2T_3^{-2}C_2S_4^2$ & $T_2^{-1}C_3S_4$ \\
\hline
\end{tabular}
\end{table}

\subsection{Space group symmetry\label{appRotation}}
The  independent group elements in the symmetry group are
 \begin{equation}\label{standardU}
  U={\cal T}^{\nu_{\cal T}} T_1^{\nu_{T_1}} T_2^{\nu_{T_2}} T_3^{\nu_{T_3}}  C_2^{\nu_{C_2}}C_3^{\nu_{C_3}}
  S_4^{\nu_{S_4}}
 \end{equation}
 where $\nu_{\cal T}\in\mathbb{Z}_2$, $\nu_{T_1,T_2,T_3}\in\mathbb{Z}$, $\nu_{C_2}\in \mathbb{Z}_2$,  $\nu_{C_3}\in \mathbb{Z}_3$, $\nu_{C_{4X}}\in \mathbb{Z}_4$. Here the operations consist of time reversal $\cal T$,
3 translations and the 24 point-group symmetry formed by $C_2, C_3, S_4$ as discussed in the main text. As will be shown below, the 3 translations $T_i$ can be generated by screw operation $S_4$ and rotation $C_3$, therefore the above labeling (\ref{standardU}) for space group elements is in fact redundant. Although it inevitably leads to redundant algebraic relations between group elements below, this labeling scheme keeps track of all algebraic relations without missing any.

The algebraic relations among the operators are
 \begin{eqnarray}\label{sgfirst}
&&
 {\cal T}^2=C_2^2=C_3^3=e,\quad
S_4^4=T_1\\
&&
 U{\cal T} = {\cal T}U,  U\in \{ T_i, C_2, C_3, S_4\}.
\\  \label{sgthird}
&&
C_2T_1=T_1^{-1}C_2,\quad  C_2T_2=T_3^{-1}C_2, \quad C_2T_3=T_2^{-1}C_2\\
&&
C_3T_1=T_2C_3, \quad C_3T_2=T_3C_3, \quad C_3T_3=T_1C_3
\\
&&
S_4T_1=T_1S_4, \quad S_4T_2=T_3S_4,\quad S_4T_3=T_2^{-1}S_4
\\ \label{sglast}
&&C_3C_2=C_2C_3^2,\quad  S_4C_2= T_1T_2^{-1}C_2S_4^3= T_2^{-1}C_2S_4^{-1} ,\\
&&
S_4C_3=T_1C_3^2S_4^3 = T_1T_3C_3^{-1}S_4^{-1},\\
&&S_4C_3^{-1} = T_1^{-1} T_3 C_2 C_3^{-1}S_4^2\label{sglastreal}
 \end{eqnarray}
These relations allows one to write operations in arbitrary sequence as the standard form  in Equation (\ref{standardU}).

Note that the commutation relation between translations $[T_i, T_j]=0$ is actually not independent due to the presence of screw rotation symmetry. Explicitly, we have
\begin{eqnarray*}
&&
T_2 = (C_2S_4^{-1})^2, \qquad
T_3 = (C_2S_4)^2\\
\Rightarrow
&&T_2T_3T_2^{-1}T_3^{-1}\\
&=&
C_2S_4^{-1}C_2S_4^{-1} T_3T_2^{-1}S_4^{-1}C_2S_4^{-1}C_2\\
&=&
(C_2S_4C_2S_4T_2T_3^{-1} S_4C_2S_4C_2)^{-1}\\
&=&
(T_2T_3^{-1}C_2S_4C_2S_4S_4 C_2S_4C_2)^{-1}\\
&=& (T_2T_3^{-1}T_3T_2^{-1})^{-1} = 1.
\end{eqnarray*}
Then sandwiching the above relation repeatedly by $C_3 (\dots) C_3 $ we have all $T_iT_jT_i^{-1}T_j^{-1} = \mathbb{I}, i,j=x,y,z$.

The constraints can be rewritten in an alternative form as
 \begin{eqnarray}\label{sgnewfirst}
&&
 {\cal T}^2=C_2^2=C_3^3=\bse,\quad
T_1^{-1}S_4^4=\bse,\\
&&
 {\cal T}^{-1} U^{-1}{\cal T}U = \bse,~~U\in [T_i, C_2, C_3, S_4]
\\
&&
T_1^{-1}C_2T_1^{-1}C_2 = T_2^{-1}C_2T_3^{-1}C_2 = T_3^{-1}C_2T_2^{-1}C_2 = \bse,
\\
&&
T_1^{-1}C_3^{-1}T_2C_3 = T_2^{-1}C_3^{-1}T_3C_3 = T_3^{-1}C_3^{-1}T_1C_3 = \bse,\\
&&
T_1^{-1}S_4^{-1}T_1S_4 = T_2^{-1}S_4^{-1}T_3S_4 = T_3^{-1}S_4^{-1}T_2^{-1}S_4 = \bse,
\\ \label{sgnewlast}
&&
(C_3C_2)^2 = T_3^{-1}T_1^{-1}(S_4C_3)^2 = T_2(S_4C_2)^2= \bse,\\
&&
T_1^{-1}T_2^{-1} T_3 C_2 (C_3^{-1}S_4)^2S_4=\bse.\label{sgnewfinal}
 \end{eqnarray}
where we used $\bse$ to denote the identity element in the symmetry group.

\subsection{$Z_2$ extension of the space group}

The classification of symmetric $Z_2$ spin liquids hosting fermionic spinon excitations on hyperkagome lattice is mathematically related to the central extension of the symmetry group $SG$, with a center $\mathcal{A}=Z_2$. Physically, fermionic spinons can transform projectively under symmetry operations due to their fractional statistics. Such a projective representation of symmetry group $SG$ with $\mathcal{A}=Z_2$ coefficient is classified by the 2nd cohomology group $H^2(SG,\mathcal{A})$ with the following short exact sequence\cite{Essin2013,Barkeshli2014,Tarantino2016}
\bea
1\rightarrow\mathcal{A}\rightarrow PSG\rightarrow SG\rightarrow1
\eea
Such a central extension was coined the ``projective symmetry group'' or $PSG$\cite{Wen2002} in the literature of slave-fermion construction of $Z_2$ spin liquids.

Therefore we classify the 2nd cohomology group $H^2(SG,Z_2)$ associated with symmetric $Z_2$ spin liquids on hyperkagome lattice, where the symmetry group is $SG=P4_132\times Z_2^{\cal T}$ in the spin-orbit-coupled system. More specifically, the group extension problem $H^2(SG,\mathcal{A})$ classifies gauge-inequivalent phase factors $\{\omega_f(\bsg_1,\bsg_2)\}$
\bea
\mathcal{R}_{\bsg_1}^{(f)}\mathcal{R}_{\bsg_2}^{(f)}=\omega_f(\bsg_1,\bsg_2)\mathcal{R}_{\bsg_1\cdot\bsg_2}^{(f)},~~~\bsg_{1,2}\in SG,~~~\omega_f(\bsg_1,\bsg_2)\in\mathcal{A}.\notag
\eea
where $\mathcal{R}^{(f)}_{\bsg_1}$ labels the symmetry operation on a single fermionic spinon $f$, associated with element $\bsg_1$ of the symmetry group. In our case of $Z_2$ spin liquids, the fusion group of the fermionic spinons is $\mathcal{A}=Z_2$ due to the $Z_2$ fusion rule $f\times f=1$, leading to the $Z_2$-valued phase factors $\{\omega_f(\bsg_1,\bsg_2)=\pm1\}$. One can always modify the symmetry operation $\mathcal{R}^{(f)}_{\bsg}$ by a local unitary transformation $\mathcal{U}^{(f)}_{\bsg}$ that form a linear representation of $SG$, i.e. redefining $\mathcal{R}_{\bsg}^{(f)}\rightarrow \mathcal{U}^{(f)}_{\bsg}\cdot\mathcal{R}_{\bsg}^{(f)}$, which will not change the universality class and topological properties of the associated spin liquid phase. These ``gauge transformations'' $\{\mathcal{U}^{(f)}_{\bsg}|\bsg\in SG\}$, however can modify the phase factors $\omega_f({\bsg_1,\bsg_2})$ by
\bea
\omega_f({\bsg_1,\bsg_2})\rightarrow \omega_f({\bsg_1,\bsg_2})\mathcal{U}^{(f)}_{\bsg_1}\mathcal{U}^{(f)}_{\bsg_2}\big[\mathcal{U}^{(f)}_{\bsg_1\cdot\bsg_2}\big]^{-1}
\eea
A typical example of such gauge transformations compatible with the $\mathcal{A}=Z_2$ fusion rule is simply a phase factor of $\mathcal{U}^{(f)}_{\bsg_1}=\pm1$. To classify distinct $Z_2$ extensions of the symmetry group, we look for different sets of phase factors $\omega_f({\bsg_1,\bsg_2})$ that cannot related by any gauge transformations.

In the following we compute the 2nd cohomology group $H^2(P4_132\times Z_2^{\cal T},Z_2)$ for our case of symmetric $Z_2$ spin liquids on hyperkagome lattice. First, notice that the whole multiplication table of symmetry group $SG=4_132\times Z_2^{\cal T}$ can be generated by algebraic relations (\ref{sgnewfirst})-(\ref{sgnewfinal}). Therefore we merely need to classify the $Z_2$-valued phase factors $\omega$'s associated with each algebraic identity in (\ref{sgnewfirst})-(\ref{sgnewfinal}), more specifically the gauge invariant ones.

Two important consistent conditions for the $Z_2$-valued phase factors are the associativity condition
\bea
\omega(\bsg_1,\bsg_2)\cdot\omega(\bsg_1\bsg_2,\bsg_3)=\omega(\bsg_2,\bsg_3)\cdot\omega(\bsg_1,\bsg_2\bsg_3)~~~~~~
\eea
and
\bea
\omega(1,\bsg)=\omega(\bsg,1)\equiv1,~~~\forall~~\bsg\in{SG}.
\eea
As is shown below, the non-symmorphic screw operation $S_4$ in symmetry group $SG=P4_132\times Z_2^{\cal T}$ provides a strong constraint on phase factors $\{\omega(\bsg_1,\bsg_2)=\pm1|\bsg_i\in SG\}\in H^2(SG,Z_2)$. It significantly reduce the number of gauge inequivalent solutions and greatly simplify the calculation.

First, we denote the $Z_2$-valued phase factors associated with each algebraic identity in (\ref{sgnewfirst})-(\ref{sgnewfinal}) as $\{\Omega_\alpha=\pm1\}$, where subscripts $\{\alpha\}$ are the same as the subscripts of $\{\eta_\alpha=\pm1\}$ in (\ref{constraintraw1})-(\ref{constraintraw:final}). This is chosen for the convenience of comparison between algebraic 2nd group cohomology described here, and physical PSG realizations in slave-fermion construction presented in Appendix \ref{app:z2 class}. In the following we briefly summarize the calculation of gauge-inequivalent solutions to $\{\Omega_\alpha=\pm1\}$. They are related to the phase factors $\{\omega(\bsg_1,\bsg_2)=\pm1|\bsg_i\in SG\}\in H^2(SG,Z_2)$ in simple relations, for example $\Omega_{\cal T}=\omega(\cal T,\cal T)$ for the first algebraic identity in (\ref{sgnewfirst}).

First of all, using gauge transformations $\{\mathcal{U}_{\bsg}^{(f)}=\pm1|\bsg\in SG\}$ we can always fix the gauge so that \bea
\Omega_3=\Omega_{3x}=\Omega_{3y}=\Omega_{4}=\Omega_{234}=1.
\eea
non-symmorphic symmetry $S_4$ with relation $\mathcal{R}^{(f)}_{T_1}=(\mathcal{R}^{(f)}_{S_4})^4$ leads to
\bea
T_1T_2T_1^{-1}T_2^{-1}=T_1T_3T_1^{-1}T_3^{-1}=e\rightarrow\Omega_{xy}=\Omega_{xz}=1.~~~~~
\eea
This excludes a large class of $Z_2$ spin liquids known as ``$\pi$-flux phases'', where spinons see a background field of $\pi$-flux per unit cell. We also have
\bea
\Omega_{4x}=\Omega_{2x}=1
\eea
where the following relations are utilized
\bea\label{sym:c2}
&(\mathcal{R}^{(f)}_{C_2})^2=\Omega_2,\\
&\label{sym:c2+s4}(\mathcal{R}^{(f)}_{S_4}\cdot\mathcal{R}^{(f)}_{C_2})^2\mathcal{R}^{(f)}_{T_2}=\Omega_{24},\\
&\label{sym:c3+s4}(\mathcal{R}^{(f)}_{S_4}\cdot\mathcal{R}^{(f)}_{C_3})^2=\Omega_{34}\cdot\mathcal{R}^{(f)}_{T_1}\cdot\mathcal{R}^{(f)}_{T_3}.
\eea
Now using rotational symmetry $(\mathcal{R}^{(f)}_{C_3})^3=\Omega_3=1$ we have
\bea
\Omega_{yz}=\Omega_{3z}=1
\eea
Similarly by 2-fold rotation $C_2$ with relation (\ref{sym:c2}) we have
\bea
\Omega_{2y}=\Omega_{2z}
\eea
Further using relation $(\mathcal{R}^{(f)}_{C_3}\cdot\mathcal{R}^{(f)}_{C_2})^2=\Omega_{23}$ we have
\bea
\Omega_{2z}=\Omega_{2x}=1
\eea
and hence $\Omega_{2x}=\Omega_{2y}=\Omega_{2z}=1$. Using (\ref{sym:c2+s4}) and $\Omega_{xy}=1$ we can further show
\bea
\Omega_{4z}=1
\eea
which also leads to
\bea
\Omega_{4y}=\Omega_{2z}=1
\eea
based on (\ref{sym:c2+s4}). Meanwhile, another algebraic identity $C_3 C_2=C_3^{-1}C_2C_3$ leads to
\bea
\Omega_{23}=\Omega_2.
\eea
Therefore in summary, without considering time reversal symmetry ${\cal T}$, there are $2^3=8$ different $Z_2$ extension (or projective representations with $Z_2$ coefficients) for the space group $P4_132$. They correspond to 3 independent coefficients (\ref{sym:c2})-(\ref{sym:c3+s4})
\bea
\{\Omega_2,\Omega_{24},\Omega_{34}=\pm1\}=H^2(P4_132,Z_2)=({\mathbb{Z}_2})^3
\eea

Now let's include time reversal symmetry. Again using relations $T_1=(S_4)^4$ and (\ref{sym:c2+s4})-(\ref{sym:c3+s4}) associated with non-symmorphic screw $S_4$ we have
\bea
\Omega_{x\cal T}=\Omega_{y\cal T}=\Omega_{z\cal T}=1
\eea
And by $(\mathcal{R}^{(f)}_{C_3})^3=1$ we have
\bea
\Omega_{3\cal T}=1.
\eea
Finally according to relation (\ref{sgnewfinal}) or (\ref{sglastreal}) we must have
\bea
\Omega_{2\cal T}=\Omega_{4\cal T}.
\eea
Therefore time reversal symmetry ${\cal T}$ only brings in one new $Z_2$-valued invariant
\bea\label{sym:c2+T}
&\mathcal{R}^{(f)}_{C_2}\mathcal{R}^{(f)}_{\cal T}=\Omega_{2\cal T}\mathcal{R}^{(f)}_{\cal T}\mathcal{R}^{(f)}_{C_2}.
\eea

As a result, the $Z_2$-valued projective representations of symmetry group $SG=P4_132\times Z_2^{\cal T}$ are classified by
\bea\notag
\{\Omega_2,\Omega_{24},\Omega_{34},\Omega_{2\cal T},\Omega_{\cal T}=\pm1\}=H^2(P4_132\times Z_2^{\cal T},Z_2)=({\mathbb{Z}_2})^5
\eea
These $2^5=32$ central extensions of our symmetry group with a $Z_2$ center correspond to the 5 independent $Z_2$-valued coefficients: as listed in (\ref{sym:c2})-(\ref{sym:c3+s4}) and (\ref{sym:c2+T}), as well as $\Omega_{\cal T}=\omega_f({\cal T,\cal T})=\pm1$.

Physically by requiring fermionic spinons to be Kramers doublets carrying spin-$1/2$ each, we must have
\bea\label{sym:T}
\Omega_{\cal T}=\omega_f({\cal T,\cal T})=-1.
\eea
Therefore among the 32 $Z_2$-valued projective representations of symmetry group $SG=P4_132\times Z_2^{\cal T}$, only half of them ($2^4=16$ projective representations) satisfying (\ref{sym:T}) are relevant for our consideration of symmetric $Z_2$ spin liquids on hyperkagome lattice.

\section{Classification of symmetric $Z_2$ spin liquids}\label{app:z2 class}

Though symmetry fractionalization class of fermionic spinons in a symmetric $Z_2$ spin liquid must belong to an element of 2nd cohomology group $H^2(SG,\mathcal{A})$, a specific lattice spin model with this symmetry group may only realize certain (but not all) elements of $H^2(SG,\mathcal{A})$. Below we concretely construct those symmetric $Z_2$ spin liquids that can be realized by the slave-fermion representation on a hyperkagome lattice. As will be shown later, among the 32 $Z_2$-valued projective representations of group $SG=P4_132\times Z_2^{\cal T}$, only 3 can be realized as symmetric $Z_2$ spin liquids in the slave-fermion representation of spin-$1/2$ particles on hyperkagome lattice.

\subsection{Projective symmetry group (PSG) of symmetric spin liquids}

Spin liquid states remain disordered and preserve the symmetry of the system. But due to the gauge redundancy  (\ref{gauget}), the mean field amplitudes are only defined up to a local SU(2) transform. This leads to the projective symmetry obeyed by the mean field Hamiltonians. Specifically, let us denote an element in the symmetry group (SG) as $U\in$SG, and its action on the operators as
\begin{equation}
U\Psi_i U^\dagger =R^\dagger_U \Psi_{U(i)}.
\end{equation}
Here $U(i)$ means the site index $i$ is changed to the transformed position $U(i)$, and $R_U$ is the spin rotation accompanying the symmetry operation. Then we see
\begin{equation}\label{psgstarting}
UH_A U^\dagger = \mbox{Tr} \left[
\left(
R_U\sigma_A R_U^\dagger \right) \Psi_{U(i)} u _{ij}^{(A)} \Psi_{U(j)}^\dagger
\right],
\end{equation}
where $A=0,x,y,z$, and $\sigma_0 = \mathbb{I}$ is the identity matrix.

We first look at $H_0$ where
\begin{equation}
UH_0U^\dagger  = \mbox{Tr} \left[ \Psi_{U(i)}u_{ij} \Psi_{U(j)}^\dagger \right].
\end{equation}
If $UH_0U^\dagger = H_0$, the (mean field) Hamiltonian is certainly invariant under symmetry operation $U$. However,
 the redundancy (\ref{gauget}) means that the equivalent class of Hamiltonians can be extended, as above equation is physically the same as
 \begin{equation}\label{gequiv}
 UH_0 U^\dagger \Leftrightarrow  \mbox{Tr}\left[ \Psi_{U(i)} G_U(U(i)) u_{ij} G_U^\dagger(U(j)) \Psi_{U(j)}^\dagger
 \right],
 \end{equation}
 where $G_U(U(i))$ denotes a local SU(2) gauge transform at site $U(i)$ for symmetry operation $U$. Then if
 \begin{equation}\label{psgcondition}
 G_U(U(i)) u_{ij} G_U^\dagger(U(j)) = u_{U(i),U(j)},
 \end{equation}
the right hand side of (\ref{gequiv}) is the same as $H_0$, and  $UH_0U^\dagger$ is equivalent to $H_0$ up to a gauge transform. The condition (\ref{psgcondition})  is often expressed as
\begin{eqnarray}\label{psgconvention}
G_UU(u_{ij}) &=& u_{ij},\\
G_U(u_{ij}) &=& G_U(U(i)) u_{ij}G_U^\dagger(U(j)),\\
U(u_{ij}) &=& u_{U^{-1}(i), U^{-1}(j)}.
\end{eqnarray}
where (\ref{psgconvention}) is the projective symmetry satisfied by the mean field ans\"{a}ts. Condition (\ref{psgconvention}) means that the mean field Hamiltonian $H_0$ only needs to satisfy a projective symmetry $G_UU(u_{ij})=u_{ij}$ if the physical Hamiltonian, expressed in terms of $\mathbf{S}_i$, satisfies a symmetry $UH_{phys}U^\dagger=H_{phys}$. Similar to $U\in$SG, the operations $G_UU$ also form a group, dubbed projetive symmetry group (PSG).  Next, we discuss the properties of the gauge transforms $G_U(i)$.

A special case is when the symmetry operation $U=e$, the identity element, i.e.
\begin{equation}\label{igg}
G_e(i) u_{ij} G_e^\dagger(j) = u_{ij}.
\end{equation}
Then the projective symmetry operation only contains a gauge transform.
The set of SU(2) matrices $\{G_e(i)\}$ satisfying (\ref{igg}) are called the invariant gauge group (IGG). Clearly, if an element $G_UU\in$PSG satisfies (\ref{psgcondition}), we can multiply it by $G_e\in$ IGG and $G_eG_UU\in$PSG still satisfies (\ref{psgcondition}). That means we have a many-to-one homomorphic mapping $h:$ PSG$\rightarrow$SG, or more precisely, SG=PSG/IGG. For IGG=$\pm1, e^{i\theta}$, SU(2), we have the PSG characterizing $\mathbb{Z}_2$, U(1) and SU(2) spin liquid states respectively.

The gauge group $G_U(i)$ describes the low energy fluctuation of the spin liquid ground state, and different choices of $G_U(i)$ gives different types of spin liquid states. The $G_U(i)$'s for various symmetry operations $U$ are not independent, as the group product rules $U_3=U_1U_2$ enforces
\begin{equation}
G_{U_3}U_3(u_{ij}) = G_{U_1}U_1 G_{U_2}U_2 (u_{ij}),
\end{equation}
or explicitly,
\begin{eqnarray}\nonumber
&& [G_{U_1}(U_1U_2(i)) G_{U_2}(U_2(i))] u_{ij}  [G_{U_1}(U_1U_2(i)) G_{U_2}(U_2(i))]^\dagger\\
&=& G_{U_3} (U_3(i)) u_{ij} G^\dagger_{U_3}(U_3(j)).
\end{eqnarray}
Therefore, the commutation rules for symmetry groups, i.e. $U_1U_2=U_2U_1$ or $U_1^{-1}U_2^{-1}U_1U_2=\mathbb{I}$ implies
\begin{eqnarray}\nonumber
&&G_{U_2}^\dagger (U_2^{-1}U_1U_2(i)) G_{U_2}^\dagger (U_1U_2(i)) G_{U_1} (U_1U_2(i)) G_{U_2}(U_2(i)) \\ \label{psgconstr}
&& = G_e(i).
\end{eqnarray}
Thus, once the IGG for $G_e(i)$ is chosen, the PSG for other symmetry operations $U$ are constrained by the commutation rules. Solving the coupled equations (\ref{psgconstr}) for all combinations of SG gives all possible PSG.

Two solutions to (\ref{psgconstr}) can be equivalent, because if $G_U(i)$ satisfy (\ref{psgcondition}) for $u_{ij}$, we have
\begin{eqnarray}
\nonumber
W_i u_{ij}W_j^\dagger &=&
\left[
W_i G_U(i)W_{U^{-1}(i)}^\dagger
\right] \left[
W_{U^{-1}(i)}u_{U^{-1}(i), U^{-1}(j)}  W^\dagger_{U^{-1}(j)}
\right]\\
&&
\left[
W_{U^{-1}(j)} G_U^\dagger (j) W_j^\dagger
\right],
\end{eqnarray}
where $W_i$ is an arbitrary SU(2) matrix. That means
\begin{equation}\label{psglocal}
\tilde{G}_U(i) = W_i G_U(i)W_{U^{-1}(i)}^\dagger
\end{equation}
 also satisfies (\ref{psgcondition}) for the ans\"{a}ts $W_i u_{ij}W_j^\dagger$. Since two ans\"{a}ts differing by an SU(2) gauge transform are equivalent, we see that $G_U$ and $\tilde{G}_U$ given by (\ref{psglocal}) are equivalent. That is, the PSG solutions $G_U(i)$ are also only defined up to a local SU(2) transform (\ref{psglocal}).

Now we look at the triplet terms $H_{x,y,z}$ that break spin rotation symmetry, i.e. in the presence of spin-orbit coupling. We similarly derive the projective symmetry condition satisfied by the triplet parts. First note that the SU(2) rotation  can be mapped into SO(3) rotation,
\begin{equation}
R_U \sigma_A R_U^\dagger = \sum_{B=x,y,z} O_{AB}^{(U)}\sigma_B
\end{equation}
where $O^{(U)}$ is an SO(3) matrix representing the rotation by $\boldsymbol{\theta}_U$ in the SU(2) matrix $R_U = e^{-i\boldsymbol{\theta}_U\cdot \boldsymbol{\sigma}/2}$, and $A, B=x,y,z$ are Cartesian indices. Then the triplet part of the mean field Hamiltonian has the transform property
\begin{eqnarray}\nonumber
&&\sum_A H_A  = \sum_{A} \mbox{Tr}\left[
\sigma_A \Psi_{U(i)} u_{ij}^{(A)} \Psi_{U(j)}^\dagger
\right],\\ \nonumber
&&\sum_A UH_A U^\dagger = \sum_{AB} O_{AB} \mbox{Tr}\left[
\sigma_B \Psi_{U(i)} u_{ij}^{(A)} \Psi_{U(j)}^\dagger
\right]\\
& \Leftrightarrow &
 \sum_{AB} O_{AB} \mbox{Tr}\left[
\sigma_B \Psi_{U(i)} G_U(U(i)) u_{ij}^{(A)} G^\dagger_U(U(j)) \Psi_{U(j)}^\dagger
\right],\qquad
\end{eqnarray}
where the last line means gauge equivalence. To restore (projective) symmetry, we require
\begin{equation}
G_U(U(i)) u_{ij}^{(A)} G_U^\dagger (U(j)) = \sum_C O^{(U)}_{AC}u^{(C)}_{U(i),U(j)} ,
\end{equation}
or it can be written as
\begin{equation}\label{psgconditiont}
G_U(U(i))\left(
\sum_{A}u^{(A)}_{ij} O^{(U)}_{AB}
\right) G_U^\dagger (U(j)) = u_{U(i),U(j)}^{(B)}
\end{equation}
Here we have used the property of orthogonal matrices $O^TO=\mathbb{I}$, or $\delta_{BC}= (O^T)_{BA} O_{AC} = O_{AB}O_{AC}$. Conditions (\ref{psgcondition}), (\ref{psgconditiont}) clearly show that given the PSG, the singlet term is decoupled from the triplet terms, while all the three triplet terms are generally mixed in Equation (\ref{psgconditiont}). For later uses, we summarize the SO(3) matrices corresponding to the $C_2, C_3, S_4$ rotations in a hyperkagome lattice:
\begin{eqnarray}\nonumber
&& O^{(C_2)} = \left(
\begin{array}{ccc}
-1 & 0 & 0\\
0  & 0 & -1\\
0 & -1 & 0
\end{array}
\right),\quad
O^{(C_3)} = \left(
\begin{array}{ccc}
0 & 0 & 1\\
1  & 0 & 0\\
0 & 1 & 0
\end{array}
\right),
\\ \label{so3matrices}
&&O^{(S_4)} = \left(
\begin{array}{ccc}
1 & 0 & 0\\
0  & 0 & -1\\
0 & 1 & 0
\end{array}
\right).
\end{eqnarray}
Note that when superposing two operations
\begin{eqnarray}\nonumber
u^{(A)}_{U_1U_2(i),U_1U_2(j)} &=&
G_{U_1}(U_1U_2(i)) G_{U_2}(U_2(i)) \\ \nonumber
&& \left(
\sum_{BC} u^{(C)}_{ij} O^{(U_2)}_{CB} O^{(U_1)}_{BA}
\right)\\ \nonumber
&& \quad G_{U_2}^\dagger (U_2(j)) G^\dagger_{U_1} (U_1U_2(j))\\
&\equiv & G_{U_1}G_{U_2} (u_{ij} O^{(U_2)}O^{(U_1)})
\end{eqnarray}
the sequence of $O^{(U_2)}O^{(U_1)}$ is opposite to $G_{U_1}G_{U_2}$.

In the above derivation, we have only considered unitary symmetry operators $U$. Now we discuss time-reversal operator which is anti-unitary
\begin{equation}
{\cal T} = i\sigma_y K,\qquad
{\cal T}^{-1} = K(-i\sigma_y) = -i\sigma_y K.
\end{equation}
where $K$ takes complex conjugation on complex number on its right, $Ki = -i K$. Written in terms of $f$-operators,
\begin{eqnarray}
{\cal T} &=& \sum_i (f_{i\downarrow}^\dagger f_{i\uparrow} - f_{i\uparrow}^\dagger f_{i\downarrow}) K.
\end{eqnarray}
That means
\begin{equation}
{\cal T}\Psi_i = i\sigma_y \Psi_i K, \qquad
\Psi_i^\dagger {\cal T}^{-1} = \Psi_i^\dagger (-i\sigma_y) K.
\end{equation}
Therefore, we have
\begin{eqnarray}\nonumber
{\cal T}H_A{\cal T}^{-1} &=& \mbox{Tr} \left[
\sigma_A^* (i\sigma_y) \Psi_i K   u_{ij}^{(A)}  \Psi_j^\dagger (-i\sigma_y K)
\right]\\ \nonumber
&=&
\mbox{Tr} \left[
\sigma_A^* (i\sigma_y) \Psi_i \left( u_{ij}^{(A)}\right) ^* \Psi_j^\dagger (-i\sigma_y)
\right]\\
&=&
\mbox{Tr}\left[
(\sigma_y\sigma_A^*\sigma_y) \Psi_i \left( u_{ij}^{(A)}\right)^* \Psi_j^\dagger
\right].
\end{eqnarray}
Following the convention in \cite{Wen2002}, we introduce the PSG as
\begin{eqnarray}\nonumber
&& {\cal T}H_A{\cal T}^{-1} \Leftrightarrow \\ \nonumber
&& \mbox{Tr} \left[
(\sigma_y\sigma_A^*\sigma_y) \Psi_i G_{\cal T}(i) (i\sigma_y) \left( u_{ij}^{(A)}\right)^* (-i\sigma_y) G_{\cal T}^\dagger (j) \Psi_j^\dagger
\right]\\
&=&
\mbox{Tr} \left[
\sigma_A \Psi_i G_{\cal T}(i) (-u_{ij}^{(A)}) G_{\cal T}^\dagger(j) \Psi_j^\dagger
\right]
\end{eqnarray}
where in the last line we have used $\sigma_y \sigma_0^* \sigma_y = \mathbb{I}$, $\sigma_y\sigma_{x,y,z}^*\sigma_y = -\sigma_{x,y,z}$, and from Equation (\ref{st}), $\sigma_y \left( u_{ij}^{(0)}\right)^* \sigma_y = -u_{ij}^{(0)}$, $\sigma_y \left( u_{ij}^{(x,y,z)}\right)^* \sigma_y = u_{ij}^{(x,y,z)}$. Thus, for time-reversal, the projective symmetry condition for singlet and triplet terms in $H_{MF}$ are the same,
\begin{equation}\label{timereversalConstr}
G_{\cal T}(i) u_{ij}^{(A)} G_{\cal T}(j)^\dagger  =  -u_{ij}^{(A)}.
\end{equation}

\subsection{Constraints of the projective symmetry group}
An element in PSG is denoted as
\begin{eqnarray}\nonumber
G_U U &=& (G_{\cal T}{\cal T})^{\nu_{\cal T}}
(G_{T_1} T_1)^{\nu_{T_1}}
(G_{T_2}T_2)^{\nu_{T_2}}
(G_{T_3}T_3)^{\nu_{T_3}}  \\
&&
(G_{C_2}C_2)^{\nu_{C_2}}
(G_{C_3}C_3)^{\nu_{C_3}}
(G_{S_4}S_4)^{\nu_{S_4}}.
\end{eqnarray}
Corresponding to the SG, the constraints in PSG with IGG=$\mathbb{Z}_2$
are
\begin{eqnarray}\label{constraintraw1}
&&
G_{\cal T}(i)^2 = \eta_{\cal T}\\
&&\label{constraintraw:c2}
G_{C_2}(C_2(i))G_{C_2}(i) = \eta_2 \\
&&\label{constraintraw:c3}
G_{C_3}(C_3^{-1}(i)) G_{C_3}(C_3(i)) G_{C_3}(i) = \eta_3\\ \nonumber
&&
G_{T_1}^{-1}(T_1S_4^{-1}(i)) G_{S_4}(T_1S_4^{-1}(i))\cdot \\ \label{G4444}
&&\qquad G_{S_4}(S_4^2(i)) G_{S_4}(S_4(i)) G_{S_4}(i) = \eta_4\\
&&\label{constraintraw:T+t1}
 G_{\cal T}^{-1}(T_1^{-1}(i)) G_{T_1}^{-1}(i) G_{\cal T}(i) G_{T_1}(i) = \eta_{x\cal T}\\
&&
 G_{\cal T}^{-1}(T_2^{-1}(i)) G_{T_2}^{-1}(i) G_{\cal T}(i) G_{T_2}(i)  = \eta_{y\cal T}\\
&&
 G_{\cal T}^{-1}(T_3^{-1}(i)) G_{T_3}^{-1}(i) G_{\cal T}(i) G_{T_3}(i)  = \eta_{z\cal T}\\
&&
  G_{\cal T}^{-1}(C_2^{-1}(i)) G_{C_2}^{-1}(i) G_{\cal T}(i) G_{C_2}(i)  = \eta_{2\cal T}\\
 &&
  G_{\cal T}^{-1}(C_3^{-1}(i)) G_{C_3}^{-1}(i) G_{\cal T}(i) G_{C_3}(i) = \eta_{3\cal T}\\
 &&\label{constraintraw:T+s4}
  G_{\cal T}^{-1}(S_4^{-1}(i)) G_{S_4}^{-1}(i) G_{\cal T}(i) G_{S_4}(i) = \eta_{4\cal T}\\
 &&\label{constraintraw:c2+t1}
 G_{T_1}^{-1}(C_2T_1^{-1}(i)) G_{C_2}^{-1}(T_1^{-1}(i)) G_{T_1}^{-1}(i) G_{C_2}(i) = \eta_{2x}\\
 &&\label{constraintraw:c2+t2}
 G_{T_2}^{-1}(C_2T_3^{-1}(i)) G_{C_2}^{-1}(T_3^{-1}(i)) G_{T_3}^{-1}((i)) G_{C_2}(i) = \eta_{2y}\\
 &&\label{constraintraw:c2+t3}
 G_{T_3}^{-1}(C_2T_2^{-1}(i)) G_{C_2}^{-1}(T_2^{-1}(i)) G_{T_2}^{-1}((i)) G_{C_2}(i) = \eta_{2z}\\
 &&\label{constraintraw:c3+t1}
 G_{T_1}^{-1}(C_3^{-1}T_2(i)) G_{C_3}^{-1}(T_2(i)) G_{T_2}(T_2(i)) G_{C_3}(i) = \eta_{3x},\\
&&\label{constraintraw:c3+t2}
G_{T_2}^{-1}(C_3^{-1}T_3(i)) G_{C_3}^{-1}(T_3(i)) G_{T_3}(T_3(i)) G_{C_3}(i)= \eta_{3y},\\
&&\label{constraintraw:c3+t3}
G_{T_3}^{-1}(C_3^{-1}T_1(i)) G_{C_3}^{-1}(T_1(i)) G_{T_1}(T_1(i)) G_{C_3}(i) =\eta_{3z},\\
 &&\label{constraintraw:s4}
 G_{T_1}^{-1}(S_4^{-1}T_1(i)) G_{S_4}^{-1}(T_1(i)) G_{T_1}(T_1(i)) G_{S_4}(i) = \eta_{4x}\\
 &&\label{constraintraw:s4+t2}
 G_{T_2}^{-1}(S_4^{-1}T_3(i)) G_{S_4}^{-1}(T_3(i)) G_{T_3}(T_3(i)) G_{S_4}(i) = \eta_{4y}\\
 &&\label{constraintraw:s4+t3}
 G_{T_3}^{-1}(S_4^{-1}T_2(i)) G_{S_4}^{-1}(T_2^{-1}(i)) G_{T_2}^{-1}(i) G_{S_4}(i) = \eta_{4z}\\
 &&\label{constraintraw3232}
 G_{C_3}(C_2(i)) G_{C_2}(C_2C_3(i)) G_{C_3}(C_3(i)) G_{C_2}(i) = \eta_{23}\\ \nonumber
 &&
 G_{T_3}^{-1}(T_3C_3^{-1}(i)) G_{T_1}^{-1}(T_1T_3C_3^{-1}(i))G_{S_4}(T_1T_3C_3^{-1}(i)) \\ \label{constraintraw4343}
 && \qquad G_{C_3}(C_3S_4(i)) G_{S_4}(S_4(i)) G_{C_3}(i) = \eta_{34}\\  \nonumber
 &&
G_{T_2}(C_2(i)) G_{S_4}(T_2^{-1}C_2(i)) G_{C_2}(C_2S_4(i))
\\
 \label{constraintrawlast}
&& \qquad\quad G_{S_4}(S_4(i)) G_{C_2}(i) = \eta_{24} \\ \nonumber
&&
G_{T_1}^{-1}(T_1S_4^{-1}(i)) G_{T_2}^{-1}(T_1T_2S_4^{-1}(i)) G_{T_3}(T_1T_2S_4^{-1}(i)) \\ \nonumber
&& \quad
G_{C_2}(T_1T_2T_3^{-1}S_4^{-1}(i)) G_{C_3}^{-1}(S_4C_3^{-1}S_4(i)) G_{S_4}(S_4C_3^{-1}S_4(i))\\
&& \quad G_{C_3}^{-1}(S_4(i)) G_{S_4}(S_4(i)) G_{S_4}(i) = \eta_{234}\label{constraintraw:final}
\end{eqnarray}
where all of the $\eta$'s take values $\pm1$. Solving these coupled constraint equations will give all possible gauge inequivalent generators in PSG, which is the task in the next two sections.

\subsection{Solution of the PSG constraints: unit-cell part}
We first focus on the unit-cell dependence of the constraints (\ref{constraintraw1})--(\ref{constraintrawlast}). The constraints can be reduced by  gauge choices:
\begin{enumerate}
\item
Consider the gauge redundancy of multiplying each $G$ with $\pm \tau_0$, which is an element in IGG. Then if a generator shows up for odd number  of times in the constraint equations, we can multiply the generator by $-\tau_0$ to change the sign of the corresponding $\eta$'s. That means we can use the gauge freedom of $G_{C_2}\rightarrow \eta_{234}=1, G_{C_3}\rightarrow \eta_3=1, G_{T_1}\rightarrow \eta_4=1, G_{T_2}\rightarrow \eta_{3x}=1, G_{T_3}\rightarrow \eta_{4y} = 1 $.
\item
Further, we can use the local $SU(2)$ gauge redundancy $G_U(i)\rightarrow W_i G_U(i)W_{U^{-1}(i)}^\dagger$ and consider $U\sim T_1, T_2, T_3$ to set
\begin{eqnarray}
\left\{\begin{array}{c}
G_{T_3} (0,0,z) = 1\\
G_{T_2} (0,y,z) = 1\\
G_{T_1}(x,y,z) = 1
\end{array} \right.
\end{eqnarray}
Then
\begin{equation}
[T_i,T_j]=0\rightarrow
\left\{
\begin{array}{c}
G_{T_1}(x,y,z,s) =1\\
G_{T_2}(x,y,z,s) = \eta_{xy}^x\\
G_{T_3}(x,y,z,s)= \eta_{xz}^x\eta_{yz}^y
\end{array} \right.
\end{equation}
Note here $T_i$ only relate the {\em same} sublattice sites in different unit cells. We will further use the local $SU(2)$ gauge choices among {\em different} sublattices in the next section.
\item We further reduce the unit-cell part of the generators.
Using the commutations between $G_{T_i}$ and $G_{C_2}$, we have
\begin{eqnarray}\label{ded:gc2}
&&G_{C_2} = \eta_{2x}^x\eta_{2y}^z\eta_{2z}^y (\eta_{xz}\eta_{xy})^{x(y+z)}A(s;x,y,z) g_{C_2}(s),\\
&& A(s;x,y,z) \equiv \left\{
\begin{array}{ll}
1: & s=1,7,9,4,8,11\\
\eta_{xy}^z\eta_{xz}^y: & s=5,6,12\\
\eta_{yz}^y: & s=2,3,10
\end{array}
\right.
\end{eqnarray}
where $g_{C_2}(s)$ is an SU(2) matrix depending only on sublattice sites $s$.
The condition $G_{C_2}^2 = \eta_2$ gives the constraint (i.e. consider the pair of sites $2\leftrightarrow 8$)
\begin{equation}
\eta_{yz}=1,\qquad
\eta_{xz}\eta_{xy} = 1\qquad
\eta_{2y}\eta_{2z} = 1
\end{equation}
Next, we use the commutations between $G_{T_i}$ and $G_{S_4}$ and obtain
\begin{eqnarray} \label{Gs4temp}
G_{S_4} = g_{S_4}(s)\eta_{4x}^x \eta_{4z}^y \times \left\{
\begin{array}{l}
\eta_{xy}^{y+z}, s = 7,10,11\\
1, s=\mbox{others}
\end{array}
\right.
\end{eqnarray}
where again $g_{S_4}(s)$ is an SU(2) matrix depending only on sublattice $s$. Consider $S_4: (1,x,y,z)\rightarrow (12, x,-z,y+1) \rightarrow (10,x+1,-y-1,-z) \rightarrow (2,x+1,z-1,-y-1)$ for $G_{S_4}^4 = 1$, we have the constraint
\begin{equation}
\eta_{xy}=1.
\end{equation}
So in sum, we have
\begin{equation}\label{grefinedti}
\eta_{xy}=\eta_{yz}=\eta_{xz}=1, \quad
G_{T_1}=G_{T_2}=G_{T_3} = 1,
\end{equation}
which greatly simplifed the further derivations. Finally, use the commutation between $G_{C_3}, G_{T_i}$, we have
\begin{eqnarray}\label{ded:gc3}
G_{C_3}(x,y,z,s) &=&  \eta_{3y}^z \eta_{3z}^x g_{C_3}(s).
\end{eqnarray}
The condition $G_{C_3}^3 =\eta_3$ gives (i.e. consider the sites $1\rightarrow 7 \rightarrow 9\rightarrow 1$)
\begin{eqnarray}
\eta_{3y}\eta_{3z} = 1.
\end{eqnarray}
\item
Correspondingly, $G_{C_2}, G_{C_3}, G_{S_4}$ in Equations (\ref{ded:gc2}), (\ref{ded:gc3}) and (\ref{Gs4temp}) are simplified to
\begin{eqnarray}
G_{C_2}(x,y,z,s) &=& \eta_{2x}^x\eta_{2y}^z\eta_{2z}^yg_{C_2}(s),  \\
G_{C_3}(x,y,z,s) &=& \eta_{3y}^z\eta_{3z}^xg_{C_3}(s),\\
G_{S_4} (x,y,z,s) &=& \eta_{4x}^{x}\eta_{4z}^y g_{S_4}(s).
\end{eqnarray}
\item
We solve the remaining unit-cell dependent part of the constraints.
The condition $(G_{C_3}G_{C_2})^2=\eta_{23}$ in (\ref{constraintraw3232}) gives (i.e. consider $1\leftrightarrow 7$)
\begin{equation}
\eta_{2x}\eta_{2y}\eta_{3z} = 1
\end{equation}
The condition $(G_{S_4}G_{C_3})^2 =  \eta_{34}$ in (\ref{constraintraw4343}) gives (i.e. consider $1\rightarrow 12 \rightarrow 11 \rightarrow 9$)
\begin{equation}
\eta_{4x}\eta_{3z} = 1
\end{equation}
These constraints: $\eta_{2x}\eta_{2y}\eta_{3z}=1, \eta_{4x}\eta_{3z}=1, \eta_{2y}\eta_{2z}=1, \eta_{3y}\eta_{3z}=1$, imply that the independent parameters are $\alpha_2, \alpha_3, \alpha_4 = \pm 1$:
\begin{eqnarray}
&&\eta_{2x} = \alpha_2\alpha_3, \quad
\eta_{2y}=\eta_{2z} = \alpha_2,\\
&&\eta_{4x}=\eta_{3y}=\eta_{3z}=\alpha_3,\quad
\eta_{4z} = \alpha_4.
\end{eqnarray}
and
\begin{eqnarray}\label{grefinedc2}
G_{C_2}(x,y,z,s) &=& \alpha_2^{x+y+z}\alpha_3^x g_{C_2}(s)\\
G_{C_3}(x,y,z,s) &=& \alpha_3^{x+z}g_{C_3}(s)\\ \label{grefineds4}
G_{S_4}(x,y,z,s) &=& \alpha_3^x\alpha_4^y g_{S_4}(s)
\end{eqnarray}
\item
Finally we look at those constraints related to time-reversal.
Use $\eta_{\cal T}$ and $\eta_{x\cal T}$--$\eta_{z\cal T}$, we have
\begin{equation}
G_{\cal T}(x,y,z,s) = \eta_{x\cal T}^x \eta_{y\cal T}^y \eta_{z\cal T}^z g_{\cal T}(s)
\end{equation}
From either ${\cal T}^{-1}S_4^{-1}{\cal T}S_4 = 1$ or ${\cal T}^{-1}C_2^{-1}{\cal T}C_2=1$, we have $\eta_{yT}\eta_{zT} = 1$; from ${\cal T}^{-1}C_3^{-1}{\cal T}C_3=1$, we have $\eta_{x{\cal T}}^{x+y}\eta_{y\cal T}^{y+z} \eta_{z\cal T}^{x+z} = 1\Rightarrow (\eta_{x\cal T}\eta_{y\cal T})^{x+y}\eta_{y\cal T}^z = 1 \Rightarrow \eta_{x\cal T}\eta_{y\cal T}=1, \eta_{y\cal T}=1$.  In sum, the independent parameter is  $\alpha_1 = \eta_{x\cal T} = \eta_{y\cal T} = \eta_{z\cal T}$, and
\begin{equation}
\label{grefinedt}
G_{\cal T} (x,y,z,s) = \alpha_1^{x+y+z} g_{\cal T}(s).
\end{equation}
\end{enumerate}
The spatial dependence of the generators are fully encoded in the parametrization in Equations (\ref{grefinedti}), (\ref{grefinedc2})--(\ref{grefineds4}), and (\ref{grefinedt}).  The constraints are therefore reduced to
\begin{eqnarray}\label{ded:t}
&&G_{\cal T}(i)^2 = \eta_{\cal T}\\  \label{ded:t2}
&&G_{\cal T}(C_2(i)) g_{C_2}^{-1}(s) G_{\cal T}(i) g_{C_2}(s) = \eta_{\cal T}\eta_{2\cal T}\\ \label{ded:t3}
&&G_{\cal T}(C_3^{-1}(i)) g_{C_3}^{-1}(s) G_{\cal T}(i) g_{C_3}(s) = \eta_{\cal T}\eta_{3\cal T}\\ \label{ded:t4}
&&G_{\cal T}(S_4^{-1}(i)) g_{S_4}^{-1}(s) G_{\cal T}(i) g_{S_4}(s) = \eta_{\cal T}\eta_{4\cal T}\\ \label{ded:c22}
&&G_{C_2}(C_2(i)) G_{C_2}(i) = \eta_2\\ \label{ded:c33}
&&G_{C_3}(C_3^{-1}(i)) G_{C_3}(C_3(i)) G_{C_3}(i) = 1\\ \label{ded:s44}
&&G_{S_4}(T_1S_4^{-1}(i)) G_{S_4}(S_4^2(i))  G_{S_4}(S_4(i)) G_{S_4}(i) = 1\\ \label{ded:3232}
 &&G_{C_3}(C_2(i)) G_{C_2}(C_2C_3(i)) G_{C_3}(C_3(i)) G_{C_2}(i) = \eta_{23}\\   \label{ded:4343}
 &&G_{S_4}(T_1T_3C_3^{-1}(i)) G_{C_3}(C_3S_4(i))  G_{S_4}(S_4(i)) G_{C_3}(i) = \eta_{34}\qquad\\ \label{ded:4242}
&&G_{S_4}(T_2^{-1}C_2(i)) G_{C_2}(C_2S_4(i))G_{S_4}(S_4(i)) G_{C_2}(i) = \eta_{24}\qquad\\ \nonumber
&&
 G_{C_2}(T_1T_2T_3^{-1}S_4^{-1}(i)) G_{C_3}^{-1} (S_4C_3^{-1}S_4(i))
\\ \label{ded:234}
&&
  G_{S_4}(S_4C_3^{-1}S_4(i)) G_{C_3}^{-1}(S_4(i)) G_{S_4}(S_4(i)) G_{S_4}(i)=1
\end{eqnarray}
We will solve the sublattice part in the next section.

\subsection{Solution of the PSG constraints: sublattice part}
Before plugging in all the sublattice sites $s$ to the above equations, we make a few simplifications. First, from
\begin{equation}
1=g_4(2)g_4(10)g_4(12)g_4(1)=\alpha_3 g_4(1)g_4(2)g_4(10)g_4(12)
\end{equation}
we see
\begin{equation}
\alpha_3=+1.
\end{equation}
And from
\begin{equation}
\eta_{24}=\alpha_4 g_4(8)g_2(7)g_4(1)g_2(2) = \alpha_2 g_4(1)g_2(2)g_4(8)g_2(7)
\end{equation}
we have
\begin{equation}
\alpha_4 \equiv \alpha_2.
\end{equation}

Now, we apply the local SU$(2)$ gauge choice $G_U(i)\rightarrow W_i G_U(i)W_{U^{-1}(i)}^\dagger$  between different sublattices to reduce the constraints. Let $U\sim C_3, S_4$, we can choose a proper gauge such that
\begin{equation}
g_{C_3}(7,9)=1,\qquad
g_{S_4}(s\ne 1,7,9)=1
\end{equation}
Then from Equations (\ref{ded:c33}) and (\ref{ded:s44}), we have
\begin{equation}
g_{C_3}(1)=1, \qquad
g_{S_4}(1, 7)=1, g_{S_4}(9)=\alpha_2
\end{equation}
Up to this point, we have used all of the SU(2) gauge redundancy. We have also solved Equation (\ref{ded:s44}) completely by specifying all $G_{S_4}$. Next we solve the remaining 9 constraints in (\ref{ded:t})--(\ref{ded:c33}) and (\ref{ded:t2})--(\ref{ded:4242}) under the chosen gauge.

\subsubsection{Solving (\ref{ded:c22}), (\ref{ded:4242}), and a part of (\ref{ded:3232})}
Now we focus on the following simplified conditions,
\begin{eqnarray}\nonumber
 \eta_2 &=&  g_{C_2}(1)g_{C_2}(7)  = \alpha_2 g_{C_2}(2)g_{C_2}(8)  \\ \nonumber
 &=& \alpha_2 g_{C_2}(3)g_{C_2}(11)=
    \alpha_2 g_{C_2}(4)g_{C_2}(10) \\   \label{ded:step11}
    &=&\alpha_2 g_{C_2}(5)g_{C_2}(12) =   \alpha_2 g_{C_2}(6)^2 =  g_{C_2}(9)^2\\
 \eta_{23} &=& g_{C_2}(1)^2 = g_{C_2}(9)g_{C_2}(7)=g_{C_2}(7)g_{C_2}(9)\\  \nonumber
\alpha_2  \eta_{24} &=& g_{C_2}(2) g_{C_2}(7)
  =g_{C_2}(8) g_{C_2}(10) =   g_{C_2}(9) g_{C_2}(11)
\\  \nonumber
 &=& \alpha_2 g_{C_2}(3) g_{C_2}(6)  = g_{C_2}(1) g_{C_2}(5)
 = g_{C_2}(4) g_{C_2}(12)\\\label{ded:step13}
\end{eqnarray}
First, from the equations for sites $(1,7,9)$ we have
\begin{eqnarray}\label{c1}
&&
g_2(9)^2 = g_2(1)g_2(7) = g_2(7)g_2(1) = \eta_2 \\
&&g_2(1)^2 =
g_2(9) g_2 (7)  =
 \eta_{23}
\end{eqnarray}
That means $\eta_{23}g_2(7)=\eta_2 g_2(1), \eta_2g_2(7)=\eta_{23}g_2(9)$, and therefore
\begin{eqnarray}
&&g_2(1)=g_2(7)=g_2(9)\\
&& g_2(1)^2=g_2(7)^2=g_2(9)^2=\eta_2
=\eta_{23}
\end{eqnarray}
So explicitly, we have the solution
\begin{equation}
\left\{
\begin{array}{ll}
\eta_2=+1: & g_2(1)=g_2(7)=g_2(9)=\lambda_1\\
\eta_2=-1: &
g_2(1)=g_2(7)=g_2(9)=\lambda_1i\tau_i
\end{array}
\right.
\end{equation}
where $\tau_i$ denotes  $\hat{n}\cdot \vec{\tau}$ for an arbitrary unit vector $\hat{n}$, and $\lambda_1=\pm1$. Here $\hat{n}$ can be rotated due to SU(2) gauge redundancy. It is the relations between this direction and the other $\hat{n}'$ for $g_{C_3}, g_{S_4}$ that are gauge independent. So here we denote it as a general $\tau_i$ and will only choose the gauge at the end.

From the solutions for sites $(1,7,9)$, by repeatedly using  Equations (\ref{ded:step11}), (\ref{ded:step13}),
we have
\begin{eqnarray}
&&
g_2(2)=g_2(5)=g_2(11)=\left\{
\begin{array}{ll}
\alpha_2\lambda_1 \eta_{24}, & \eta_2=+1\\
\alpha_2\lambda_1\eta_{24}(-i\tau_i), & \eta_2=-1
\end{array}
\right.\\
&&
g_2(3)=g_2(8)=g_2(12)=\left\{
\begin{array}{ll}
\lambda_1 \eta_{24}, & \eta_2=+1\\
\lambda_1\eta_{24}(-i\tau_i), & \eta_2=-1
\end{array}
\right.\quad\\
&&
g_2(4)=g_2(10)=\left\{
\begin{array}{ll}
\alpha_2\lambda_1, & \eta_2=+1\\
\alpha_2\lambda_1(i\tau_i), & \eta_2=-1
\end{array}
\right.,\\
&&
g_2(6) = \left\{
\begin{array}{ll}
\lambda_1, & \eta_2=+1\\
\lambda_1 (i\tau_i), & \eta_2 = -1
\end{array}
\right.
\end{eqnarray}
Further, $\alpha_2\eta_2=g_2(4)g_2(10)=g_2(6)^2$ gives
\begin{equation}
\alpha_2=+1
\end{equation}
These results completely solve the constraints concerning $\eta_2, \eta_{24}$ in (\ref{ded:c22}) (\ref{ded:4242}). Now we note that  $\lambda_1$  is just the overall $\pm$ sign for $g_{C_2}(s)$, which is an element in IGG. This gauge choice will be fixed by Equation (\ref{ded:234}) later as we have used the $\mathbb{Z}_s$ freedom of $G_{C_2}$ to set $\eta_{234}=1$ at the beginning. Also note that up to this point, we already have $\alpha_2=\alpha_3=\alpha_4=+1$, so
\begin{equation}
G_{C_2}=g_{C_2}(s), \quad
G_{C_3}=g_{C_3}(s), \quad
G_{S_4}=g_{S_4}(s)
\end{equation}
 and there is no spatial dependence. It simplifies the following analysis of the remaining sublattice-dependent equations.

\subsubsection{Solving (\ref{ded:c33}), (\ref{ded:4343})}
Focus next on the relations
\begin{eqnarray}\nonumber
1 &=&
 g_{C_3}(8) g_{C_3}(6) g_{C_3}(2)=g_{C_3}(3) g_{C_3}(4) g_{C_3}(5)\\
 &=&g_{C_3}(10) g_{C_3}(11) g_{C_3}(12),\\\nonumber
  \eta_{34} &=&   g_{C_3}(11)
 = g_{C_3}(12)^2
 =  g_{C_3}(2)
 =  g_{C_3}(3) g_{C_3}(8)
 \\
 &=&  g_{C_3}(5)   =   g_{C_3}(4)^2
 = g_{C_3}(6) g_{C_3}(10)
\end{eqnarray}
We can directly read
\begin{equation}
g_{C_3}(11)=\eta_{34},\qquad
g_{C_3}(2)= \eta_{34},\qquad
g_{C_3}(5)=\eta_{34}
\end{equation}
Then for $ \eta_{34}=+1$:
\begin{eqnarray}\nonumber
 && g_{C_3}(12)=\lambda_2\rightarrow g_{C_3}(10)=\lambda_2 \rightarrow g_{C_3}(6)=\lambda_2\rightarrow \\
&&g_{C_3}(8)=\lambda_2\rightarrow g_{C_3}(3)=\lambda_2 \rightarrow g_{C_3}(4)= \lambda_2
\end{eqnarray}
When $ \eta_{34}=-1$,
\begin{eqnarray}\nonumber
 &&  g_{C_3}(12) =\lambda_2i\tau_j\rightarrow g_{C_3}(10)=\lambda_2 i\tau_j \rightarrow g_{C_3}(6)=\lambda_2 i\tau_j \qquad \\
&&
\rightarrow g_{C_3}(8)=\lambda_2 i\tau_j \rightarrow g_{C_3}(3)=\lambda_2  i\tau_j  \rightarrow g_{C_3}(4)= \lambda_2  i\tau_j
\end{eqnarray}
where $\lambda_2=\pm1$.
These results completely solve (\ref{ded:c33}) and (\ref{ded:4343}). Again, we will see $\lambda_2$ can be fixed uniquely by (\ref{ded:234}).

\subsubsection{Solving (\ref{ded:3232}), (\ref{ded:234}) }
Plug in the above results for $g_{C_2}(s), g_{C_3}(s)$ to (\ref{ded:3232}), we find the additional constraint
\begin{equation}
 \left\{
\begin{array}{ll}
1 =\eta_{24}, & \eta_{34}=1, \eta_2=1\\
-1 =\eta_{24}, & \eta_{34}=1,\eta_2=-1\\
1 =-\eta_{24}, & \eta_{34}=-1, \eta_2=1\\
-1 =-\eta_{24}\tau_j\tau_i\tau_j\tau_i, & \eta_{34}=-1, \eta_{2}=-1.
\end{array}
\right.
\end{equation}
where $\tau_i, \tau_j$ are used in $g_{C_2}, g_{C_3}$ respectively. That means $\eta_{24}$ is not a free parameter. We can summarize the results obtained so far for $g_{C_2}, g_{C_3}, g_{S_4}$
\begin{eqnarray}\nonumber
\mbox{$\checkmark$Case 1: } &&
\eta_2=+1, \eta_{34}=+1, \eta_{24}=+1;\\ \nonumber
&&g_{C_2}(s) = \lambda_1,\\
&&\left\{
\begin{array}{l}
g_{C_3}(1,2,5,7,9,11) = 1\\
g_{C_3}(3,4,6,8,10,12)=\lambda_2
\end{array}
\right.
\\ \nonumber
\mbox{$\times$Case 2: } &&
\eta_2=+1, \eta_{34}=-1, \eta_{24}=-1;\\ \nonumber
&&
\left\{
\begin{array}{l}
g_{C_2}(1,4,6,7,9,10)=+\lambda_1\\
g_{C_2}(2,3,5,8,11,12)=-\lambda_1
\end{array}
\right.\\
&&\left\{
\begin{array}{l}
g_{C_3}(1,7,9)=-g_{C_3}(2,5,11)=1\\
g_{C_3}(3,4,6,8,10,12)=\lambda_2 i\tau_i, \forall \tau_i
\end{array}
\right.
\\ \nonumber
\mbox{$\times$Case 3: } && \eta_2=-1, \eta_{34}=+1, \eta_{24}=-1;\\ \nonumber
&&g_{C_2}(s) =\lambda_1 i\tau_i , \forall \tau_i\\
&&\left\{
\begin{array}{l}
g_{C_3}(1,2,5,7,9,11) = 1\\
g_{C_3}(3,4,6,8,10,12)=\lambda_2
\end{array}
\right.
\\ \nonumber
\mbox{$\checkmark$Case 4: } &&
\eta_2=-1, \eta_{34}=-1, \eta_{24}=+1;\\ \nonumber
&&
\left\{
\begin{array}{l}
g_{C_2}(1,4,6,7,9,10)=+\lambda_1 i\tau_i\\
g_{C_2}(2,3,5,8,11,12)=-\lambda_1 i\tau_i
\end{array}
\right.\\
&&\left\{
\begin{array}{l}
g_{C_3}(1,7,9)=-g_{C_3}(2,5,11)=1\\
g_{C_3}(3,4,6,8,10,12)=\lambda_2 i\tau_j\qquad
(\tau_j=\tau_i)
\end{array}
\right.
\\ \nonumber
\mbox{$\times$Case 5: } &&
\eta_2=-1, \eta_{34}=-1, \eta_{24}=-1\\ \nonumber
&&
g_{C_2}(s)= \lambda_1 i\tau_i\\
&& \left\{
\begin{array}{l}
g_{C_3}(1,7,9)=-g_{C_3}(2,5,11)=+1\\
g_{C_3}(3,4,6,8,10,12) = \lambda_2 i\tau_j \qquad (\tau_j\ne \tau_i)
\end{array}
\right.
\end{eqnarray}

Next, Equation (\ref{ded:234}) can be simplified to $g_{C_2}(i) = g_{C_3}(S_4^2(i)) g_{C_3}(C_2C_3^{-1}(i))$, i.e. $g_{C_2}(1) = g_{C_3}(10)g_{C_3}(1)$. It provides the further constraint
\begin{equation}
\lambda_1=\lambda_2 \equiv +1,
\end{equation}
and show that only Case 1 and Case 4 stand. Equations concerning other sites give the same constraints.

\subsubsection{Solving (\ref{ded:t}), (\ref{ded:t2}), (\ref{ded:t3}) and (\ref{ded:t4}) for  $g_{\cal T}(s)$}
The equations can be written explicitly as
\begin{eqnarray} \label{tempt}
 \eta_{\cal T} &=& g_{\cal T}(s)^2 \\  \nonumber
 \eta_{\cal T}\eta_{2\cal T} &=&
g_{\cal T}(A)g_{C_2}^{-1}(B)g_{\cal T}(B)g_{C_2}(B)\\ \nonumber
&&
\mbox{[For (A,B) = (7,1), (1,7), (9,9)]}\\   \nonumber
&=&
 \alpha_1 g_{\cal T}(A)g_{C_2}^{-1}(B)g_{\cal T}(B)g_{C_2}(B)\\ \nonumber
&&
 \mbox{[For (A,B) = (8,2), (2,8), (11,3), (3,11), }\\ \label{temp2t}
 && \mbox{(4,10), (10,4), (12,5), (5,12), (6,6)]} \\  \nonumber
    \eta_{\cal T}\eta_{3\cal T} &=& g_{\cal T}(A) g_{C_3}^{-1}(B) g_{\cal T} (B) g_{C_3}(B),\\ \label{temp3t}
    &&  \mbox{ [(1,7,9), (2,6,8), (5,4,3), (12,11,10)]}\\ \nonumber
    \eta_{\cal T}\eta_{4\cal T} &=&
    g_{\cal T}(2)g_{\cal T}(1) = \alpha_1 g_{\cal T}(1) g_{\cal T}(12)\\ \nonumber
    &&
    = \alpha_1 g_{\cal T}(12)g_{\cal T}(10)  = \alpha_1 g_{\cal T}(10)g_{\cal T}(2) \\ \nonumber
    &=&
    g_{\cal T}(7)g_{\cal T}(8) = g_{\cal T}(8) g_{\cal T}(4)\\ \nonumber
    && =
    g_{\cal T}(4) g_{\cal T}(5) = \alpha_1 g_{\cal T}(5)g_{\cal T}(7)\\ \nonumber
    &=&
    g_{\cal T}(3)g_{\cal T}(6) = \alpha_1 g_{\cal T}(6)g_{\cal T}(11)\\  \label{temp4t}
    && =
    \alpha_1 g_{\cal T}(11)g_{\cal T}(9) = \alpha_1 g_{\cal T}(9) g_{\cal T}(3)
\end{eqnarray}
Equation (\ref{temp3t}) means $(A,B)=(1,7), (7,9), (9,1)$, and etc. Now use (\ref{temp4t}),
\begin{eqnarray*}
&&\left(g_{\cal T}(2)g_{\cal T}(1) \right) \left(g_{\cal T}(12) g_{\cal T}(10)\right) = \eta_{\cal T}\eta_{4\cal T}\cdot \eta_{\cal T}\eta_{4\cal T}\alpha_1 = \alpha_1\\
&=&
g_{\cal T}(2)\left( g_{\cal T}(1) g_{\cal T}(12)\right) g_{\cal T}(10)
= g_{\cal T}(2) \alpha_1\eta_{\cal T}\eta_{4\cal T}g_{\cal T}(10) = 1
\end{eqnarray*}
So
\begin{equation}
\alpha_1 = 1,\qquad G_{\cal T}(x,y,z,s) = g_{\cal T}(s),
\end{equation}
as expected.

{\bf (a)} For $\eta_{\cal T}=+1$, $g_{\cal T}(s)= \gamma_s$, where $\gamma_s=\pm1$ depends on sublattice sites. From (\ref{temp4t}) we have
\begin{equation}\label{4tlots}
\left\{
\begin{array}{l}
\gamma_1 = \gamma_2\eta_{4\cal T} = \eta_{4\cal T}\gamma_{12} = \gamma_{10}\\
\gamma_7 = \gamma_5\eta_{4\cal T} = \gamma_8 \eta_{4\cal T} = \gamma_4\\
\gamma_9 = \gamma_{11}\eta_{4\cal T} = \gamma_3\eta_{4\cal T} = \gamma_6
\end{array}
\right.
\end{equation}
From (\ref{temp3t}),
$
\eta_{3\cal T} = \gamma_1\gamma_7= \gamma_7\gamma_9 = \gamma_9\gamma_1.
$
That means
\begin{equation}
\eta_{3\cal T}=+1,
\end{equation}
as there is no way to choose $\gamma_{1,7,9}$ such that $\eta_{3\cal T}=-1$. Then we have
\begin{eqnarray}\nonumber
&&\gamma_1=\gamma_7=\gamma_9,\quad
\gamma_2=\gamma_6=\gamma_8,\\
&&\gamma_3=\gamma_4=\gamma_5, \qquad
\gamma_{12}=\gamma_{11}=\gamma_{10}.
\end{eqnarray}
Plug back to (\ref{4tlots}), we see
\begin{equation}
\eta_{4\cal T}=+1,
\end{equation}
and all $g_{\cal T}(s)$ are the same. Equation (\ref{temp2t}) gives $\eta_{2\cal T}=+1$. In sum, we can choose a gauge to set
\begin{equation}
\eta_{2\cal T}=\eta_{3\cal T}=\eta_{4\cal T}=+1, \qquad
g_{\cal T}(s) = 1.
\end{equation}

{\bf (b)} For $\eta_{\cal T}=-1$, $g_{\cal T}(s) = \gamma_s i\tau_k^{(s)}$, where $\gamma_s=\pm1$ can be further constrained by other equations. Here $\tau_k^{(s)}=\hat{n}_s\cdot \vec{\tau}$, where $\hat{n}_s$ is a unit vector  along different directions for different sublattices. From (\ref{temp3t}), use $g_{C_3}(1,7,9)=1$, we have for $g_{\cal T}(1,7,9)$,
\begin{eqnarray}
&& \tau_k^{(1)}=\tau_k^{(7)}=\tau_k^{(9)}\equiv \tau_k,\\
&&-\eta_{3\cal T}=-\gamma_1\gamma_7= -\gamma_7\gamma_9=-\gamma_9\gamma_1
\end{eqnarray}
Then we similarly have $\eta_{3\cal T}=+1$. Choose the gauge
\begin{equation}\label{old:3}
g_{\cal T}(1)=g_{\cal T}(7)=g_{\cal T}(9)= i\tau_k
\end{equation}
where $\tau_k$ is one of the three Pauli matrices $\tau_1, \tau_2,\tau_3$.
Use (\ref{temp4t}), we have all the other $g_{\cal T}(s)$:
\begin{eqnarray}\label{old:4}
&&g_{\cal T}(1,4,6,7,9,10) =  i\tau_k,\\ \label{old:5}
&&g_{\cal T}(2,3,5,8,11,12)   = \eta_{4\cal T} i\tau_k.
\end{eqnarray}
Here all of the $g_{\cal T}(s)$ involve the same  $\tau_k$. The relation between these $\tau_k$ and the $\tau_i, \tau_j$ used in $g_{C_2}, g_{C_3}$ is specified by the conditions (\ref{temp2t}) (\ref{temp3t}). First we use any pair of sites in (\ref{temp2t}),
\begin{eqnarray}
-\eta_{2\cal T}&=&i\tau_k\left(\begin{array}{l}
1\\
-i\tau_i
\end{array}\right)
i\tau_k \left(\begin{array}{l}
1\\
i\tau_i
\end{array}\right)=
\left(\begin{array}{l}
-1\\
-\tau_k\tau_i\tau_k\tau_i
\end{array}\right)
\end{eqnarray}
where the upper and lower row in the bracket stand for $\eta_2=+1$ and $\eta_2=-1$ respectively.
Thus,
\begin{eqnarray}\label{old:6}
&&\eta_{2}=+1: \eta_{2\cal T}=+1,\\ \label{old:7}
&& \eta_{2}=-1:
\left\{
\begin{array}{ll}
\eta_{2\cal T}=+1, & \tau_k=\tau_i\\
\eta_{2\cal T}=-1, & \tau_k\ne\tau_i
\end{array}
\right.
\end{eqnarray}
 Thus, the relative relation between the Pauli matrices used in $g_{\cal T}(s)\sim\tau_k$ and $g_{C_2}(s)\sim\tau_i$ is fixed.
Then from (\ref{temp3t}),
\begin{equation}
-1= \eta_{4\cal T}i\tau_k\left(
\begin{array}{l}
1\\
-  i\tau_j
\end{array}
\right) i\tau_k \left(
\begin{array}{l}
1\\
  i\tau_j
\end{array}
\right) = \eta_{4\cal T}
\left(
\begin{array}{l}
-1\\
-\tau_k\tau_j\tau_k\tau_j
\end{array}
\right)
\end{equation}
where upper and lower row in the bracket stand for $\eta_{34}=+1$ and $\eta_{34}=-1$ respectively.
So
\begin{equation}\label{old:8}
\eta_{34}=+1: \eta_{4\cal T}=+1,\qquad
\eta_{34}=-1: \left\{
\begin{array}{ll}
\eta_{4\cal T}=+1 & \tau_k=\tau_j\\
\eta_{4\cal T}=-1 & \tau_k\ne \tau_j
\end{array}
\right.
\end{equation}
 Thus, the relative relation for the Pauli matrix used in $g_{C_3}(s)\sim\tau_j$ and $g_{\cal T}(s)\sim\tau_k$ is also fixed.

\subsection{Summary  and PSG solutions in a site-independent gauge}
We first summarize all of the results in the gauge used in previous sections, and then make gauge transforms to obtain better forms for the analysis of mean field ans\"{a}ts. The states are specified by the $\mathbb{Z}_2$ numbers $\eta_{\cal T}, \eta_{2\cal T}, \eta_{4\cal T}, \eta_2, \eta_{34}, \eta_{24} = \pm1$; they are not completely independent and only certain combinations are viable. There are totally 5 PSG solutions, labeled 1(a), 1(b) for $\eta_{\cal T}=+1$ and 2(a), 2(b), 2(c) for $\eta_{\cal T}=-1$. Choosing specific gauges for $\tau_i, \tau_j, \tau_k$ used for $g_{C_2}, g_{C_3}, g_{\cal T}$, we have
\begin{enumerate}
\item
$\eta_{\cal T}=\eta_{2\cal T}=\eta_{4\cal T}=+1$, $g_{\cal T}=1$
\begin{enumerate}
\item
$\eta_2=\eta_{34}=\eta_{24}=+1$, $g_{C_2}(s) = g_{C_3}(s) = +1$
\item
$\eta_2=\eta_{34}=-\eta_{24}=-1$,
\begin{eqnarray*}
&& g_{C_2}(1,4,6,7,9,10)=-g_{C_2}(2,3,5,8,11,12)=i\tau_3\\
&& g_{C_3}(1,7,9)=-g_{C_3}(2,5,11)=1\\
&& g_{C_3}(3,4,6,8,10,12)=i\tau_3
\end{eqnarray*}
\end{enumerate}
\item
$\eta_{\cal T}=-1$,
\begin{enumerate}
\item
$\eta_2=\eta_{34}=\eta_{24}=+1, \eta_{2\cal T}=\eta_{4\cal T}=+1$. Then $g_{\cal T}(s)=i\tau_2$, $g_{C_2}(s)=g_{C_3}(s)=1$.
\item
$\eta_2=\eta_{34}=-\eta_{24}=-1$, $\eta_{2\cal T}=\eta_{4\cal T}=+1$,
\begin{eqnarray*}
&& g_{\cal T}(s) = i\tau_2\\
&& g_{C_2}(1,4,6,7,9,10)=-g_{C_2}(2,3,5,8,11,12) = i\tau_2\\
&& \left\{
\begin{array}{l}
g_{C_3}(1,7,9)=-g_{C_3}(2,5,11) = 1\\
g_{C_3}(3,4,6,8,10,12) = i\tau_2
\end{array}
\right.
\end{eqnarray*}
\item
$\eta_2=\eta_{34}=-\eta_{24}=-1$, $\eta_{2\cal T}=\eta_{4\cal T}=-1$.
\begin{eqnarray*}
&& g_{\cal T}(1,4,6,7,9,10)=-g_{\cal T}(2,3,5,8,11,12)=i\tau_2\\
&& g_{C_2}(1,4,6,7,9,10) = -g_{C_2}(2,3,5,8,11,12) = i\tau_3\\
&& \left\{
\begin{array}{l}
g_{C_3}(1,7,9)=-g_{C_3}(2,5,11)=1\\
g_{C_3}(3,4,6,8,10,12) = i\tau_3
\end{array}
\right.
\end{eqnarray*}
\end{enumerate}
\end{enumerate}
Note that there is no unit-cell dependence so all the $g_U$'s are equal to $G_U$'s.

We would like to make gauge transforms $G_U(i)\rightarrow W_i G_U(i) W_{U^{-1}(i)}^\dagger$ to further eliminate site dependence for the PSG. It turns out that such gauge choice exists for all   cases,
\begin{enumerate}
\item
\begin{enumerate}
\item
No change
\item
$W_{2,5,11}=-W_{3,8,12}=i\tau_3, W_{4,6,10}=-W_{1,7,9}=1$.
\end{enumerate}
\item
\begin{enumerate}
\item
No change.
\item
$W_{2,5,11}=-W_{3,8,12}=i\tau_2, W_{4,6,10}=-W_{1,7,9}=1$.
\item
$W_{2,5,11}=-W_{3,8,12}=i\tau_3, W_{4,6,10}=-W_{1,7,9}=1$.
\end{enumerate}
\end{enumerate}
Then we have the PSG solutions summarized in Table \ref{tab:psg}.

\section{Classification of symmetric $U(1)$ spin liquids}\label{app:u1 class}

In this section we classify symmetric $U(1)$ spin liquids of spin-$1/2$ particles on hyperkagome lattice in the slave-fermion construction. In the slave-fermion mean-field ansatz of these $U(1)$ spin liquids, there are only non-zero hopping amplitudes while all pairing amplitudes vanish identically. In other words the wavefunctions of $U(1)$ spin liquids are projected fermi liquids or projected insulators, in comparison to projected superconductors in the case of $Z_2$ spin liquids. In terms of classification, the major difference between $Z_2$ and $U(1)$ spin liquids lies in their invariant gauge group (IGG): $Z_2$ spin liquids have an IGG=$\{G_e({\bf r})=\pm1\}=Z_2$, while $U(1)$ spin liquids have an IGG=$\{G_e({\bf r})=e^{i\theta\tau_3}|0\leq\theta<2\pi\}=U(1)$.

As shown in Ref.\cite{Wen2002}, in the so-called ``canonical gauge'' where all mean-field amplitudes are hoppings in the $U(1)$ spin liquid, any gauge rotations that keep the hopping ansatz invariant must take the form of
\bea\label{u(1) psg:general form}
G_{\hat g}({\bf r})=(i\tau^1)^{n_{\hat g}}e^{i\phi_{\hat g}({\bf r})\tau^3},~~~n_{\hat g}=0,1,~~~\forall~\hat g\in SG.
\eea
This is because the hopping flux through each plaquette may be reversed ($n_g=1$) by a symmetry operation $g$. In the case of $U(1)$ spin liquids, the constraints (\ref{constraintraw1})-(\ref{constraintraw:final}) for symmetry operations $\{G_g({\bf r})|g\in SG\}$ still hold but with $U(1)$-valued coefficients $\eta_\alpha=e^{i\theta_\alpha\tau_3}$ on the r.h.s. of the equations, as compared to $Z_2$ case. Below we outline the calculations of gauge-inequivalent solutions to these constraint equations.

First of all, non-symmorphic screw symmetry $S_4$ with constraints (\ref{constraintraw:s4}) and (\ref{constraintraw4343})-(\ref{constraintrawlast}) leads to
\bea
n_{T_1}=n_{T_2}=n_{T_3}=0.
\eea
in (\ref{u(1) psg:general form}). Therefore non-symmorphic symmetries generically rule out a large class of $U(1)$ spin liquids called ``staggered-flux phases''\cite{Wen2002}. Moreover 3-fold rotation $C_3$ with constraint (\ref{constraintraw:c3}) leads to
\bea
n_{C_3}=0.
\eea
and constraint (\ref{constraintraw:final}) further determines
\bea
n_{C_2}=n_{S_4}=0,1.
\eea

Using translations $T_i$ and 3-fold rotation $C_3$ with constraints (\ref{constraintraw:c3+t1})-(\ref{constraintraw:c3+t3}), we can always choose a gauge so that
\bea
&G_{T_1}({\bf r},s)\equiv1,~~~G_{T_2}(x,y,z,s)=e^{i\varphi_{xy}x\tau_3},\\
&G_{T_3}(x,y,z,s)=e^{i\varphi_{xy}(y-x)\tau_3},~~~G_{C_3}(x,y,z,s)=e^{i x(y-z)\varphi_{xy}\tau_3}.\notag
\eea
By using constraint (\ref{constraintraw:s4}) it's straightforward to show that
\bea
\varphi_{xy}=0\mod2\pi.
\eea
and hence
\bea\label{u(1) sym:translation+c3}
G_{T_i}({\bf r},s)=G_{C_3}({\bf r},s)\equiv1.
\eea
similar to the $Z_2$ spin liquid case.

Next we consider 2-fold rotation $C_2$ and 4-fold screw $S_4$. Using constraints (\ref{constraintraw:c2+t1})-(\ref{constraintraw:c2+t3}) and (\ref{constraintraw:s4})-(\ref{constraintraw:s4+t3}) we can always choose a gauge so that
\bea\notag
&G_{C_2}({\bf r},s)=(i\tau_1)^{n_2}e^{i\phi_2(s)\tau_3},~~~G_{S_4}({\bf r},s)=(i\tau_1)^{n_4}e^{i\phi_4(s)\tau_3},\\
&n_2=n_4=0,1.
\eea
Due to constraints (\ref{constraintraw:s4}) and (\ref{constraintraw:s4+t3}), by gauge fixing one can always trivialize the sublattice part and show
\bea\label{u(1) sym:s4}
G_{S_4}({\bf r},s)\equiv (i\tau_1)^{n_4},~~~n_4=0,1.
\eea
Furthermore, constraints (\ref{constraintraw:c2}), (\ref{constraintraw3232}), (\ref{constraintrawlast}) and (\ref{constraintraw:final}) also lead to
\bea\label{u(1) sym:c2}
G_{C_2}({\bf r},s)\equiv (i\tau_1)^{n_2},~~~n_2=n_4=0,1.
\eea
Now that spatial symmetries are all fixed, we consider time reversal operation in the end. Regarding that each fermionic spinon carries spin-$1/2$ and forms a Kramers doublet, we require that $n_{\cal T}=1$ for our physical system in the canonical gauge. It's straightforward to show that
\bea\label{u(1) sym:trs}
G_{\cal T}\equiv i\tau_1.
\eea
from constraints (\ref{constraintraw:T+t1})-(\ref{constraintraw:T+s4}), by a proper gauge fixing. Therefore there are only two distinct spin-$1/2$ $U(1)$ spin liquids with fermionic spinons on hyperkagome lattice, with symmetry operations
\bea
&G_{T_i}({\bf r},s)=G_{C_3}({\bf r},s)\equiv1,~~~G_{\cal T}\equiv i\tau_1,\\
&G_{C_2}({\bf r},s)=G_{S_4}({\bf r},s)\equiv (i\tau_1)^{n_4},~~~n_4=0,1.\label{u(1) psg:summary}
\eea
with IGG=$\{e^{i\theta\tau_3}|0\leq\theta<2\pi\}=U(1)$. We label them as ``$U1^{n_4}$'' states. In the notation of Ref.\cite{Lawler2008}, $U1^0$ state is referred as ``$U(1)$-uniform state'' while $U1^1$ state is referred as ``$U(1)$-staggered state''.

It's instructive to understand the relation between the 2 symmetric $U(1)$ states and the 3 symmetric $Z_2$ spin liquids classified in Appendix \ref{app:z2 class}. Quite generally, a $Z_2$ spin liquid can be obtained by breaking the $U(1)$ gauge group in a $U(1)$ spin liquid through the Anderson-Higgs mechanism. Specifically, which $Z_2$ spin liquid is connected to $U1^0$ (or $U1^1$) state via such a continuous phase transition induced by pairing (Higgs) terms? While this issue can be addressed by investigating their mean-field ansatz explicitly, here we identify the neighboring $Z_2$ states of each $U(1)$ state from an algebraic viewpoint by looking at their projective symmetry operations.

Once the IGG=$U(1)$ is broken down to its subgroup IGG=$Z_2$, the gauge rotation associated with each symmetry generator is fixed only up to a $\pm1$ sign. In particular, each gauge rotation $G_g({\bf r},s)$ in (\ref{u(1) psg:summary}) for $U(1)$ states can be accompanied by an extra $i\tau_3$ rotation in a neighboring $Z_2$ state. Simple algebra immediately leads to the results summarized in TABLE \ref{tab:psg}: 2(a) and 2(c) states lie in the neighborhood of $U1^0$ state, while 2(b) and 2(c) states are in proximity to $U1^1$ state. In particular, both $U(1)$ states can be driven into $Z_2$ state 2(c) by continuous phase transitions.

\section{Mean-field ansatz of $Z_2$ spin liquids}\label{app:mf ansatz}

 \subsection{Terms in mean field Hamiltonian\label{appendix:MFterms}}
 The following are mean field Hamiltonians (\ref{hmfPsi1}) expanded in terms of slave-fermion operators $f_{i\alpha}$:
\begin{eqnarray}
\nonumber
H_0 &=& is_0 (f_{i\uparrow}f_{j\uparrow}^\dagger + f_{i\downarrow}^\dagger f_{j\downarrow} + f_{i\downarrow}f_{j\downarrow}^\dagger + f_{i\uparrow}^\dagger f_{j\uparrow})\\ \nonumber
&& + s_3
(f_{i\uparrow}f_{j\uparrow}^\dagger-f_{i\downarrow}^\dagger f_{j\downarrow} +f_{i\downarrow}f_{j\downarrow}^\dagger - f_{i\uparrow}^\dagger f_{j\uparrow})\\ \nonumber
&& + s_1
(f_{i\uparrow}f_{j\downarrow} + f_{i\downarrow}^\dagger f_{j\uparrow}^\dagger - f_{i\downarrow}f_{j\uparrow} - f_{i\uparrow}^\dagger f_{j\downarrow}^\dagger)\\
&& + is_2
(-f_{i\uparrow}f_{j\downarrow}+ f_{i\downarrow}^\dagger f_{j\uparrow}^\dagger + f_{i\downarrow}f_{j\uparrow} - f_{i\uparrow}^\dagger f_{j\downarrow}^\dagger)\\ \nonumber
H_z &=& t^z_0
(f_{i\uparrow}f_{j\uparrow}^\dagger + f_{i\downarrow}^\dagger f_{j\downarrow} - f_{i\downarrow}f_{j\downarrow}^\dagger - f_{i\uparrow}^\dagger f_{j\uparrow})\\ \nonumber
&& + it^z_3
(f_{i\uparrow}f_{j\uparrow}^\dagger - f_{i\downarrow}^\dagger f_{j\downarrow} - f_{i\downarrow}f_{j\downarrow}^\dagger + f_{i\uparrow}^\dagger f_{j\uparrow})\\ \nonumber
&& + it^z_1
(f_{i\uparrow}f_{j\downarrow} + f_{i\downarrow}^\dagger f_{j\uparrow}^\dagger + f_{i\downarrow}f_{j\uparrow} + f_{i\uparrow}^\dagger f_{j\downarrow}^\dagger)\\
&& - t^z_2
(-f_{i\uparrow}f_{j\downarrow} + f_{i\downarrow}^\dagger f_{j\uparrow}^\dagger - f_{i\downarrow}f_{j\uparrow} + f_{i\uparrow}^\dagger f_{j\downarrow}^\dagger)\\ \nonumber
H_x &=& t^x_0
(f_{\i\downarrow}f_{j\uparrow}^\dagger - f_{i\uparrow}^\dagger f_{j\downarrow} + f_{i\uparrow}f_{j\downarrow}^\dagger - f_{i\downarrow}^\dagger f_{j\uparrow}) \\ \nonumber
&& + it^x_3
(f_{i\downarrow}f_{j\uparrow}^\dagger + f_{i\uparrow}^\dagger f_{j\downarrow} + f_{i\uparrow}f_{j\downarrow}^\dagger + f_{i\downarrow}^\dagger f_{j\uparrow})\\ \nonumber
&& + it^x_1
(f_{i\downarrow}f_{j\downarrow} - f_{i\uparrow}^\dagger f_{j\uparrow}^\dagger -f_{i\uparrow}f_{j\uparrow}+ f_{i\downarrow}^\dagger f_{j\downarrow}^\dagger)\\
&& - t^x_2
(-f_{i\downarrow}f_{j\downarrow} - f_{i\uparrow}^\dagger f_{j\uparrow}^\dagger + f_{i\uparrow}f_{j\uparrow} + f_{i\downarrow}^\dagger f_{j\downarrow}^\dagger)\\ \nonumber
H_y &=& it^y_0
(-f_{i\downarrow}f_{j\uparrow}^\dagger + f_{i\uparrow}^\dagger f_{j\downarrow}  + f_{i\uparrow}f_{j\downarrow}^\dagger - f_{i\downarrow}^\dagger f_{j\uparrow}) \\
\nonumber
&& - t^y_3 (-f_{i\downarrow}f_{j\uparrow}^\dagger -f_{i\uparrow}^\dagger f_{j\downarrow} + f_{i\uparrow}f_{j\downarrow}^\dagger + f_{i\downarrow}^\dagger f_{j\uparrow})\\ \nonumber
&& -t^y_1 (-f_{i\downarrow}f_{j\downarrow} + f_{i\uparrow}^\dagger f_{j\uparrow}^\dagger - f_{i\uparrow}f_{j\uparrow} + f_{i\downarrow}^\dagger f_{j\downarrow}^\dagger )\\
&& - it^y_2
(f_{i\downarrow}f_{j\downarrow} + f_{i\uparrow}^\dagger f_{j\uparrow}^\dagger + f_{i\uparrow}f_{j\uparrow} + f_{i\downarrow}^\dagger f_{j\downarrow}^\dagger)
\end{eqnarray}

\subsection{Constraints for mean field amplitudes} \label{appendix:mf}
The PSG solutions are summarized in Table \ref{tab:psg}, from which we can derive the mean field amplitudes. First we look at constraints on possible mean field amplitude. As discussed in the main text, if a symmetry operation takes a bond back to itself, it may constrain the possible MF terms allowed in the bond.

 Firstly, note time-reversal ${\cal T}(\{u_{ij}\})=\{-u_{ij}\}$.  When $\eta_{\cal T}=+1$, $G_{\cal T}(s;x,y,z)=g_{\cal T}(s)=1$, so from (\ref{timereversalConstr})
\begin{equation}
-u_{ij}=G_{\cal T}(i)u_{ij} G_{\cal T}^\dagger (j)=u_{ij}.
\end{equation}
That means the mean field amplitudes in all bonds must vanish. Thus, in the following we focus only on the $\eta_{\cal T}=-1$ case. Then $G_{\cal T}(s;x,y,z)=g_{\cal T}(s) = i\tau_2$, implying
\begin{equation}
-u_{ij}=G_{\cal T}(i) u_{ij} G_{\cal T}(j)^\dagger = \tau_2 u_{ij} \tau_2.
\end{equation}
It means
\begin{equation}\label{def:uij}
u_{ij}=a_{ij}\tau_1 + b_{ij}\tau_3.
\end{equation}
for both singlet and triplet terms.

Secondly, we consider the constraints given by spatial symmetry operations. Let $j=(0,0,0,1)$, and $i=(x,y,z,s)$ in $u_{ij}$, where $s=1,2,3,6,12$ involves only onsite and nearest neighbour amplitudes. As it turns out, the only symmetry operation that gives constraint is
\begin{eqnarray}
C_2C_3:&& (x,y,z,1)\rightarrow (-z,-y,-x,1),\\
&& (0,0,0,1)\rightarrow(0,0,0,1)
\end{eqnarray}
None of the other operations take the sublattices $(s,1)$ to either $(s,1)$ or $(1,s)$. {\em So the self-consistency only constrains the possible onsite terms and does not constrain the nearest neighbour terms.} The reason is that there are 24 symmetry group elements connecting different sublattices; meanwhile, the hyperkagome lattice is made of corner-sharing triangles, so there are also exactly $\frac{4\times12}{2}=24$ different nearest neighbour bonds per unit cell. So 24 symmetry operation will take one nearest neighbour bond to 24 other different nearest neighbour bonds, and there is no remaining symmetry operation to give consistency constraints for N.N. bonds. (In comparison, there are 12 sublattice sites, and half of the symmetry operations give onsite consistency constraints). Then we have
 the PSG constraint for onsite amplitude
\begin{eqnarray}\nonumber
u_{(\mathbf{0},1),(\mathbf{0},1)}
&=&
g_{C_2}g_{C_3}(1) u_{(\mathbf{0},1),(\mathbf{0},1)} (g_{C_2}g_{C_3}(1))^\dagger\\
&=&g_{C_2} u_{(\mathbf{0},1),(\mathbf{0},1)} g_{C_2}^\dagger,
\end{eqnarray}
where we have adopted the gauge in Table \ref{tab:psg}. This gives the onsite constraints  in Table \ref{tab:mfsinglet} and \ref{tab:mftriplet}. There is no constraint for nearest neighbour amplitude as discussed above, so in all cases hopping and pairing for nearest neighbour bonds are all allowed.
Further apply (\ref{psgcondition}) and consider all SG $\{U=C_2^{\nu_{C_2}} C_3^{\nu_{C_3}} S_4^{\nu_{S_4}}| \nu_{C_2}\in \mathbb{Z}_2, \nu_{C_3}\in \mathbb{Z}_3, \nu_{S_4}\in \mathbb{Z}_4\}$, we have the mean field amplitudes up to nearest neighbour bonds as summarized in Table \ref{tab:mfsinglet} and \ref{tab:mftriplet}. Mean field amplitudes for next nearest neighbours and etc. can be similarly obtained.

\subsection{Mean Field Calculations\label{appendix:MFcal}}
Written in terms of the slave fermions, the Heisenberg term becomes
\begin{eqnarray}\label{spinmodel0}
 \mathbf{S}_i\cdot \mathbf{S}_j &=& \frac{1}{4} \left(2f_{i\alpha}^\dagger f_{j\beta}^\dagger f_{j\alpha} f_{i\beta}  - f_{i\alpha}^\dagger f_{j\beta}^\dagger f_{j\beta} f_{i\alpha} \right);
\end{eqnarray}
the DM  terms $d_\mu = \mathbf{e}_\mu \cdot (\mathbf{S}_i\times \mathbf{S}_j)$ become
\begin{eqnarray}\nonumber
d_x &=&\frac{i}{4} ( \duuu + \uddd - \dddu - \uuud \\  \nonumber
&& - \uduu - \dudd + \uudu + \ddud)\\
\quad
 \\\nonumber
d_y &=& \frac{1}{4} (\uduu - \dudd + \uudu - \ddud \\  \nonumber
&& -\duuu +\uddd + \dddu - \uuud) \\
\quad\\
d_z &=& \frac{i}{2} \left(
f_{i\uparrow}^\dagger f_{j\downarrow}^\dagger f_{j\uparrow}f_{i\downarrow} - f_{i\downarrow}^\dagger f_{j\uparrow}^\dagger f_{j\downarrow}f_{i\uparrow}
\right)
\end{eqnarray}
the Kitaev terms become
\begin{eqnarray}
\kappa_x \equiv S_i^x S_j^x &=& \frac{1}{4} (f_{i\downarrow}^\dagger f_{i\uparrow} + f_{i\uparrow}^\dagger f_{i\downarrow})(f_{j\downarrow}^\dagger f_{j\uparrow} + f_{j\uparrow}^\dagger f_{j\downarrow})
\\
\kappa_y \equiv  S_i^y S_j^y &=&- \frac{1}{4} (f_{i\downarrow}^\dagger f_{i\uparrow} - f_{i\uparrow}^\dagger f_{i\downarrow})(f_{j\downarrow}^\dagger f_{j\uparrow} - f_{j\uparrow}^\dagger f_{j\downarrow})
\\
\kappa_z\equiv S_i^z S_j^z &=& \frac{1}{4} (f_{i\uparrow}^\dagger f_{i\uparrow} - f_{i\downarrow}^\dagger f_{i\downarrow}) (f_{j\uparrow}^\dagger f_{j\uparrow} - f_{j\downarrow}^\dagger f_{j\downarrow})
\end{eqnarray}
the $\Gamma$ terms $\gamma_\mu = S_i^\nu S_j^\rho + S_i^\rho S_j^\nu$ (where $\mu,\nu,\rho$ form a cyclic) become
\begin{eqnarray} \nonumber
\gamma_x &=& \frac{i}{4} (\duuu + \uddd - \dddu - \uuud \\  \nonumber
&& +\uduu + \dudd - \uudu - \ddud)\\ \quad
\\ \nonumber
\gamma_y &=& \frac{1}{4} (\uduu - \dudd + \uudu - \ddud\\  \nonumber
&& +\duuu - \uddd - \dddu + \uuud)\\ \quad
\\ \label{spinmodel9}
\gamma_z &=& \frac{i}{2} (\dduu - \uudd)
\end{eqnarray}

Next, we make the observation that we only need to compute the mean field energy on one bond for the spin liquid states, and all other bonds will yield the same energy. This is because the physical Hamiltonian on different bonds are related by the spin rotation $R$ accompanying the symmetry operation $U$; meanwhile, the mean field ansats $H_A$ on different bonds are related by the same spin rotations $R$. Thus, computing the mean field energy in different bonds just correspond to computing the same energy with different quantization axes. Thus, we can conveniently choose to compute the bond connecting sublattices $(m_i=2, n_j=3)$,
\begin{eqnarray}\nonumber
\frac{\langle H\rangle}{N_{\mbox{\scriptsize site}}} &=& 2\left(J\langle \mathbf{S}_i \cdot \mathbf{S}_j\rangle
+ \sum_{\mu =x,y,z} D^\mu \langle d_\mu \rangle + K^\mu \langle \kappa_\mu\rangle + \Gamma^\mu \langle \gamma_\mu \rangle
 \right).\\ \label{mftotalappendix}
\end{eqnarray}
The factor 2 comes from the fact that each site has 2 nearest neighbour bonds (without double counting).
In the following we will focus on the bond $(2,3)$ and drop the sublattice indices. Also, due to translation symmetry we will neglect the unit cell indices $i,j$.

Now we consider the mean field decomposition for the possible terms (\ref{spinmodel0})--(\ref{spinmodel9}) in the spin model. Note that the mean field ansats  respect such symmetry, which means the mean field decomposition have equivalent terms. From ${\cal T}(f_\uparrow, f_\downarrow) = (f_\downarrow^\dagger, -f_\uparrow^\dagger)$, we have
\begin{eqnarray}
&&\rho_{\alpha\beta} \equiv \langle f_{i\alpha}^\dagger f_{j\beta}\rangle = \left(
\begin{array}{cc}
\langle f_{i\uparrow}^\dagger f_{j\uparrow}\rangle &
\langle f_{i\uparrow}^\dagger f_{j\downarrow}\rangle\\
\langle f_{i\downarrow}^\dagger f_{j\uparrow}\rangle &
\langle f_{i\downarrow}^\dagger f_{j\downarrow}\rangle
\end{array}
\right)= \left(
\begin{array}{cc}
\rho_1 & \rho_2 \\
-\rho^*_2 & \rho_1^*
\end{array}
\right), \quad \\
&&
\Delta_{\alpha\beta} \equiv \langle f_{i\alpha}^\dagger f_{j\beta}^\dagger \rangle = \left(
\begin{array}{cc}
\langle f_{i\uparrow}^\dagger f_{j\uparrow}^\dagger\rangle &
\langle f_{i\uparrow}^\dagger f_{j\downarrow}^\dagger\rangle\\
\langle f_{i\downarrow}^\dagger f_{j\uparrow}^\dagger\rangle &
\langle f_{i\downarrow}^\dagger f_{j\downarrow}^\dagger\rangle
\end{array}
\right) =  \left(
\begin{array}{cc}
\Delta_2 & \Delta_1\\
-\Delta_1^* & \Delta_2^*
\end{array}
\right).\quad
\end{eqnarray}
Here Re$\rho_1$ and Re$\Delta_1$ are the singlet MF amplitudes, and others are triplet amplitudes.
We can further separate the singlet and triplet parts
\begin{eqnarray}
&&\rho_1 = \rho_s + i\rho_z,\qquad
\rho_2 = \rho_y+i\rho_x\\
&& \Delta_1 = \Delta_s +i\Delta_z\qquad
\Delta_2 = \Delta_x+i\Delta_y
\end{eqnarray}

With all the above information, we can write the mean field energy per site (\ref{mftotalappendix}) as follows,
\begin{eqnarray}
{\cal E}& \equiv & \frac{\langle H\rangle}{N_{\mbox{\scriptsize site}}} = {\cal E}_J + {\cal E}_D + {\cal E}_K + {\cal E}_\Gamma.
\end{eqnarray}
The Heisenberg term reads
\begin{eqnarray}
 \nonumber
{\cal E}_J = 2J\langle \mathbf{S}_i\cdot \mathbf{S}_j\rangle &=&  -J\left[
3(\mbox{Re}\rho_1)^2 -(\mbox{Im}\rho_1)^2 - |\rho_2|^2
\right]\\ \label{mfEJ}
&& - J\left[
3(\mbox{Re}\Delta_1)^2 - (\mbox{Im}\Delta_1)^2  - |\Delta_2|^2
\right]\quad
\end{eqnarray}
Thus, it is clear that the Heisenberg term will favor singlet hopping Re$\rho_1 = \frac{1}{2}(\langle f_{i\uparrow}^\dagger f_{j\uparrow}\rangle + \langle f_{i\downarrow}^\dagger f_{j\downarrow}\rangle)$ and singlet pairing Re$\Delta_1 = \frac{1}{2} (\langle f_{i\uparrow}^\dagger f_{j\downarrow}^\dagger\rangle + \langle f_{j\downarrow}f_{i\uparrow}\rangle)$, and suppress the triplet terms. Further, the DM terms read
\begin{eqnarray}\nonumber
{\cal E}_D &=&
4D_x[\rho_x\rho_s - \Delta_y\Delta_s]  + 4D_y[\Delta_x\Delta_s + \rho_s\rho_y]\\
&& +4D_z\left[ \Delta_s \Delta_z + \rho_s\rho_z
\right],
\end{eqnarray}
from which we see the linear coupling of singlet and triplet terms in all $D_x, D_y , D_z$ terms. On the other hand, in Kitaev terms
\begin{eqnarray}\nonumber
{\cal E}_K
&=&  K^x \left[ \Delta_x^2-\Delta_y^2-\Delta_s^2+\Delta_z^2 +\rho_y^2-\rho_x^2-\rho_s^2+\rho_z^2
\right]
\\ \nonumber
&& - K^y \left[
 \Delta_x^2-\Delta_y^2+\Delta_s^2-\Delta_z^2 -\rho_y^2+\rho_x^2-\rho_s^2+\rho_z^2
\right]
\\
&& + K^z \left[
 \Delta_x^2+\Delta_y^2-\Delta_s^2-\Delta_z^2 +\rho_y^2+\rho_x^2-\rho_s^2-\rho_z^2
\right]\quad
\end{eqnarray}
and the transverse Ising terms
\begin{eqnarray}\nonumber
{\cal E}_\Gamma &=&
-4\Gamma^x (\Delta_x\Delta_z - \rho_z\rho_y)
+ 4\Gamma^y (\Delta_y\Delta_z - \rho_z\rho_x)
\\
&& + 4\Gamma^z (\Delta_x\Delta_y + \rho_x\rho_y),
\end{eqnarray}
we see that the singlet and triplet terms are completely decoupled.

\begin{figure}
\begin{center}
\includegraphics[width=8cm]{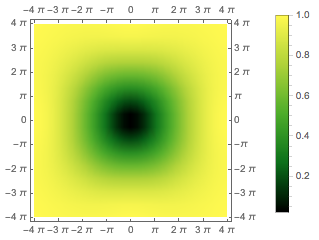}\\
\includegraphics[width=8cm]{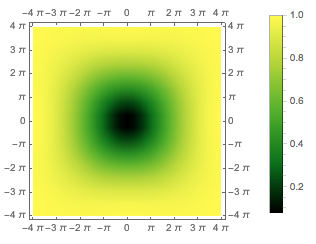}\\
\includegraphics[width=8.5cm]{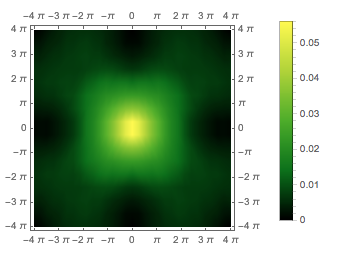}\\
\end{center}
\caption{ The (normalized) static structure factor $S(\mathbf{q})$ for isotropic spin liquid (top, MF parameters $s_3=0.16456, t^y_3=0, \mu=-0.078$),  anisotropic spin liquid (middle, MF parameters $s_3=0.16584, t^y_3=0.01672, \mu=-0.079$), and their difference (bottom). Here $q_x-q_y$ planes are plotted, and due to the $C_3$ rotation symmetry, $q_y-q_z, q_x-q_z$ planes have the same intensity. It shows that the minimum of $S(\mathbf{q})$ is shallower for the anisotropic spin liquid state, and the distribution has a  tendency towards anisotropy.
\label{fig:sq}}
\end{figure}

To evaluate the mean field amplitudes for states written in $\bbk$-space, we note the unit-cell translational invariance and perform the Fourier transform,
\begin{eqnarray}
\nonumber
\langle f_{i\mu}^\dagger f_{j\nu}\rangle &=& \frac{1}{N_{\mbox{\scriptsize cell}}} \sum_i \frac{1}{N_{\mbox{\scriptsize cell}}}\sum_{\bbk \bbk'} e^{i\bbk (\bbr_i + \boldsymbol{\delta}_m)} e^{-i\bbk' (\bbr_j + \boldsymbol{\delta}_n)} \langle f_{m\bbk \mu}^\dagger f_{n\bbk' \nu}\rangle\\
&=&
\frac{1}{N_{\mbox{\scriptsize cell}}} \sum_\bbk \langle f_{m\bbk\mu}^\dagger f_{n\bbk\nu}\rangle e^{i\bbk(\boldsymbol\delta_m-\boldsymbol\delta_n)}\\ \nonumber
\langle f_{i\mu}^\dagger f_{j\nu}^\dagger \rangle &=& \frac{1}{N_{\mbox{\scriptsize cell}}}\sum_i \frac{1}{N_{\mbox{\scriptsize cell}}}\sum_{\bbk \bbk'} e^{i\bbk (\bbr_i + \boldsymbol{\delta}_m)} e^{i\bbk' (\bbr_j + \boldsymbol{\delta}_n)} \langle f_{m\bbk \mu}^\dagger f_{n\bbk' \nu}\rangle\\
&=&
\frac{1}{N_{\mbox{\scriptsize cell}}} \sum_\bbk \langle f_{m\bbk\mu}^\dagger f^\dagger_{n,-\bbk\nu}\rangle e^{i\bbk(\boldsymbol\delta_m-\boldsymbol\delta_n)}
\end{eqnarray}
Here the Fourier transform convention is
\begin{eqnarray}
f_i &=& \frac{1}{\sqrt{N_{\mbox{\scriptsize cell}}}} \sum_\bbk e^{-i\bbk\cdot\boldsymbol R_i}f_\bbk\\
f_\bbk &=& \frac{1}{\sqrt{N_{\mbox{\scriptsize cell}}}} \sum_\bbk e^{i\bbk\cdot\boldsymbol R_i}f_i.
\end{eqnarray}
 where $\boldsymbol R_i = \bbr_i + \boldsymbol \delta_m$, $\bbr_i$ is the unit cell coordinate, while $\boldsymbol{\delta}_m$ is the sublattice displacement for sublattice $m$. $N_{\mbox{\scriptsize cell}} = N_{\mbox{\scriptsize site}}/12$ is the number of cubic unit cells.

It is also useful to adopt a self-consistency formalism, where the mean field Hamiltonian written in the basis $(f_{i\uparrow}, f_{i\downarrow}^\dagger, f_{i\downarrow}, -f_{i\uparrow}^\dagger)$ reads (for the $(i,j)$ block)
\begin{eqnarray}\nonumber
H_J^{\mbox{\scriptsize MF}} &=& -\frac{J}{4}
\left(
\begin{array}{cccc}
2\rho_1 + \rho_1^* & 2\Delta_1+\Delta_1^* & -\rho_2^* & \Delta_2^*\\
2\Delta_1+\Delta^*_1 & -(2\rho_1+\rho_1^*) & \Delta_2^* & \rho_2^*\\
\rho_2 & -\Delta_2 & 2\rho_1^* +\rho_1 & 2\Delta_1^* + \Delta_1\\
- \Delta_2 & -\rho_2 & 2\Delta_1^* + \Delta_1 & -(2\rho_1^*+\rho_1)
\end{array}
\right)\\
H_{DM}^{\mbox{\scriptsize}MF} &=&  \frac{iD}{2}
\left(
\begin{array}{cccc}
-\rho_1 & -\Delta_1 & 0&0\\
-\Delta_1 & \rho_1 &0&0\\
0&0& \rho_1^* & \Delta_1^*\\
0&0& \Delta_1^* & -\rho_1^*
\end{array}
\right)
\end{eqnarray}
\begin{eqnarray}
H_K^{\mbox{\scriptsize MF}} &=&
\frac{K}{4}
\left(
\begin{array}{cccc}
-\rho_1^* & -\Delta_1^* & \rho_2^* & -\Delta_2^*\\
-\Delta_1^* & \rho_1^* & -\Delta_2^* & -\rho_2^*\\
-\rho_2 & \Delta_2 & -\rho_1 & -\Delta_1\\
\Delta_2 & \rho_2 & -\Delta_1 & \rho_1
\end{array}
\right)
\\
H_\Gamma^{\mbox{\scriptsize MF}} &=& \frac{i\Gamma}{2}
\left(
\begin{array}{cccc}
0 & 0 & -\rho_2 & \Delta_2\\
0&0& \Delta_2 & \rho_2\\
-\rho_2^* & \Delta_2^* &0&0\\
\Delta_2^* & \rho_2^* &0&0
\end{array}
\right)
\end{eqnarray}
Thus, the consistency equations are (consider bond $(i\in2, j\in3)$)
\begin{eqnarray}
s_3 + it^z_3 &=&
-\frac{J}{4}(2\rho_1+\rho_1^*) - \frac{iD}{2}\rho_1  - \frac{K}{4}\rho_1^*
\\
s_1+it^z_1 &=&
-\frac{J}{4}(2\Delta_1+\Delta_1^*) - \frac{iD}{2}\Delta_1 - \frac{K}{4}\Delta_1^*
\\
t^y_3+it^x_3 &=& \frac{J}{4}\rho_2^* + \frac{K}{4} \rho_2^* -\frac{i\Gamma}{2}\rho_2
\\
t^y_1+it^x_1 &=& -\frac{J}{4}\Delta_2^* -\frac{K}{4}\Delta_2^* + \frac{i\Gamma}{2}\Delta_2
\end{eqnarray}

\subsection{Static spin structure factor}
Here we compute the static spin structure factor
\begin{equation}
S(\mathbf{q}) = \frac{1}{2N_s} \sum_{ij} e^{i\mathbf{q}\cdot (\bbr_i-\bbr_j)} \langle \mathbf{S}_i \cdot \mathbf{S}_j\rangle.
\end{equation}
for the  ground state that could be induced by Heisenberg-DM interactions, see Fig.\ref{fig:sq}. The parameters are the same as in Fig. \ref{fig:fsballs} in the main text.
Compared with the magnetically ordered states \cite{Mizoguchi2016}, the structure factor does not have any peaks. Instead, it shows a smooth intensity that has a single minimum at the center for both isotropic and anisotropic states. The anisotropic state has a slightly shallower minimum at the center. Also, for the anisotropic spin liquid, the intensity  of $S(\mathbf{q})$ shows a tendancy towards anisotropic distributions.

\begin{thebibliography}{44}
\expandafter\ifx\csname natexlab\endcsname\relax\def\natexlab#1{#1}\fi
\expandafter\ifx\csname bibnamefont\endcsname\relax
  \def\bibnamefont#1{#1}\fi
\expandafter\ifx\csname bibfnamefont\endcsname\relax
  \def\bibfnamefont#1{#1}\fi
\expandafter\ifx\csname citenamefont\endcsname\relax
  \def\citenamefont#1{#1}\fi
\expandafter\ifx\csname url\endcsname\relax
  \def\url#1{\texttt{#1}}\fi
\expandafter\ifx\csname urlprefix\endcsname\relax\def\urlprefix{URL }\fi
\providecommand{\bibinfo}[2]{#2}
\providecommand{\eprint}[2][]{\url{#2}}

\bibitem[{\citenamefont{Balents}(2010)}]{Balents2010}
\bibinfo{author}{\bibfnamefont{L.}~\bibnamefont{Balents}},
  \emph{\bibinfo{title}{Spin liquids in frustrated magnets}},
  \bibinfo{journal}{Nature} \textbf{\bibinfo{volume}{464}},
  \bibinfo{pages}{199} (\bibinfo{year}{2010}).

\bibitem[{\citenamefont{Lee}(2014)}]{Lee2014a}
\bibinfo{author}{\bibfnamefont{P.~A.} \bibnamefont{Lee}},
  \emph{\bibinfo{title}{Quantum spin liquid: a tale of emergence from
  frustration}}, \bibinfo{journal}{Journal of Physics: Conference Series}
  \textbf{\bibinfo{volume}{529}}, \bibinfo{pages}{012001}
  (\bibinfo{year}{2014}).

\bibitem[{\citenamefont{{Savary} and {Balents}}(2016)}]{Savary2016}
\bibinfo{author}{\bibfnamefont{L.}~\bibnamefont{{Savary}}} \bibnamefont{and}
  \bibinfo{author}{\bibfnamefont{L.}~\bibnamefont{{Balents}}},
  \emph{\bibinfo{title}{{Quantum Spin Liquids}}}, \bibinfo{journal}{ArXiv
  e-prints} \eprint{1601.03742} (\bibinfo{year}{2016}).

\bibitem[{\citenamefont{{Zhou} et~al.}(2016)\citenamefont{{Zhou}, {Kanoda}, and
  {Ng}}}]{Zhou2016}
\bibinfo{author}{\bibfnamefont{Y.}~\bibnamefont{{Zhou}}},
  \bibinfo{author}{\bibfnamefont{K.}~\bibnamefont{{Kanoda}}}, \bibnamefont{and}
  \bibinfo{author}{\bibfnamefont{T.-K.} \bibnamefont{{Ng}}},
  \emph{\bibinfo{title}{{Quantum Spin Liquid States}}}, \bibinfo{journal}{ArXiv
  e-prints} \eprint{1607.03228} (\bibinfo{year}{2016}).

\bibitem[{\citenamefont{Okamoto et~al.}(2007)\citenamefont{Okamoto, Nohara,
  Aruga-Katori, and Takagi}}]{Okamoto2007}
\bibinfo{author}{\bibfnamefont{Y.}~\bibnamefont{Okamoto}},
  \bibinfo{author}{\bibfnamefont{M.}~\bibnamefont{Nohara}},
  \bibinfo{author}{\bibfnamefont{H.}~\bibnamefont{Aruga-Katori}},
  \bibnamefont{and} \bibinfo{author}{\bibfnamefont{H.}~\bibnamefont{Takagi}},
  \emph{\bibinfo{title}{Spin-Liquid State in the S=1/2 Hyperkagome
  Antiferromagnet $Na_{4}Ir_{3}O_{8}$}}, \bibinfo{journal}{Phys. Rev. Lett.}
  \textbf{\bibinfo{volume}{99}}, \bibinfo{pages}{137207}
  (\bibinfo{year}{2007}).

\bibitem[{\citenamefont{Singh et~al.}(2013)\citenamefont{Singh, Tokiwa, Dong,
  and Gegenwart}}]{Singh2013}
\bibinfo{author}{\bibfnamefont{Y.}~\bibnamefont{Singh}},
  \bibinfo{author}{\bibfnamefont{Y.}~\bibnamefont{Tokiwa}},
  \bibinfo{author}{\bibfnamefont{J.}~\bibnamefont{Dong}}, \bibnamefont{and}
  \bibinfo{author}{\bibfnamefont{P.}~\bibnamefont{Gegenwart}},
  \emph{\bibinfo{title}{Spin liquid close to a quantum critical point in
  Na${}_{4}$Ir${}_{3}$O${}_{8}$}}, \bibinfo{journal}{Phys. Rev. B}
  \textbf{\bibinfo{volume}{88}}, \bibinfo{pages}{220413}
  (\bibinfo{year}{2013}).

\bibitem[{\citenamefont{Hopkinson et~al.}(2007)\citenamefont{Hopkinson, Isakov,
  Kee, and Kim}}]{Hopkinson2007}
\bibinfo{author}{\bibfnamefont{J.~M.} \bibnamefont{Hopkinson}},
  \bibinfo{author}{\bibfnamefont{S.~V.} \bibnamefont{Isakov}},
  \bibinfo{author}{\bibfnamefont{H.-Y.} \bibnamefont{Kee}}, \bibnamefont{and}
  \bibinfo{author}{\bibfnamefont{Y.~B.} \bibnamefont{Kim}},
  \emph{\bibinfo{title}{Classical Antiferromagnet on a Hyperkagome Lattice}},
  \bibinfo{journal}{Phys. Rev. Lett.} \textbf{\bibinfo{volume}{99}},
  \bibinfo{pages}{037201} (\bibinfo{year}{2007}).

\bibitem[{\citenamefont{Lawler et~al.}(2008{\natexlab{a}})\citenamefont{Lawler,
  Kee, Kim, and Vishwanath}}]{Lawler2008a}
\bibinfo{author}{\bibfnamefont{M.~J.} \bibnamefont{Lawler}},
  \bibinfo{author}{\bibfnamefont{H.-Y.} \bibnamefont{Kee}},
  \bibinfo{author}{\bibfnamefont{Y.~B.} \bibnamefont{Kim}}, \bibnamefont{and}
  \bibinfo{author}{\bibfnamefont{A.}~\bibnamefont{Vishwanath}},
  \emph{\bibinfo{title}{Topological Spin Liquid on the Hyperkagome Lattice of
  ${\mathrm{Na}}_{4}{\mathrm{Ir}}_{3}{\mathrm{O}}_{8}$}},
  \bibinfo{journal}{Phys. Rev. Lett.} \textbf{\bibinfo{volume}{100}},
  \bibinfo{pages}{227201} (\bibinfo{year}{2008}{\natexlab{a}}).

\bibitem[{\citenamefont{Zhou et~al.}(2008)\citenamefont{Zhou, Lee, Ng, and
  Zhang}}]{Zhou2008a}
\bibinfo{author}{\bibfnamefont{Y.}~\bibnamefont{Zhou}},
  \bibinfo{author}{\bibfnamefont{P.~A.} \bibnamefont{Lee}},
  \bibinfo{author}{\bibfnamefont{T.-K.} \bibnamefont{Ng}}, \bibnamefont{and}
  \bibinfo{author}{\bibfnamefont{F.-C.} \bibnamefont{Zhang}},
  \emph{\bibinfo{title}{Na$_4$Ir$_3$O$_8$ as a 3D Spin Liquid with Fermionic
  Spinons}}, \bibinfo{journal}{Phys. Rev. Lett.}
  \textbf{\bibinfo{volume}{101}}, \bibinfo{pages}{197201}
  (\bibinfo{year}{2008}).

\bibitem[{\citenamefont{Lawler et~al.}(2008{\natexlab{b}})\citenamefont{Lawler,
  Paramekanti, Kim, and Balents}}]{Lawler2008}
\bibinfo{author}{\bibfnamefont{M.~J.} \bibnamefont{Lawler}},
  \bibinfo{author}{\bibfnamefont{A.}~\bibnamefont{Paramekanti}},
  \bibinfo{author}{\bibfnamefont{Y.~B.} \bibnamefont{Kim}}, \bibnamefont{and}
  \bibinfo{author}{\bibfnamefont{L.}~\bibnamefont{Balents}},
  \emph{\bibinfo{title}{Gapless Spin Liquids on the Three-Dimensional
  Hyperkagome Lattice of Na$_4$Ir$_3$O$_8$}}, \bibinfo{journal}{Phys. Rev.
  Lett.} \textbf{\bibinfo{volume}{101}}, \bibinfo{pages}{197202}
  (\bibinfo{year}{2008}{\natexlab{b}}).

\bibitem[{\citenamefont{Chen and Kim}(2013)}]{Chen2013c}
\bibinfo{author}{\bibfnamefont{G.}~\bibnamefont{Chen}} \bibnamefont{and}
  \bibinfo{author}{\bibfnamefont{Y.~B.} \bibnamefont{Kim}},
  \emph{\bibinfo{title}{Anomalous enhancement of the Wilson ratio in a quantum
  spin liquid: The case of Na${}_{4}$Ir${}_{3}$O${}_{8}$}},
  \bibinfo{journal}{Phys. Rev. B} \textbf{\bibinfo{volume}{87}},
  \bibinfo{pages}{165120} (\bibinfo{year}{2013}).

\bibitem[{\citenamefont{Singh and Oitmaa}(2012)}]{Singh2012}
\bibinfo{author}{\bibfnamefont{R.~R.~P.} \bibnamefont{Singh}} \bibnamefont{and}
  \bibinfo{author}{\bibfnamefont{J.}~\bibnamefont{Oitmaa}},
  \emph{\bibinfo{title}{High-temperature series expansion study of the
  Heisenberg antiferromagnet on the hyperkagome lattice: Comparison with
  Na${}_{4}$Ir${}_{3}$O${}_{8}$}}, \bibinfo{journal}{Phys. Rev. B}
  \textbf{\bibinfo{volume}{85}}, \bibinfo{pages}{104406}
  (\bibinfo{year}{2012}).

\bibitem[{\citenamefont{Buessen and Trebst}(2016)}]{Buessen2016}
\bibinfo{author}{\bibfnamefont{F.~L.} \bibnamefont{Buessen}} \bibnamefont{and}
  \bibinfo{author}{\bibfnamefont{S.}~\bibnamefont{Trebst}},
  \emph{\bibinfo{title}{Competing magnetic orders and spin liquids in two- and
  three-dimensional kagome systems: Pseudofermion functional renormalization
  group perspective}}, \bibinfo{journal}{Phys. Rev. B}
  \textbf{\bibinfo{volume}{94}}, \bibinfo{pages}{235138}
  (\bibinfo{year}{2016}).

\bibitem[{\citenamefont{Chen and Balents}(2008)}]{Chen2008}
\bibinfo{author}{\bibfnamefont{G.}~\bibnamefont{Chen}} \bibnamefont{and}
  \bibinfo{author}{\bibfnamefont{L.}~\bibnamefont{Balents}},
  \emph{\bibinfo{title}{Spin-orbit effects in Na$_4$Ir$_3$O$_8$: A hyper-kagome
  lattice antiferromagnet}}, \bibinfo{journal}{Phys. Rev. B}
  \textbf{\bibinfo{volume}{78}}, \bibinfo{pages}{094403}
  (\bibinfo{year}{2008}).

\bibitem[{\citenamefont{Dally et~al.}(2014)\citenamefont{Dally, Hogan, Amato,
  Luetkens, Baines, Rodriguez-Rivera, Graf, and Wilson}}]{Dally2014}
\bibinfo{author}{\bibfnamefont{R.}~\bibnamefont{Dally}},
  \bibinfo{author}{\bibfnamefont{T.}~\bibnamefont{Hogan}},
  \bibinfo{author}{\bibfnamefont{A.}~\bibnamefont{Amato}},
  \bibinfo{author}{\bibfnamefont{H.}~\bibnamefont{Luetkens}},
  \bibinfo{author}{\bibfnamefont{C.}~\bibnamefont{Baines}},
  \bibinfo{author}{\bibfnamefont{J.}~\bibnamefont{Rodriguez-Rivera}},
  \bibinfo{author}{\bibfnamefont{M.~J.} \bibnamefont{Graf}}, \bibnamefont{and}
  \bibinfo{author}{\bibfnamefont{S.~D.} \bibnamefont{Wilson}},
  \emph{\bibinfo{title}{Short-Range Correlations in the Magnetic Ground State
  of Na$_4$Ir$_3$O$_8$}}, \bibinfo{journal}{Phys. Rev. Lett.}
  \textbf{\bibinfo{volume}{113}}, \bibinfo{pages}{247601}
  (\bibinfo{year}{2014}).

\bibitem[{\citenamefont{Shockley et~al.}(2015)\citenamefont{Shockley, Bert,
  Orain, Okamoto, and Mendels}}]{Shockley2015}
\bibinfo{author}{\bibfnamefont{A.~C.} \bibnamefont{Shockley}},
  \bibinfo{author}{\bibfnamefont{F.}~\bibnamefont{Bert}},
  \bibinfo{author}{\bibfnamefont{J.-C.} \bibnamefont{Orain}},
  \bibinfo{author}{\bibfnamefont{Y.}~\bibnamefont{Okamoto}}, \bibnamefont{and}
  \bibinfo{author}{\bibfnamefont{P.}~\bibnamefont{Mendels}},
  \emph{\bibinfo{title}{Frozen State and Spin Liquid Physics in
  Na$_4$Ir$_3$O$_8$: An NMR Study}}, \bibinfo{journal}{Phys. Rev. Lett.}
  \textbf{\bibinfo{volume}{115}}, \bibinfo{pages}{047201}
  (\bibinfo{year}{2015}).

\bibitem[{\citenamefont{Shindou}(2016)}]{Shindou2016}
\bibinfo{author}{\bibfnamefont{R.}~\bibnamefont{Shindou}},
  \emph{\bibinfo{title}{Nature of the possible magnetic phases in a frustrated
  hyperkagome iridate}}, \bibinfo{journal}{Phys. Rev. B}
  \textbf{\bibinfo{volume}{93}}, \bibinfo{pages}{094419}
  (\bibinfo{year}{2016}).

\bibitem[{\citenamefont{Mizoguchi et~al.}(2016)\citenamefont{Mizoguchi, Hwang,
  Lee, and Kim}}]{Mizoguchi2016}
\bibinfo{author}{\bibfnamefont{T.}~\bibnamefont{Mizoguchi}},
  \bibinfo{author}{\bibfnamefont{K.}~\bibnamefont{Hwang}},
  \bibinfo{author}{\bibfnamefont{E.~K.-H.} \bibnamefont{Lee}},
  \bibnamefont{and} \bibinfo{author}{\bibfnamefont{Y.~B.} \bibnamefont{Kim}},
  \emph{\bibinfo{title}{Generic model for the hyperkagome iridate
  ${\text{Na}}_{4}{\text{Ir}}_{3}{\text{O}}_{8}$ in the local-moment regime}},
  \bibinfo{journal}{Phys. Rev. B} \textbf{\bibinfo{volume}{94}},
  \bibinfo{pages}{064416} (\bibinfo{year}{2016}).

\bibitem[{\citenamefont{You et~al.}(2012)\citenamefont{You, Kimchi, and
  Vishwanath}}]{You2012a}
\bibinfo{author}{\bibfnamefont{Y.-Z.} \bibnamefont{You}},
  \bibinfo{author}{\bibfnamefont{I.}~\bibnamefont{Kimchi}}, \bibnamefont{and}
  \bibinfo{author}{\bibfnamefont{A.}~\bibnamefont{Vishwanath}},
  \emph{\bibinfo{title}{Doping a spin-orbit Mott insulator: Topological
  superconductivity from the Kitaev-Heisenberg model and possible application
  to (Na${}_{2}$/Li${}_{2}$)IrO${}_{3}$}}, \bibinfo{journal}{Phys. Rev. B}
  \textbf{\bibinfo{volume}{86}}, \bibinfo{pages}{085145}
  (\bibinfo{year}{2012}).

\bibitem[{\citenamefont{Shindou et~al.}(2011)\citenamefont{Shindou, Yunoki, and
  Momoi}}]{Shindou2011}
\bibinfo{author}{\bibfnamefont{R.}~\bibnamefont{Shindou}},
  \bibinfo{author}{\bibfnamefont{S.}~\bibnamefont{Yunoki}}, \bibnamefont{and}
  \bibinfo{author}{\bibfnamefont{T.}~\bibnamefont{Momoi}},
  \emph{\bibinfo{title}{Projective studies of spin nematics in a quantum
  frustrated ferromagnet}}, \bibinfo{journal}{Phys. Rev. B}
  \textbf{\bibinfo{volume}{84}}, \bibinfo{pages}{134414}
  (\bibinfo{year}{2011}).

\bibitem[{\citenamefont{Dodds et~al.}(2013)\citenamefont{Dodds, Bhattacharjee,
  and Kim}}]{Dodds2013}
\bibinfo{author}{\bibfnamefont{T.}~\bibnamefont{Dodds}},
  \bibinfo{author}{\bibfnamefont{S.}~\bibnamefont{Bhattacharjee}},
  \bibnamefont{and} \bibinfo{author}{\bibfnamefont{Y.~B.} \bibnamefont{Kim}},
  \emph{\bibinfo{title}{Quantum spin liquids in the absence of spin-rotation
  symmetry: Application to herbertsmithite}}, \bibinfo{journal}{Phys. Rev. B}
  \textbf{\bibinfo{volume}{88}}, \bibinfo{pages}{224413}
  (\bibinfo{year}{2013}).

\bibitem[{\citenamefont{Parameswaran et~al.}(2013)\citenamefont{Parameswaran,
  Turner, Arovas, and Vishwanath}}]{Parameswaran2013}
\bibinfo{author}{\bibfnamefont{S.~A.} \bibnamefont{Parameswaran}},
  \bibinfo{author}{\bibfnamefont{A.~M.} \bibnamefont{Turner}},
  \bibinfo{author}{\bibfnamefont{D.~P.} \bibnamefont{Arovas}},
  \bibnamefont{and}
  \bibinfo{author}{\bibfnamefont{A.}~\bibnamefont{Vishwanath}},
  \emph{\bibinfo{title}{Topological order and absence of band insulators at
  integer filling in non-symmorphic crystals}}, \bibinfo{journal}{Nat Phys}
  \textbf{\bibinfo{volume}{9}}, \bibinfo{pages}{299} (\bibinfo{year}{2013}).

\bibitem[{\citenamefont{Watanabe et~al.}(2015)\citenamefont{Watanabe, Po,
  Vishwanath, and Zaletel}}]{Watanabe2015}
\bibinfo{author}{\bibfnamefont{H.}~\bibnamefont{Watanabe}},
  \bibinfo{author}{\bibfnamefont{H.~C.} \bibnamefont{Po}},
  \bibinfo{author}{\bibfnamefont{A.}~\bibnamefont{Vishwanath}},
  \bibnamefont{and} \bibinfo{author}{\bibfnamefont{M.}~\bibnamefont{Zaletel}},
  \emph{\bibinfo{title}{Filling constraints for spin-orbit coupled insulators
  in symmorphic and nonsymmorphic crystals}}, \bibinfo{journal}{Proceedings of
  the National Academy of Sciences} \textbf{\bibinfo{volume}{112}},
  \bibinfo{pages}{14551} (\bibinfo{year}{2015}).

\bibitem[{\citenamefont{Lee et~al.}(2016)\citenamefont{Lee, Hermele, and
  Parameswaran}}]{Lee2016}
\bibinfo{author}{\bibfnamefont{S.}~\bibnamefont{Lee}},
  \bibinfo{author}{\bibfnamefont{M.}~\bibnamefont{Hermele}}, \bibnamefont{and}
  \bibinfo{author}{\bibfnamefont{S.~A.} \bibnamefont{Parameswaran}},
  \emph{\bibinfo{title}{Fractionalizing glide reflections in two-dimensional
  ${Z}_{2}$ topologically ordered phases}}, \bibinfo{journal}{Phys. Rev. B}
  \textbf{\bibinfo{volume}{94}}, \bibinfo{pages}{125122}
  (\bibinfo{year}{2016}).

\bibitem[{\citenamefont{{Lu}}(2016)}]{Lu2016b}
\bibinfo{author}{\bibfnamefont{Y.-M.} \bibnamefont{{Lu}}},
  \emph{\bibinfo{title}{{Symmetry protected gapless $Z\_2$ spin liquids}}},
  \bibinfo{journal}{ArXiv e-prints} \eprint{1606.05652} (\bibinfo{year}{2016}).

\bibitem[{\citenamefont{Wen}(2002)}]{Wen2002}
\bibinfo{author}{\bibfnamefont{X.-G.} \bibnamefont{Wen}},
  \emph{\bibinfo{title}{Quantum orders and symmetric spin liquids}},
  \bibinfo{journal}{Phys. Rev. B} \textbf{\bibinfo{volume}{65}},
  \bibinfo{pages}{165113} (\bibinfo{year}{2002}).

\bibitem[{\citenamefont{Koteswararao et~al.}(2014)\citenamefont{Koteswararao,
  Kumar, Khuntia, Bhowal, Panda, Rahman, Mahajan, Dasgupta, Baenitz, Kim
  et~al.}}]{Koteswararao2014}
\bibinfo{author}{\bibfnamefont{B.}~\bibnamefont{Koteswararao}},
  \bibinfo{author}{\bibfnamefont{R.}~\bibnamefont{Kumar}},
  \bibinfo{author}{\bibfnamefont{P.}~\bibnamefont{Khuntia}},
  \bibinfo{author}{\bibfnamefont{S.}~\bibnamefont{Bhowal}},
  \bibinfo{author}{\bibfnamefont{S.~K.} \bibnamefont{Panda}},
  \bibinfo{author}{\bibfnamefont{M.~R.} \bibnamefont{Rahman}},
  \bibinfo{author}{\bibfnamefont{A.~V.} \bibnamefont{Mahajan}},
  \bibinfo{author}{\bibfnamefont{I.}~\bibnamefont{Dasgupta}},
  \bibinfo{author}{\bibfnamefont{M.}~\bibnamefont{Baenitz}},
  \bibinfo{author}{\bibfnamefont{K.~H.} \bibnamefont{Kim}},
  \bibnamefont{et~al.}, \emph{\bibinfo{title}{Magnetic properties and heat
  capacity of the three-dimensional frustrated $S=\frac{1}{2}$ antiferromagnet
  PbCuTe$_2$O$_6$}}, \bibinfo{journal}{Phys. Rev. B}
  \textbf{\bibinfo{volume}{90}}, \bibinfo{pages}{035141}
  (\bibinfo{year}{2014}).

\bibitem[{\citenamefont{Khuntia et~al.}(2016)\citenamefont{Khuntia, Bert,
  Mendels, Koteswararao, Mahajan, Baenitz, Chou, Baines, Amato, and
  Furukawa}}]{Khuntia2016}
\bibinfo{author}{\bibfnamefont{P.}~\bibnamefont{Khuntia}},
  \bibinfo{author}{\bibfnamefont{F.}~\bibnamefont{Bert}},
  \bibinfo{author}{\bibfnamefont{P.}~\bibnamefont{Mendels}},
  \bibinfo{author}{\bibfnamefont{B.}~\bibnamefont{Koteswararao}},
  \bibinfo{author}{\bibfnamefont{A.~V.} \bibnamefont{Mahajan}},
  \bibinfo{author}{\bibfnamefont{M.}~\bibnamefont{Baenitz}},
  \bibinfo{author}{\bibfnamefont{F.~C.} \bibnamefont{Chou}},
  \bibinfo{author}{\bibfnamefont{C.}~\bibnamefont{Baines}},
  \bibinfo{author}{\bibfnamefont{A.}~\bibnamefont{Amato}}, \bibnamefont{and}
  \bibinfo{author}{\bibfnamefont{Y.}~\bibnamefont{Furukawa}},
  \emph{\bibinfo{title}{Spin Liquid State in the 3D Frustrated Antiferromagnet
  PbCuTe$_{2}${O}$_{6}$: NMR and Muon Spin Relaxation Studies}},
  \bibinfo{journal}{Phys. Rev. Lett.} \textbf{\bibinfo{volume}{116}},
  \bibinfo{pages}{107203} (\bibinfo{year}{2016}).

\bibitem[{\citenamefont{Singh and Huse}(2008)}]{Singh2008}
\bibinfo{author}{\bibfnamefont{R.~R.~P.} \bibnamefont{Singh}} \bibnamefont{and}
  \bibinfo{author}{\bibfnamefont{D.~A.} \bibnamefont{Huse}},
  \emph{\bibinfo{title}{Triplet and singlet excitations in the valence bond
  crystal phase of the kagome lattice Heisenberg model}},
  \bibinfo{journal}{Phys. Rev. B} \textbf{\bibinfo{volume}{77}},
  \bibinfo{pages}{144415} (\bibinfo{year}{2008}).

\bibitem[{\citenamefont{Reuther et~al.}(2014)\citenamefont{Reuther, Lee, and
  Alicea}}]{Reuther2014}
\bibinfo{author}{\bibfnamefont{J.}~\bibnamefont{Reuther}},
  \bibinfo{author}{\bibfnamefont{S.-P.} \bibnamefont{Lee}}, \bibnamefont{and}
  \bibinfo{author}{\bibfnamefont{J.}~\bibnamefont{Alicea}},
  \emph{\bibinfo{title}{Classification of spin liquids on the square lattice
  with strong spin-orbit coupling}}, \bibinfo{journal}{Phys. Rev. B}
  \textbf{\bibinfo{volume}{90}}, \bibinfo{pages}{174417}
  (\bibinfo{year}{2014}).

\bibitem[{\citenamefont{Essin and Hermele}(2013)}]{Essin2013}
\bibinfo{author}{\bibfnamefont{A.~M.} \bibnamefont{Essin}} \bibnamefont{and}
  \bibinfo{author}{\bibfnamefont{M.}~\bibnamefont{Hermele}},
  \emph{\bibinfo{title}{Classifying fractionalization: Symmetry classification
  of gapped $Z_{2}$ spin liquids in two dimensions}}, \bibinfo{journal}{Phys.
  Rev. B} \textbf{\bibinfo{volume}{87}}, \bibinfo{pages}{104406}
  (\bibinfo{year}{2013}).

\bibitem[{\citenamefont{{Barkeshli} et~al.}(2014)\citenamefont{{Barkeshli},
  {Bonderson}, {Cheng}, and {Wang}}}]{Barkeshli2014}
\bibinfo{author}{\bibfnamefont{M.}~\bibnamefont{{Barkeshli}}},
  \bibinfo{author}{\bibfnamefont{P.}~\bibnamefont{{Bonderson}}},
  \bibinfo{author}{\bibfnamefont{M.}~\bibnamefont{{Cheng}}}, \bibnamefont{and}
  \bibinfo{author}{\bibfnamefont{Z.}~\bibnamefont{{Wang}}},
  \emph{\bibinfo{title}{{Symmetry, Defects, and Gauging of Topological
  Phases}}}, \bibinfo{journal}{ArXiv e-prints} \eprint{1410.4540}
  (\bibinfo{year}{2014}).

\bibitem[{\citenamefont{Tarantino et~al.}(2016)\citenamefont{Tarantino,
  Lindner, and Fidkowski}}]{Tarantino2016}
\bibinfo{author}{\bibfnamefont{N.}~\bibnamefont{Tarantino}},
  \bibinfo{author}{\bibfnamefont{N.~H.} \bibnamefont{Lindner}},
  \bibnamefont{and}
  \bibinfo{author}{\bibfnamefont{L.}~\bibnamefont{Fidkowski}},
  \emph{\bibinfo{title}{Symmetry fractionalization and twist defects}},
  \bibinfo{journal}{New Journal of Physics} \textbf{\bibinfo{volume}{18}},
  \bibinfo{pages}{035006} (\bibinfo{year}{2016}).

\bibitem[{\citenamefont{O'Brien et~al.}(2016)\citenamefont{O'Brien, Hermanns,
  and Trebst}}]{OBrien2016}
\bibinfo{author}{\bibfnamefont{K.}~\bibnamefont{O'Brien}},
  \bibinfo{author}{\bibfnamefont{M.}~\bibnamefont{Hermanns}}, \bibnamefont{and}
  \bibinfo{author}{\bibfnamefont{S.}~\bibnamefont{Trebst}},
  \emph{\bibinfo{title}{Classification of gapless ${\mathbb{Z}}_{2}$ spin
  liquids in three-dimensional Kitaev models}}, \bibinfo{journal}{Phys. Rev. B}
  \textbf{\bibinfo{volume}{93}}, \bibinfo{pages}{085101}
  (\bibinfo{year}{2016}).

\bibitem[{\citenamefont{Kitaev}(2006)}]{Kitaev2006}
\bibinfo{author}{\bibfnamefont{A.}~\bibnamefont{Kitaev}},
  \emph{\bibinfo{title}{Anyons in an exactly solved model and beyond}},
  \bibinfo{journal}{Annals of Physics} \textbf{\bibinfo{volume}{321}},
  \bibinfo{pages}{2} (\bibinfo{year}{2006}).

\bibitem[{\citenamefont{Burnell and Nayak}(2011)}]{Burnell2011}
\bibinfo{author}{\bibfnamefont{F.~J.} \bibnamefont{Burnell}} \bibnamefont{and}
  \bibinfo{author}{\bibfnamefont{C.}~\bibnamefont{Nayak}},
  \emph{\bibinfo{title}{SU(2) slave fermion solution of the Kitaev honeycomb
  lattice model}}, \bibinfo{journal}{Phys. Rev. B}
  \textbf{\bibinfo{volume}{84}}, \bibinfo{pages}{125125}
  (\bibinfo{year}{2011}).

\bibitem[{\citenamefont{Chen et~al.}(2012)\citenamefont{Chen, Essin, and
  Hermele}}]{Chen2012a}
\bibinfo{author}{\bibfnamefont{G.}~\bibnamefont{Chen}},
  \bibinfo{author}{\bibfnamefont{A.}~\bibnamefont{Essin}}, \bibnamefont{and}
  \bibinfo{author}{\bibfnamefont{M.}~\bibnamefont{Hermele}},
  \emph{\bibinfo{title}{Majorana spin liquids and projective realization of
  SU(2) spin symmetry}}, \bibinfo{journal}{Phys. Rev. B}
  \textbf{\bibinfo{volume}{85}}, \bibinfo{pages}{094418}
  (\bibinfo{year}{2012}).

\bibitem[{\citenamefont{Fang and Fu}(2015)}]{Fang2015a}
\bibinfo{author}{\bibfnamefont{C.}~\bibnamefont{Fang}} \bibnamefont{and}
  \bibinfo{author}{\bibfnamefont{L.}~\bibnamefont{Fu}},
  \emph{\bibinfo{title}{New classes of three-dimensional topological
  crystalline insulators: Nonsymmorphic and magnetic}}, \bibinfo{journal}{Phys.
  Rev. B} \textbf{\bibinfo{volume}{91}}, \bibinfo{pages}{161105}
  (\bibinfo{year}{2015}).

\bibitem[{\citenamefont{Shiozaki et~al.}(2015)\citenamefont{Shiozaki, Sato, and
  Gomi}}]{Shiozaki2015}
\bibinfo{author}{\bibfnamefont{K.}~\bibnamefont{Shiozaki}},
  \bibinfo{author}{\bibfnamefont{M.}~\bibnamefont{Sato}}, \bibnamefont{and}
  \bibinfo{author}{\bibfnamefont{K.}~\bibnamefont{Gomi}},
  \emph{\bibinfo{title}{${Z}_{2}$ topology in nonsymmorphic crystalline
  insulators: M\"obius twist in surface states}}, \bibinfo{journal}{Phys. Rev.
  B} \textbf{\bibinfo{volume}{91}}, \bibinfo{pages}{155120}
  (\bibinfo{year}{2015}).

\bibitem[{\citenamefont{Alexandradinata
  et~al.}(2016)\citenamefont{Alexandradinata, Wang, and
  Bernevig}}]{Alexandradinata2016}
\bibinfo{author}{\bibfnamefont{A.}~\bibnamefont{Alexandradinata}},
  \bibinfo{author}{\bibfnamefont{Z.}~\bibnamefont{Wang}}, \bibnamefont{and}
  \bibinfo{author}{\bibfnamefont{B.~A.} \bibnamefont{Bernevig}},
  \emph{\bibinfo{title}{Topological Insulators from Group Cohomology}},
  \bibinfo{journal}{Phys. Rev. X} \textbf{\bibinfo{volume}{6}},
  \bibinfo{pages}{021008} (\bibinfo{year}{2016}).

\bibitem[{\citenamefont{{Varjas} et~al.}(2016)\citenamefont{{Varjas}, {de
  Juan}, and {Lu}}}]{Varjas2016}
\bibinfo{author}{\bibfnamefont{D.}~\bibnamefont{{Varjas}}},
  \bibinfo{author}{\bibfnamefont{F.}~\bibnamefont{{de Juan}}},
  \bibnamefont{and} \bibinfo{author}{\bibfnamefont{Y.-M.} \bibnamefont{{Lu}}},
  \emph{\bibinfo{title}{{Space group constraints on weak indices in topological
  insulators}}}, \bibinfo{journal}{ArXiv e-prints} \eprint{1603.04450}
  (\bibinfo{year}{2016}).

\bibitem[{\citenamefont{{Qi} and {Cheng}}(2016)}]{Qi2016}
\bibinfo{author}{\bibfnamefont{Y.}~\bibnamefont{{Qi}}} \bibnamefont{and}
  \bibinfo{author}{\bibfnamefont{M.}~\bibnamefont{{Cheng}}},
  \emph{\bibinfo{title}{{Classification of symmetry fractionalization in gapped
  $\mathbb Z\_2$ spin liquids}}}, \bibinfo{journal}{ArXiv e-prints}
  \eprint{1606.04544} (\bibinfo{year}{2016}).

\bibitem[{\citenamefont{Kimchi and Vishwanath}(2014)}]{Kimchi2014a}
\bibinfo{author}{\bibfnamefont{I.}~\bibnamefont{Kimchi}} \bibnamefont{and}
  \bibinfo{author}{\bibfnamefont{A.}~\bibnamefont{Vishwanath}},
  \emph{\bibinfo{title}{Kitaev-Heisenberg models for iridates on the
  triangular, hyperkagome, kagome, fcc, and pyrochlore lattices}},
  \bibinfo{journal}{Phys. Rev. B} \textbf{\bibinfo{volume}{89}},
  \bibinfo{pages}{014414} (\bibinfo{year}{2014}).

\bibitem[{\citenamefont{Micklitz and Norman}(2010)}]{Micklitz2010a}
\bibinfo{author}{\bibfnamefont{T.}~\bibnamefont{Micklitz}} \bibnamefont{and}
  \bibinfo{author}{\bibfnamefont{M.~R.} \bibnamefont{Norman}},
  \emph{\bibinfo{title}{Spin Hamiltonian of hyper-kagome
  ${\text{Na}}_{4}{\text{Ir}}_{3}{\text{O}}_{8}$}}, \bibinfo{journal}{Phys.
  Rev. B} \textbf{\bibinfo{volume}{81}}, \bibinfo{pages}{174417}
  (\bibinfo{year}{2010}).

\end{thebibliography}

\end{document}